\documentclass[11pt]{article}

\usepackage[margin=0.8in]{geometry}
\usepackage{float}
\usepackage{parskip}
\usepackage{graphicx}
\usepackage{hyperref}
\usepackage{amsmath}
\usepackage{amsfonts}
\usepackage{amsthm}
\usepackage{amssymb}
\usepackage{thmtools}
\usepackage{soul}
\usepackage{listings}
\usepackage{xcolor}
\usepackage{lipsum}

\DeclareRobustCommand{\bbone}{\text{\usefont{U}{bbold}{m}{n}1}}
\newcommand{\E}[1]{\mathbb{E}\left[#1\right]}
\newcommand{\EU}[1]{\mathbb{E}^\mathbb{U}\left[#1\right]}
\newcommand{\VU}[1]{\operatorname{Var}^\mathbb{U}\left[#1\right]}

\declaretheoremstyle[
    spaceabove=0.5\topsep,
    spacebelow=1\topsep,
    name=Proof,
    headfont=\itshape, 
    numbered=no,
    qed=\qedsymbol
]{proofsty}
\declaretheorem[style=proofsty]{myproof}

\theoremstyle{proposition}
\newtheorem{proposition}{Proposition}
\newtheorem*{remark}{Remark}

\newtheoremstyle{named}{}{}{\itshape}{}{\bfseries}{.}{.5em}{#3}
\theoremstyle{named}
\newtheorem*{namedthm}{}

\definecolor{codegreen}{rgb}{0,0.6,0}
\definecolor{codegray}{rgb}{0.5,0.5,0.5}
\definecolor{codepurple}{rgb}{0.58,0,0.82}
\definecolor{backcolour}{rgb}{0.95,0.95,0.92}

\lstdefinestyle{mystyle}{
    backgroundcolor=\color{backcolour},
    commentstyle=\color{codegreen},
    keywordstyle=\color{magenta},
    numberstyle=\tiny\color{codegray},
    stringstyle=\color{codepurple},
    basicstyle=\ttfamily\footnotesize,
    breakatwhitespace=false,         
    breaklines=true,                 
    captionpos=b,                    
    keepspaces=true,                 
    numbers=left,                    
    numbersep=5pt,                  
    showspaces=false,                
    showstringspaces=false,
    showtabs=false,                  
    tabsize=2
}

\title{Saddle-Point Approach to Large-Time Volatility Smile}
\author{Chun Yat Yeung\thanks{Department of IEOR, Columbia University, \href{mailto:cy2623@columbia.edu}{\texttt{cy2623@columbia.edu}}}, Ali Hirsa\thanks{Department of IEOR, Columbia University, \href{mailto:ah2347@columbia.edu}{\texttt{ah2347@columbia.edu}}}}
\date{}

\begin{document}

\maketitle

\begin{abstract}
\noindent
We extend upon the saddle-point equation presented in \cite{gat-convhes} to derive large-time model-implied volatility smiles, providing its theoretical foundation and studying its applications in classical models. As long as characteristic function fulfills a Lévy-type scaling behavior in large time, the approach allows us to study analytically the large-time smile behaviors under specific models, and moreover, to reach a very wide class of arbitrage-free model-inspired parametrizations, in the same manner as stochastic-volatility-inspired (SVI).
\end{abstract}

\tableofcontents

\section{Introduction} \label{sec:intro}

From \cite{gat-convhes}, the stochastic-volatility-inspired (SVI) parameterization of implied volatility surface was devised at Merrill Lynch in 1999, exhibiting two properties that led to its subsequent popularity with practitioners: (1) variance smile at each time slice becomes linear in the wings as log-strike $|k| \rightarrow \infty$, which is consistent with Lee's bound; (2) calendar arbitrage is easily removable via penalization procedure. It was later proved that SVI is the exact solution for implied variance under Heston model as time to expiry $T \rightarrow \infty$, and in the paper, a saddle-point equation was presented to verify the large-time Heston variance solution.

Here, we start with the saddle-point equation and provide its theoretical foundation, showing that variance smile obtained from the equation is well-defined and closely connected to moment properties of log-spot density. This observation leads us to a model-free at-the-money moment expansion of the large-time smile in log-strike $k$. We then study its applications in classical models, including Heston, variance-gamma (VG), bilateral-gamma (BG), CGMY and Merton-jump, deriving the model-implied variance smiles, from which arbitrage-free smile parametrizations may be motivated.

This paper is organized as follows. In sections 2 and 3, we derive the saddle-point equation for large-time implied variance under general characteristic function, and prove that the obtained variance solution is well-defined. In section 4, we consider an expansion to the smile in log-strike and represent the coefficients by moments of log-spot density under some Esscher measure. In section 5, for classical models, we give analytical or approximate formulas for the smiles, case by case. In section 6, we demonstrate and analyze the calibration of an arbitrage-free parametrization inspired from bilateral-gamma model. In section 7, we conclude.


\section{Saddle-Point Equation} \label{sec:sdlpt}

Consider a call option $C(K,T)$ with strike $K$ and time to expiry $T$. Spot price $S_t \in \mathbb{R}^+$ evolves according to some stochastic process under the risk-neutral measure $\mathbb{Q}$, initially set at $S_0$. All expectations without a superscript are with respect to $\mathbb{Q}$. We assume zero interest rate and dividend yield, or equivalently we work in a forward measure and model the forward price. Define log-spot $X_T = \log(S_T/S_0) \in \mathbb{R}$ and log-strike $k = \log(K/S_0) \in \mathbb{R}$. We have the martingale condition for log-spot: $\E{e^{X_T}} = \E{S_T/S_0} = 1$, and we consider processes whose density of $X_T$ has support over the whole real line i.e.\ terminal spot $S_T$ can end up anywhere in $\mathbb{R}^+$.

From \cite{gat-volsurf}, Lewis equation states that under characteristic function $\phi_T(u)=\E{e^{iuX_T}}, u \in \mathbb{R}$, the call price is given by
\begin{equation*} C(K,T) = S_0 - \frac{\sqrt{S_0K}}{2\pi} \int_{-\infty}^\infty \frac{du}{u^2+\frac{1}{4}} e^{-iuk} \phi_T\left(u-\frac{i}{2}\right) . \end{equation*} 

A Black-Scholes (BS) model with volatility $\sigma$ has characteristic function $\phi_T^{BS}(u) = e^{-\frac{1}{2}u(u+i)\sigma^2T}$ and call price function $C_{BS}$. For each strike $K$, we \textit{quote} $C(K,T)$ in $C_{BS}(K,T,\sigma)$, thus
\begin{equation} \int_{-\infty}^\infty \frac{du}{u^2+\frac{1}{4}} e^{-iuk} \phi_T\left(u-\frac{i}{2}\right) = \int_{-\infty}^\infty \frac{du}{u^2+\frac{1}{4}} e^{-iuk} e^{-\frac{1}{2}\left(u^2+\frac{1}{4}\right)\sigma^2T} \label{eq:lewis} \end{equation} 
where $\sigma = \sigma(k,T)$ is the \textit{implied volatility} of our interest.\footnote{Throughout this paper, we use $\sigma(k,T)$ as the Black-Scholes implied volatility at log-strike $k$ and time to expiry $T$. The volatility smile emerges because we are using a wrong (reference) model, the Black-Scholes, to \textit{quote} options prices driven by stochastic processes different from Black-Scholes diffusion.} From now on, we will omit its dependence on $T$ for simplicity.

Consider the class of models in which characteristic function $\phi_T(u-i/2) \sim e^{-\psi(u)T}$ for large $T$, a Lévy-type scaling behavior.\footnote{Physically this means, at large time, log-spot $X_T$ evolves like a Lévy process, so that time $T$ in characteristic exponent factors out.} By definition, this is satisfied by all Lévy processes, and some path-dependent processes that forget about its initial states over time e.g.\ Heston.

Define time-scaled log-strike $x=k/T \in \mathbb{R}$, abbreviated ``\textit{strike}'' below. As $x$ remains finite, this corresponds to the case of large strike and large time. Substituting the large-time $\phi_T$ into LHS we get
\begin{equation*} \int_{-\infty}^\infty \frac{du}{u^2+\frac{1}{4}} e^{-(iux+\psi(u))T} . \end{equation*} 

Using Taylor-expansion for the exponent around an arbitrary complex point $\tilde{u}$:
\begin{equation*} ix\tilde{u} + ix(u-\tilde{u}) + \psi(\tilde{u}) + \psi'(\tilde{u})(u-\tilde{u}) + \frac{\psi''(\tilde{u})}{2}(u-\tilde{u})^2 + O(u-\tilde{u})^3 . \end{equation*} 

Requiring $\psi'(\tilde{u}) = -ix$ (now keep in mind $\tilde{u}=\tilde{u}(x)$) to kill linear term, with $\tilde{u}$ a \textit{saddle-point} making the $u$-derivative of exponent $iux+\psi(u)$ vanish, LHS simplifies to
\begin{equation*} \int_{-\infty}^\infty \frac{du}{u^2+\frac{1}{4}} e^{-(i\tilde{u}x+\psi(\tilde{u}))T - \frac{\psi''(\tilde{u})T}{2}(u-\tilde{u})^2 - T \cdot O(u-\tilde{u})^3} \approx \frac{e^{-(i\tilde{u}x+\psi(\tilde{u}))T}}{\tilde{u}^2+\frac{1}{4}} \sqrt{\frac{2\pi}{\psi''(\tilde{u})T}} . \end{equation*} 

The approximation that $e^{-(iux+\psi(u))T}$ takes the form of a Gaussian in $u$ is valid when (1) $i\tilde{u}x+\psi(\tilde{u})$ is real, (2) $\psi''(\tilde{u})>0$, (3) $T$ is large, because for $u$ far away from $\tilde{u}$ and $T$ large, $e^{- T \cdot O(u-\tilde{u})^3} \rightarrow 0$ -- tails flatten to zero; for $u$ close to $\tilde{u}$, constant and quadratic term in $e^{-(i\tilde{u}x+\psi(\tilde{u}))T - \frac{\psi''(\tilde{u})T}{2}(u-\tilde{u})^2 - T \cdot O(u-\tilde{u})^3}$ dominate. This approximation is known as the Laplace method.

For Black-Scholes, $\psi_{BS}(u) = \frac{1}{2} \left(u^2+\frac{1}{4}\right) v$ where $v = \sigma^2$, and noting $\psi'_{BS}(u) = uv$ and $\psi''_{BS}(u) = v$, by solving $\psi'_{BS}(\tilde{u}_{BS}) = -ix$ we get
\begin{equation} \tilde{u}_{BS} = -\frac{ix}{v} . \label{eq:bsutilde} \end{equation}

Thus our saddle-point condition:
\begin{align*}
\frac{e^{-(i\tilde{u}x+\psi(\tilde{u}))T}}{\tilde{u}^2+\frac{1}{4}} \sqrt{\frac{2\pi}{\psi''(\tilde{u})T}} &\sim \frac{e^{-(i\tilde{u}_{BS}x+\psi_{BS}(\tilde{u}_{BS}))T}}{\tilde{u}_{BS}^2+\frac{1}{4}} \sqrt{\frac{2\pi}{\psi''_{BS}(\tilde{u}_{BS})T}} \\
&= \frac{e^{-\left(\frac{x^2}{v}+\frac{v}{2}\left(\frac{1}{4}-\left(\frac{x}{v}\right)^2\right)\right)T}}{\frac{1}{4}-\left(\frac{x}{v}\right)^2} \sqrt{\frac{2\pi}{vT}} \\
&\approx 4 \exp\left( -\left(\frac{v}{8}+\frac{x^2}{2v}\right)T \right) \sqrt{\frac{2\pi}{vT}} .
\end{align*} 

Constant terms are of similar orders which are dominated by $e^{-UT}$ for some quantity $U$ and we make exponents equal to get the \textbf{saddle-point equation}:
\begin{equation} \omega(x) \equiv i\tilde{u}(x) \cdot x + \psi(\tilde{u}(x)) = \frac{v(x)}{8}+\frac{x^2}{2v(x)} , \label{eq:sdlpt} \end{equation} 
a quadratic equation with \textit{model-specific} $\omega(x)$, with solution
\begin{equation} v(x) = 4 \left( \omega(x) + \left(\bbone_{x \in (x_-,x_+)}-\bbone_{x \in \mathbb{R}\backslash(x_-,x_+)}\right) \sqrt{\omega(x)^2 - \frac{x^2}{4}} \right) . \label{eq:vx0} \end{equation} 

Denote $|\bar{\omega}(x)| \equiv \sqrt{\omega(x)^2 - \frac{x^2}{4}}$, vanishing at $x_\pm$ which solve $\omega(x) = \pm x/2$, chosen to fulfill $\bar{\omega}((x_-,x_+)) < 0$ and $\bar{\omega}(\mathbb{R}\backslash(x_-,x_+)) > 0$. Note that $x_\pm$ also solve $v(x)=\pm2x$. Our variance smile may be rewritten as
\begin{equation} v(x) = 4 (\omega(x) - \bar{\omega}(x)) . \label{eq:vx1} \end{equation} 

As $\omega(x)$, $\bar{\omega}(x)$ carry the dimension of variance, we call them ``\textit{variance quantities}'', and we call $v(x)$, the central entity studied, ``\textit{variance smile}'', ``\textit{the smile}'', or loosely, ``\textit{the volatility smile}''.

We may express large-time asymptotic characteristic function in form $e^{-\psi(u)T}$ to reach a very wide class of model-inspired parametrizations.

\subsection*{In a Nutshell} \label{sec:nutshell}

Denote time-scaled log-strike $x=k/T$ and variance $v(x)$, implied from characteristic function $\phi_T(u)=\E{e^{iuX_T}}$ where log-spot $X_T=\log(S_T/S_0)$. Our saddle-point procedure reads:
\begin{enumerate}
\item evaluate characteristic function $\phi_T(u-i/2) \equiv e^{-\psi(u)T}$ to get $\psi(u)$;
\item compute saddle-point $\tilde{u}$ which fulfills $\psi'(\tilde{u}) = -ix$;
\item evaluate $\psi(\tilde{u})$ thus $\omega(x) \equiv i\tilde{u} \cdot x + \psi(\tilde{u})$ and $|\bar{\omega}(x)| \equiv \sqrt{\omega(x)^2 - \frac{x^2}{4}}$;
\item if $\bar{\omega}(x)$ has an analytic form, full smile is explicitly given by $v(x) = 4\left( \omega(x) - \bar{\omega}(x) \right)$; otherwise, implicitly given by equation $v(x)/8+x^2/2v(x)=\omega(x)$.
\end{enumerate}

\begin{remark}
In deriving the quadratic saddle-point equation, we matched the leading order term in $T$. We also present a sketch of another version of the equation including terms of order $T^{-1}$ and higher, still with $T$ large, but the solution is non-trivial. The series expansion to the first order in $T^{-1}$ that we obtain agrees with \cite{jac-asyhes}, whose result specializes to Heston model. See appendix \ref{apdx:sdl}, A Fuller Saddle-Point Equation, for the discussion.
\end{remark}

\begin{remark}
We also present an analog saddle-point equation for small time by a change of variable $u'=uT$ to the integral in Lewis equation, still assuming the Lévy-type scaling $\phi_T(u-i/2) \sim e^{-\psi(u)T}$. Surprisingly, the equations for small time and large time bear a similar structure, up to a time scaling. See appendix \ref{apdx:sdlsmallT}, Saddle-Point Equation for Small Time, for the discussion.
\end{remark}


\section{Variance Smile} \label{sec:varsmile}

Here we prove that the Gaussian approximation is valid and that the variance solution thus obtained is real and non-negative.

\subsection{Gaussian Approximation}

Our large-time Gaussian approximation in Lewis equation is valid for the following conditions: (1) $\omega(x) \equiv i\tilde{u}x+\psi(\tilde{u})$ is real, or equivalently $i\tilde{u}$ and $\psi(\tilde{u})$ are real, and (2) $\psi''(\tilde{u})>0$. We now prove that these are always true from cumulant properties.

\begin{proposition}
If log-spot $X_T$ has support over $\mathbb{R}$, then $i\tilde{u}$ and $\psi(\tilde{u})$ are real, and $\psi''(\tilde{u})>0$.
\end{proposition}

\begin{myproof}
In large time, recall our characteristic function
\begin{equation*} \phi_T\left(u-\frac{i}{2}\right) = \E{e^{\left(iu+\frac{1}{2}\right)X_T}} = e^{-\psi(u)T} \end{equation*}
thus $\psi(u)$ is a cumulant:
\begin{equation*} \psi(u) = - \frac{1}{T} \log \E{e^{\left(iu+\frac{1}{2}\right)X_T}} . \end{equation*}

Differentiating,
\begin{equation*} \psi'(u) = - \frac{1}{T} \frac{\E{iX_Te^{\left(iu+\frac{1}{2}\right)X_T}}}{\E{e^{\left(iu+\frac{1}{2}\right)X_T}}} . \end{equation*}

$\tilde{u}$ satisfies
\begin{equation*} \psi'(\tilde{u}) = - \frac{1}{T} \frac{\E{iX_Te^{\left(i\tilde{u}+\frac{1}{2}\right)X_T}}}{\E{e^{\left(i\tilde{u}+\frac{1}{2}\right)X_T}}} = -ix \end{equation*}
so
\begin{equation*} \E{X_Te^{\left(\hat{u}+\frac{1}{2}\right)X_T}} = k \E{e^{\left(\hat{u}+\frac{1}{2}\right)X_T}} \end{equation*}
where we define $\hat{u} \equiv i\tilde{u}$. This is an equation in $\hat{u}$, it begs the question does a real $\hat{u}$ exist and is it unique?

Rewrite this in a more insightful form: define \textit{Esscher measure} $\mathbb{U}$ with a change of measure
\begin{equation} \frac{d\mathbb{U}}{d\mathbb{Q}} = \frac{e^{\left(\hat{u}+\frac{1}{2}\right)X_T}}{\E{e^{\left(\hat{u}+\frac{1}{2}\right)X_T}}} \label{eq:esscher} \end{equation}
where $\mathbb{Q}$ is our risk-neutral measure which we have been working with. Then
\begin{equation} \EU{X_T} = k \label{eq:uhateq} \end{equation}
thus $\hat{u}(k)$ defines a measure under which expectation of log-spot is exactly log-strike. For $\mathbb{U}$ to be a properly defined measure, $\hat{u}$ has to be real -- does it exist and is it unique?

Intuitively, $\hat{u}$ translates log-spot density -- imagine Gaussian $f_{X_T}(x) \stackrel{\mathbb{Q}}{\sim} e^{-(x+1/2)^2/2}$, then under $\mathbb{U}$, $f_{X_T}(x) \stackrel{\mathbb{U}}{\sim} e^{-(x-\hat{u})^2/2}$ and tanslation of density corresponds to translation of mean, so we can always suitably choose (solve for) $\hat{u}$ s.t.\ mean exactly matches log-strike i.e.\ unique real $\hat{u}$ exists.

Now we make rigorous why $\hat{u}$, equivalently $\bar{u}=\hat{u}+1/2$, always exists and is unique. Rewrite the expectation equation as
\begin{equation*} k = \EU{X_T} = \frac{\E{X_Te^{\bar{u}X_T}}}{\E{e^{\bar{u}X_T}}} = \frac{\partial}{\partial\bar{u}} \log \E{e^{\bar{u}X_T}} . \end{equation*}

This is the first derivative of cumulant (under risk-neutral measure, not Esscher), which spans the support of $X_T \in \mathbb{R}$ (thus existence of at least a real root), and monotonically (thus one unique root $\hat{u}$), as its second derivative is variance -- always positive.

With real $\hat{u}$, $\mathbb{U}$ is a properly defined measure equivalent to $\mathbb{Q}$.

Now,
\begin{equation*} \psi(\tilde{u}) = - \frac{1}{T} \log \E{e^{\left(\hat{u}+\frac{1}{2}\right)X_T}} \in \mathbb{R} . \end{equation*}

For our saddle-point approximation to be valid, we also need $\psi''(\tilde{u}) > 0$:
\begin{align*}
\psi''(\tilde{u}) &= - \frac{1}{T} \left( - \frac{\E{X_T^2e^{\left(i\tilde{u}+\frac{1}{2}\right)X_T}}}{\E{e^{\left(i\tilde{u}+\frac{1}{2}\right)X_T}}} - \frac{\E{iX_Te^{\left(i\tilde{u}+\frac{1}{2}\right)X_T}}}{\left(\E{e^{\left(i\tilde{u}+\frac{1}{2}\right)X_T}}\right)^2} \E{iX_Te^{\left(i\tilde{u}+\frac{1}{2}\right)X_T}} \right) \\
&= \frac{1}{T} \left( \frac{\E{X_T^2e^{\left(i\tilde{u}+\frac{1}{2}\right)X_T}}}{\E{e^{\left(i\tilde{u}+\frac{1}{2}\right)X_T}}} - \left( \frac{\E{X_Te^{\left(i\tilde{u}+\frac{1}{2}\right)X_T}}}{\E{e^{\left(i\tilde{u}+\frac{1}{2}\right)X_T}}} \right)^2 \right) \\
&= \frac{1}{T} \left( \frac{\E{X_T^2e^{\left(\hat{u}+\frac{1}{2}\right)X_T}}}{\E{e^{\left(\hat{u}+\frac{1}{2}\right)X_T}}} - \left( \frac{\E{X_Te^{\left(\hat{u}+\frac{1}{2}\right)X_T}}}{\E{e^{\left(\hat{u}+\frac{1}{2}\right)X_T}}} \right)^2 \right) \\
&= \frac{1}{T} \left( \EU{X_T^2} - \left( \EU{X_T} \right)^2 \right) \\
&= \frac{\VU{X_T}}{T}
\end{align*}
i.e.\ $\psi''(\tilde{u})$ is log-spot variance under $\mathbb{U}$ -- as expected, because differentiated cumulant gives central moments.

Thus if $X_T$ is supported over $\mathbb{R}$, as long as we can write down the characteristic function, for large $T$, our saddle-point condition is always valid.
\end{myproof}

\begin{remark}
An Esscher transform modifies a density function $f(x)$ of random variable $X$ by an exponential $e^{hx}$ and here, we modify the risk-neutral measure $\mathbb{Q}$ of stochastic process $X_T$ by an exponential $e^{(\hat{u}+1/2)X_T}$, thus we name the resultant measure $\mathbb{U}$ ``Esscher measure''.
\end{remark}

\subsection{Existence of Variance Solution}

Our variance solution $v(x)$ is well-defined if $|\bar{\omega}(x)| \equiv \sqrt{\omega(x)^2 - \frac{x^2}{4}}$ is real and bounded above by $\omega(x)$.

\begin{proposition}
$\omega(x)$ tangentially touches $|x/2|$ at points $x_\pm$ -- equivalently $\omega(x) \mp x/2$ and its derivative vanish at some $x_\pm$ -- and its second derivative is positive everywhere. Thus, $v(x) \geq 0$ is well-defined.
\end{proposition}

\begin{myproof}
Consider first derivatives of $\psi(\tilde{u}(x))$ and $\hat{u}(x)$ wrt.\ strike $x$:
\begin{align*}
&\psi(\tilde{u}(x)) = - \frac{1}{T} \log \E{e^{\left(\hat{u}+\frac{1}{2}\right)X_T}} \Rightarrow \psi'(\tilde{u}(x)) = - \frac{1}{T} \frac{\E{X_Te^{\left(\hat{u}+\frac{1}{2}\right)X_T}}}{\E{e^{\left(\hat{u}+\frac{1}{2}\right)X_T}}} \hat{u}'(x) = -x\hat{u}'(x) \\
&k = \frac{\E{X_Te^{\left(\hat{u}+\frac{1}{2}\right)X_T}}}{\E{e^{\left(\hat{u}+\frac{1}{2}\right)X_T}}} \Rightarrow T = \hat{u}'(x) \EU{(X_T-k)^2} \Rightarrow \hat{u}'(x) = \frac{T}{\EU{(X_T-k)^2}} .
\end{align*}

We write
\begin{equation} \omega(x) \equiv x \hat{u}(x) + \psi(\tilde{u}(x)) \label{eq:omega} \end{equation}
thus derivatives 
\begin{align*}
\omega'(x) &= \hat{u}(x) + x\hat{u}'(x) + \psi'(\tilde{u}(x)) = \hat{u}(x) \\
\omega''(x) &= \hat{u}'(x) = \frac{T}{\EU{(X_T-k)^2}} > 0
\end{align*}
so second derivative is everywhere positive, and
\begin{equation*} \partial_x \left( \omega(x) \mp x/2 \right) = \hat{u}(x) \mp \frac{1}{2} \end{equation*}
which vanishes at some $x_\pm$,
\begin{equation} \hat{u}(x_\pm) = \pm \frac{1}{2} . \label{eq:xpm} \end{equation}

Does such $x_\pm$ always exist? Yes. Recall that $\hat{u}$ is roughly the log-spot density shift (via Esscher measure $\mathbb{U}$) s.t.\ expectation of log-spot $\EU{X_T}$ exactly matches log-strike $k$, equivalently $x$, thus $\hat{u}$ is monotonic increasing in $x$, spanning $\mathbb{R}$. We can always find such an $x_\pm$ s.t.\ $\hat{u}(x_\pm)$ exactly matches $\pm 1/2$.

Lastly, $\omega(x) \mp x/2$ vanishes at $x_\pm$ as
\begin{equation*} \omega(x_\pm) \mp \frac{x_\pm}{2} = x_\pm \hat{u}(x_\pm) - \frac{1}{T} \log \E{e^{\left(\hat{u}(x_\pm)+\frac{1}{2}\right)X_T}} \mp \frac{x_\pm}{2} = \pm \frac{x_\pm}{2} \mp \frac{x_\pm}{2} - \frac{1}{T} \log \E{e^{\left(\pm\frac{1}{2}+\frac{1}{2}\right)X_T}} = 0 \end{equation*}
as $\E{e^{X_T}} = 1$ by martingale condition.

So, $\omega(x) \geq |x/2|$ for all strikes $x$, due to convexity, and $|\bar{\omega}(x)| \equiv \sqrt{\omega(x)^2-x^2/4}$ is real and bounded above by $\omega(x)$. Our variance smile 
\begin{equation*} v(x) = 4 (\omega(x) - \bar{\omega}(x)) \geq 0 \end{equation*}
is always well-defined.
\end{myproof}

\begin{remark}
Readers may refer to numerical experiments of Heston and variance-gamma, resp.\ sections \ref{sec:hesnum} and \ref{sec:vgnum}, for graphical illustrations of this proposition.
\end{remark}


\section{Moment Expansion}

We show that \textit{derivatives of $\psi(\tilde{u})$ and $\bar{u}$ are connected to central moments of log-spot under Esscher measure}, from which model-free moment expansions of variance quantity $\omega T$ and variance smile $v$ are obtained.\footnote{Here we have a slight abuse of notation with $f(x)$ and $f(k)$ meaning quantity $f$ \textit{written} as a function of resp.\ $x$ and $k$.}

Define
\begin{equation} \bar{u} = \hat{u} + \frac{1}{2} . \label{eq:ubar} \end{equation}
In terms of $\bar{u}$, we rewrite our cumulant
\begin{equation} \psi(\tilde{u}) = - \frac{1}{T} \log \E{e^{\bar{u}X_T}} , \label{eq:psi} \end{equation}
and saddle-point equation as follows:
\begin{equation} k = \E{ X_T \frac{e^{\bar{u}X_T}}{\E{e^{\bar{u}X_T}}} } . \label{eq:ubareq} \end{equation}

First consider derivatives of $\psi(\tilde{u})$:
\begin{align*}
\partial_x \psi(\tilde{u}) &= - \frac{1}{T} \frac{\E{X_Te^{\bar{u}X_T}}}{\E{e^{\bar{u}X_T}}} \bar{u}' = -x\bar{u}' \\
\Rightarrow \partial_x^{n+1} \psi(\tilde{u}) &= - \partial_x^n (x\bar{u}') = -n\bar{u}^{(n)}-x\bar{u}^{(n+1)} .
\end{align*}

As long as we know derivatives $\partial_x^n \bar{u}$, we can compute all derivatives of $\psi(\tilde{u})$. Equivalently we compute $\partial_k^n \bar{u}$, with strike derivatives related by $\partial_x = T\partial_k$. 

Next consider derivatives of $\bar{u}(k)$. Differentiate saddle-point equation to get
\begin{equation*} \partial_k^n k = \E{ X_T \partial_k^n \frac{e^{\bar{u}(k)X_T}}{\E{e^{\bar{u}(k)X_T}}} } \end{equation*}
where $k$-dependence is stressed. Note when $n = 0$, we have $k = \EU{X_T}$.

For $n = 1$ and $n = 2$ we do explicit calculations. For $n = 1$,
\begin{align*}
1 &= \E{ X_T \partial_k \frac{e^{\bar{u}X_T}}{\E{e^{\bar{u}X_T}}} } \\
&= \E{ X_T \left( \frac{X_Te^{\bar{u}X_T}\bar{u}'}{\E{e^{\bar{u}X_T}}} - \frac{e^{\bar{u}X_T}}{(\E{e^{\bar{u}X_T}})^2} \E{X_Te^{\bar{u}X_T}}\bar{u}' \right) } \\
&= \bar{u}' \left( \EU{X_T^2} - (\EU{X_T})^2 \right) \\
&= \bar{u}' \EU{(X_T-k)^2}
\end{align*}
so
\begin{equation*} \bar{u}'(k) = \frac{1}{\EU{(X_T-k)^2}} . \end{equation*}

For $n = 2$,
\begin{align*}
0 &= \E{ X_T \partial_k^2 \frac{e^{\bar{u}X_T}}{\E{e^{\bar{u}X_T}}} } \\
&= \E{ X_T \left( \frac{\partial_k^2 e^{\bar{u}X_T}}{\E{e^{\bar{u}X_T}}} + 2 \partial_k e^{\bar{u}X_T} \partial_k \frac{1}{\E{e^{\bar{u}X_T}}} + e^{\bar{u}X_T} \partial_k^2 \frac{1}{\E{e^{\bar{u}X_T}}} \right) }
\end{align*}
which we have
\begin{align*}
\partial_k e^{\bar{u}X_T} &= X_T e^{\bar{u}X_T} \bar{u}' \\
\partial_k^2 e^{\bar{u}X_T} &= X_T^2 e^{\bar{u}X_T} (\bar{u}')^2 + X_T e^{\bar{u}X_T} \bar{u}'' \\
\partial_k \frac{1}{\E{e^{\bar{u}X_T}}} &= - \frac{\E{X_Te^{\bar{u}X_T}}}{(\E{e^{\bar{u}X_T}})^2} \bar{u}' \\
\partial_k^2 \frac{1}{\E{e^{\bar{u}X_T}}} &= \frac{2(\E{X_Te^{\bar{u}X_T}})^2}{(\E{e^{\bar{u}X_T}})^3} (\bar{u}')^2 - \frac{1}{(\E{e^{\bar{u}X_T}})^2} \left[ \E{X_T^2e^{\bar{u}X_T}}(\bar{u}')^2 + \E{X_T e^{\bar{u}X_T}} \bar{u}'' \right] .
\end{align*}

Combining,
\begin{align*}
0 &= \E{ X_T \left( \frac{\partial_k^2 e^{\bar{u}X_T}}{
\E{e^{\bar{u}X_T}}} + 2 \partial_k e^{\bar{u}X_T} \partial_k \frac{1}{\E{e^{\bar{u}X_T}}} + e^{\bar{u}X_T} \partial_k^2 \frac{1}{\E{e^{\bar{u}X_T}}} \right) } \\
&= \mathbb{E}\left[ X_T \left\{ \frac{X_T^2 e^{\bar{u}X_T} (\bar{u}')^2 + X_T e^{\bar{u}X_T} \bar{u}''}{\E{e^{\bar{u}X_T}}} + 2 \left(X_T e^{\bar{u}X_T} \bar{u}'\right) \left(- \frac{\E{X_Te^{\bar{u}X_T}}}{(\E{e^{\bar{u}X_T}})^2} \bar{u}'\right) + \right. \right. \\
&\quad \left. \left. e^{\bar{u}X_T} \left( \frac{2(\E{X_Te^{\bar{u}X_T}})^2}{(\E{e^{\bar{u}X_T}})^3} (\bar{u}')^2 - \frac{\E{X_T^2e^{\bar{u}X_T}}(\bar{u}')^2 + \E{X_T e^{\bar{u}X_T}} \bar{u}''}{(\E{e^{\bar{u}X_T}})^2} \right) \right\} \right] \\
&= \left( \EU{X_T^2} - (\EU{X_T})^2 \right) \bar{u}'' + \left( \EU{X_T}^3 - 3 \EU{X_T^2} \EU{X_T} + 2 (\EU{X_T})^3 \right) (\bar{u}')^2 \\
&= \EU{(X_T-k)^2} \bar{u}'' + \EU{(X_T-k)^3} (\bar{u}')^2 .
\end{align*}

Solving,
\begin{equation*} \bar{u}''(k) = -\frac{\EU{(X_T-k)^3}}{(\EU{(X_T-k)^2})^3} .\end{equation*}

Calculations for $n = 3$ are obtained similarly:
\begin{equation*} 0 = \EU{(X_T-k)^2} \bar{u}''' + 3 \EU{(X_T-k)^3} \bar{u}' \bar{u}'' + \left( \EU{(X_T-k)^4} - 3 (\EU{(X_T-k)^2})^2 \right) (\bar{u}')^3 \end{equation*}
solved to give
\begin{equation*} \bar{u}'''(k) = \frac{3 (\EU{(X_T-k)^2})^3 + 3 (\EU{(X_T-k)^3})^2 - \EU{(X_T-k)^2} \EU{(X_T-k)^4}}{(\EU{(X_T-k)^2})^5} . \end{equation*}

Casting back to $x$, we have
\begin{align*}
\bar{u}'(x) &= T \frac{1}{\EU{(X_T-k)^2}} \\
\bar{u}''(x) &= - T^2 \frac{\EU{(X_T-k)^3}}{(\EU{(X_T-k)^2})^3} \\
\bar{u}'''(x) &= T^3 \frac{3 (\EU{(X_T-k)^2})^3 + 3 (\EU{(X_T-k)^3})^2 - \EU{(X_T-k)^2} \EU{(X_T-k)^4}}{(\EU{(X_T-k)^2})^5} .
\end{align*}

Note that an Esscher measure depends on $\bar{u}$, which in turn is a function of $k$. At-the-money (ATM) where $k = 0$, choose $\bar{u}=\bar{u}_0$ s.t.\ $\EU{X_T}=0$ -- now $\mathbb{U}$ is the ATM Esscher measure, which we specialize to.

Thus a \textit{moment expansion of $\bar{u}$ to third order}
\begin{equation} \bar{u}(k) = \bar{u}_0 + \frac{1}{\EU{X_T^2}} k - \frac{1}{2} \frac{\EU{X_T^3}}{(\EU{X_T^2})^3} k^2 + \frac{1}{6} \frac{3 (\EU{X_T^2})^3 + 3 (\EU{X_T^3})^2 - \EU{X_T^2} \EU{X_T^4}}{(\EU{X_T^2})^5} k^3 ... \label{eq:ubarexp} \end{equation}

Differentiation of $\psi$ yields
\begin{align*}
\partial_x \psi(\tilde{u}_0) &= -x\bar{u}' = 0 \quad \text{saddle-point condition} \\
\partial_x^2 \psi(\tilde{u}_0) &= -\bar{u}'-x\bar{u}'' = -T\frac{1}{\EU{X_T^2}} \\
\partial_x^3 \psi(\tilde{u}_0) &= -2\bar{u}''-x\bar{u}''' = 2T^2\frac{\EU{X_T^3}}{(\EU{X_T^2})^3} \\
\partial_x^4 \psi(\tilde{u}_0) &= -3\bar{u}'''-x\bar{u}'''' = -3T^3\frac{3 (\EU{X_T^2})^3 + 3 (\EU{X_T^3})^2 - \EU{X_T^2} \EU{X_T^4}}{(\EU{X_T^2})^5} .
\end{align*}

At-the-money, we have
\begin{equation*} \psi(\tilde{u}_0) \equiv \frac{\psi_0}{T} = -\frac{1}{T}\log \E{e^{\bar{u}_0X_T}} . \end{equation*}

Thus a \textit{moment expansion of $\psi$ to forth order}
\begin{equation} \psi(k) = \frac{\psi_0}{T} - \frac{1}{2} \frac{1}{\EU{X_T^2}} \frac{k^2}{T} + \frac{1}{3} \frac{\EU{X_T^3}}{(\EU{X_T^2})^3} \frac{k^3}{T} - \frac{1}{8} \frac{3 (\EU{X_T^2})^3 + 3 (\EU{X_T^3})^2 - \EU{X_T^2} \EU{X_T^4}}{(\EU{X_T^2})^5} \frac{k^4}{T} ... \label{eq:psiexp} \end{equation}

\subsection{Interpretation of $\bar{u}$}

By our moment expansion for $\bar{u}$, for small $x_\pm$ and by (\ref{eq:xpm}),
\begin{equation*} \bar{u}(x_\pm) \approx \bar{u}_0 + \frac{T}{\EU{X_T^2}} x_\pm \equiv \frac{1}{2} \pm \frac{1}{2} \end{equation*}
which we approximate
\begin{equation} x_\pm \approx \frac{\EU{X_T^2}}{T} \left( \frac{1}{2} \pm \frac{1}{2} - \bar{u}_0 \right) \approx \pm \frac{\EU{X_T^2}}{2T} \sim \pm \frac{\sigma^2}{2} \label{eq:xpmapprox} \end{equation}
as $\bar{u}_0 \approx 1/2$, from Itô correction, and $\sigma$ is some characteristic volatility in model e.g.\ in Heston $\sigma$ is the long-run mean volatility $\sqrt{\bar{v}}$.

To reason about this, consider BS diffusion: log-spot $X_T = -\sigma^2T/2 + \sigma Z_T$ and denote total variance $w = \sigma^2T$ so density $f_{X_T}(x) \stackrel{\mathbb{Q}}{\sim} e^{-(x+w/2)^2/2w}$. Under ATM Esscher measure defined by $\bar{u}_0$, $f_{X_T}(x) \stackrel{\mathbb{U}}{\sim} e^{-(x+w/2-w\bar{u}_0)^2/2w}$. For $\EU{X_T} = 0$, we demand $\bar{u}_0=1/2$ -- this is exact. At-the-money, if the terminal log-spot density is not too far off from BS diffusion, approximation $\bar{u}_0 \approx 1/2$ is not unreasonable.

As shown in section \ref{sec:leemoment}, there is a one-to-one correspondence between $\hat{u}(\pm\infty)$, or equivalently $\bar{u}(\pm\infty)$, and call/put-wing skew (slope) $\beta_\pm$, by Lee's moment formula.

\subsection{Derivatives of $\bar{u}$}

For $n = 1$ and $n=2$, we have the following expressions respectively
\begin{align*}
1 &= \EU{(X_T-k)^2} \bar{u}' \\
0 &= \EU{(X_T-k)^2} \bar{u}'' + \EU{(X_T-k)^3} (\bar{u}')^2 .
\end{align*}

Note each term carries dimension $1/k^{n-1}$. To generalize this line, let us back out a bit and consider our saddle-point equation in an alternative form
\begin{equation*} \partial_k \log \E{e^{\bar{u}X_T}} = k \bar{u}' \end{equation*}
so its $n^{th}$ derivative
\begin{equation*} \partial_k^{n+1} \log \E{e^{\bar{u}X_T}} = \partial_k^n (k \bar{u}') = n \bar{u}^{(n)} + k \bar{u}^{(n+1)} , \quad n \geq 0 \end{equation*}
from which we recursively solve $\bar{u}^{(n)}$ in terms of $\bar{u}^{(j)}$, $j \leq n-1$, and collect its coefficients into central moments. Derivatives of cumulant can be computed from Faà di Bruno's formula for derivatives of function composition (of $\log \cdot$ and $\E{e^{\bar{u}(\cdot)X_T}}$).

\subsection{Moment Expansion of $\omega$}

Variance quantity $\omega$ in our smile expands as
\begin{align*}
\omega(k) &\equiv \hat{u}(k)\frac{k}{T} + \psi(k) \\
&= \frac{k}{T} \left( \bar{u}_0 - \frac{1}{2} + \frac{1}{\EU{X_T^2}} k - \frac{1}{2} \frac{\EU{X_T^3}}{(\EU{X_T^2})^3} k^2 + \frac{1}{6} \frac{3 (\EU{X_T^2})^3 + 3 (\EU{X_T^3})^2 - \EU{X_T^2} \EU{X_T^4}}{(\EU{X_T^2})^5} k^3 ... \right) + \\
&\quad \frac{\psi_0}{T} - \frac{1}{2} \frac{1}{\EU{X_T^2}} \frac{k^2}{T} + \frac{1}{3} \frac{\EU{X_T^3}}{(\EU{X_T^2})^3} \frac{k^3}{T} - \frac{1}{8} \frac{3 (\EU{X_T^2})^3 + 3 (\EU{X_T^3})^2 - \EU{X_T^2} \EU{X_T^4}}{(\EU{X_T^2})^5} \frac{k^4}{T} ...
\end{align*}
so
\begin{equation} \omega(k)T = \psi_0 + \left(\bar{u}_0 - \frac{1}{2}\right) k + \frac{1}{2} \frac{1}{\EU{X_T^2}} k^2 - \frac{1}{6} \frac{\EU{X_T^3}}{(\EU{X_T^2})^3} k^3 + \frac{1}{24} \frac{3 (\EU{X_T^2})^3 + 3 (\EU{X_T^3})^2 - \EU{X_T^2} \EU{X_T^4}}{(\EU{X_T^2})^5} k^4 ... \label{eq:omegaexp} \end{equation}

Denote $\sigma_0^2T \equiv \EU{X_T^2}$, and we rewrite the smile in terms of normalized strike
\begin{align*}
\omega(k)T &= \psi_0 + \left(\bar{u}_0 - \frac{1}{2}\right) \sigma_0\sqrt{T} \left(\frac{k}{\sigma_0\sqrt{T}}\right) + \frac{1}{2} \left(\frac{k}{\sigma_0\sqrt{T}}\right)^2 - \frac{1}{6} \frac{\EU{X_T^3}}{(\sigma_0\sqrt{T})^3} \left(\frac{k}{\sigma_0\sqrt{T}}\right)^3 + \\
&\quad \frac{1}{24} \frac{3 (\EU{X_T^2})^3 + 3 (\EU{X_T^3})^2 - \EU{X_T^2} \EU{X_T^4}}{(\sigma_0\sqrt{T})^6} \left(\frac{k}{\sigma_0\sqrt{T}}\right)^4 ...
\end{align*}
with dimensionless coefficients.

Typically, if model produces heavy left-tail, say under Heston model, the mode of log-spot density is biased to the right, so we require a smaller Esscher shift to match the expectation of log-spot to zero -- thus $\bar{u}_0 < 1/2$, a negatively skewed $\omega(k)T$.

\subsection{Moment Expansion of Smile $v$}

We now obtain a moment expansion of $\bar{\omega}(k)T$ hence smile $v(k)$.

We defined $|\bar{\omega}(x)| \equiv \sqrt{\omega(x)^2-x^2/4}$, and variance smile is given by 
\begin{equation*} v(x) \sim 4(\omega(x)-\bar{\omega}(x)) . \end{equation*}

Express total variance in terms of log-strike
\begin{equation*} w(k) \equiv v(k)T \sim 4(\omega(k)T-\bar{\omega}(k)T) \end{equation*}
with moment expansion
\begin{equation*} \omega(k)T = \psi_0 - \left(\frac{1}{2} - \bar{u}_0\right) k + \frac{1}{2} \frac{1}{\EU{X_T^2}} k^2 - \frac{1}{6} \frac{\EU{X_T^3}}{(\EU{X_T^2})^3} k^3 + \frac{1}{24} \frac{3 (\EU{X_T^2})^3 + 3 (\EU{X_T^3})^2 - \EU{X_T^2} \EU{X_T^4}}{(\EU{X_T^2})^5} k^4 ... \end{equation*}

$|\bar{\omega}(k)T|$ expands ATM as
\begin{align*}
|\bar{\omega}(k)T| &= -\bar{\omega}(k)T = \psi_0 - \left(\frac{1}{2} - \bar{u}_0\right) k + \frac{1}{2} \left(\frac{1}{\EU{X_T^2}}-\frac{1}{4\psi_0}\right) k^2 - \frac{1}{6} \left(\frac{\EU{X_T^3}}{(\EU{X_T^2})^3}-\frac{6\bar{u}_0-3}{8\psi_0^2}\right) k^3 + \\
&\quad \frac{1}{24} \left(\frac{3 (\EU{X_T^2})^3 + 3 (\EU{X_T^3})^2 - \EU{X_T^2} \EU{X_T^4}}{(\EU{X_T^2})^5}+\frac{3}{2\psi_0^2 \EU{X_T^2}}-\frac{3(16\bar{u}_0^2-16\bar{u}_0+5)}{16\psi_0^3}\right) k^4 ...
\end{align*}

Therefore, we obtain full variance smile moment expansion
\begin{equation} \begin{aligned}
w(k) &= 8 \left\{ \psi_0 - \left(\frac{1}{2} - \bar{u}_0\right) k + \frac{1}{2} \left(\frac{1}{\EU{X_T^2}}-\frac{1}{8\psi_0}\right) k^2 - \frac{1}{6} \left(\frac{\EU{X_T^3}}{(\EU{X_T^2})^3}-\frac{6\bar{u}_0-3}{16\psi_0^2}\right) k^3 + \right. \\
&\quad \left. \frac{1}{24} \left(\frac{3 (\EU{X_T^2})^3 + 3 (\EU{X_T^3})^2 - \EU{X_T^2} \EU{X_T^4}}{(\EU{X_T^2})^5}+\frac{3}{4\psi_0^2 \EU{X_T^2}}-\frac{3(16\bar{u}_0^2-16\bar{u}_0+5)}{32\psi_0^3}\right) k^4 ... \right\} .
\end{aligned} \label{eq:wexp} \end{equation}

Notably, variance skew is difference of ATM Esscher shift between model's $\bar{u}_0$ and BS's $1/2$; variance kurtosis is inverse difference between ATM Esscher log-spot variance and ATM implied variance. They are measures of deviation from BS diffusion progressively to higher orders.
\begin{equation} \begin{aligned}
\frac{\partial w}{\partial k} &= -8 \left( \frac{1}{2} - \bar{u}_0 \right) \\
\frac{\partial^2 w}{\partial k^2} &= 8 \left( \frac{1}{\EU{X_T^2}}-\frac{1}{w(0)} \right)
\end{aligned} \label{eq:wderiv} \end{equation}
Under BS flat variance $w_0$ and Esscher shift $\bar{u}_0=1/2$, we have exactly $w(k)=w_0$. Note that by writing $w(k) \equiv v(k)T$, we have the inverse scaling in variance skew: $\partial v/\partial k \sim 1/T$, a well-known result for Lévy models, and we have assumed large-time log-spot behavior tends to Lévy.

A practical use is, we fit a polynomial, say quartic, to ATM and from coefficients we imply the moments. But the moments are under ATM Esscher measure, which makes them less useful. If log-spot density is close enough to Gaussian, moments under the risk-neutral or Esscher measure are close, as effect of the Esscher measure change becomes merely a translation. But at least Esscher moments give an order-of-magnitude estimates of true implied moments.

\subsection{Relation to Lee's Moment Formula} \label{sec:leemoment}

Lee presented his moment formula establishing the connection between wing skews and finite moments in the spot density \cite{lee-moment}. The moment formula reads:

\begin{namedthm}[Moment Formula]
On the call side, let call-wing skew be $\displaystyle \beta_+ \equiv \limsup_{x\to\infty} \frac{v(x)}{|x|}$ and $\tilde{p} \equiv \sup\left\{p:\E{S_T^{p+1}}<\infty\right\}$. Then, $\beta_+ \in [0,2]$ and
\begin{equation*} \frac{\beta_+}{8} + \frac{1}{2\beta_+} - \frac{1}{2} = \tilde{p} . \end{equation*}
On the put side, let put-wing skew be $\displaystyle \beta_- \equiv \limsup_{x\to-\infty} \frac{v(x)}{|x|}$ and $\tilde{q} \equiv \sup\left\{q:\E{S_T^{-q}}<\infty\right\}$. Then, $\beta_- \in [0,2]$ and
\begin{equation*} \frac{\beta_-}{8} + \frac{1}{2\beta_-} - \frac{1}{2} = \tilde{q} . \end{equation*}
\end{namedthm}

Taking derivatives on strike $x$ for the saddle-point equation (\ref{eq:sdlpt}), we get
\begin{equation*} \frac{x}{v(x)} + \left( \frac{1}{8} - \frac{1}{2} \left(\frac{x}{v(x)}\right)^2 \right) v'(x) = \omega'(x) = \hat{u}(x) , \end{equation*}
which is valid for all strikes $x$. For example, when $x=0$ at-the-money, $v'(0)=8\hat{u}(0)$, consistent with the derivative (\ref{eq:wderiv}) obtained from moment expansion.

By the moment formula, at large strike, the dominant term in variance $v(x)$ must be linear, and we write
\begin{equation*} v(x) \sim \begin{cases} \beta_+ x & \text{as } x \rightarrow +\infty \\ -\beta_- x & \text{as } x \rightarrow -\infty \end{cases} \end{equation*}
thus $x/v(x) \sim \pm 1/\beta_\pm$ and $v'(x) \sim \pm \beta_\pm$.

Plugging back in,
\begin{equation} \frac{\beta_\pm}{8} + \frac{1}{2\beta_\pm} = \pm \hat{u}(\pm\infty) \label{eq:betapm} , \end{equation}
where we have the explicit closed form for (model-specific) $\hat{u}(x)$ so $\beta_\pm$ can be computed exactly. We provide the bounds of $\hat{u}(x)$ for each model discussed in section \ref{sec:model}.

This also connects to the finite moments of spot density via the following relations:
\begin{align*}
\tilde{p} &= \hat{u}(+\infty) - \frac{1}{2} \\
\tilde{q} &= -\hat{u}(-\infty) - \frac{1}{2} .
\end{align*}


\section{Large-Time Model-Implied Smile} \label{sec:model}

We apply the saddle-point approach outlined in section \ref{sec:sdlpt}, by beginning with the characteristic functions for log-spot $X_T$ under the risk-neutral measure (so $X_T$ is compensated s.t.\ $\phi_T(-i)=\E{e^{X_T}}=1$), to derive large-time model-implied volatility smiles of some classical models, as concrete examples. In each sub-section, we sketch the models and state the corresponding characteristic functions, then proceed to calculate these entities: $\psi(u)$, $\hat{u}(x)$, $\psi(\tilde{u}(x))$, and $\omega(x)$. For Heston stochastic volatility and variance-gamma models, we also present $\bar{\omega}(x)$ and $v(x)$. Notations in each sub-section are independent.

\subsection{Heston Stochastic Volatility Model} \label{sec:heston}

We derive the large-time Heston variance smile $v(x)$ as a function of time-scaled log-strike $x$.

Heston model is characterized by diffusion processes in spot $(S_t)_{t \geq 0}$ and variance $(v_t)_{t \geq 0}$ with correlated Brownian motions $Z_t$ and $W_t$, parametrized by initial variance $v$, long-run mean variance $\bar{v}$, mean-reversion rate $\lambda$, volatility of volatility $\eta$ and correlation $\rho$, suitably constraint\footnote{In models that follow, as we only care about the characteristic functions and the model details do not affect our saddle-point calculations, we will only lay out (handwave) the model parameters without worrying about parameter domains, provided that they are ``\textit{suitably constraint}'' in the usual context. For example, under Heston, we demand strictly positive $v,\bar{v},\lambda,\eta$ and $\rho \in [-1,1]$, with Feller condition $2\lambda\bar{v} \geq \eta^2$ s.t.\ $v_t$ never touches zero almost surely.}:
\begin{align*}
&dS_t = \sqrt{v_t} S_t dZ_t \\
&dv_t = \lambda (\bar{v}-v_t) dt + \eta \sqrt{v_t} dW_t \\
&dZ_t dW_t = \rho dt
\end{align*}
with characteristic function $\phi_T(u) = \exp\left(\mathcal{C}_T(u)\bar{v}+\mathcal{D}_T(u)v\right)$ where
\begin{align*}
\mathcal{C}_T(u) &= \lambda \left( r_-T - \frac{2}{\eta^2} \log \left(\frac{1-ge^{-dT}}{1-g}\right) \right) \\
\mathcal{D}_T(u) &= r_- \frac{1-e^{-dT}}{1-ge^{-dT}}
\end{align*}
in which
\begin{align*}
\alpha &= - \frac{u^2}{2} - \frac{iu}{2} \\
\beta &= \lambda - \rho\eta iu \\
\gamma &= \frac{\eta^2}{2} \\
d &= \sqrt{\beta^2 - 4\alpha\gamma} \\
r_\pm &= \frac{\beta \pm d}{2\gamma} \\
g &= \frac{r_+}{r_-}
\end{align*}
following \cite{gat-volsurf}. For derivation of characteristic function of log-spot under Heston model, see \cite{hir-comp}.

We require $\tilde{u}$ and $\psi(\tilde{u})$ under Heston in order to derive the smile. For large $T$, characteristic function $\phi_T$ factorizes into $e^{-\psi(u)T}$:
\begin{equation*} \phi_T\left(u-\frac{i}{2}\right) \sim e^{\lambda\bar{v}r_-(u-i/2)T} \equiv e^{-\psi(u)T} \end{equation*}
thus
\begin{equation*} \psi(u) = -\lambda\bar{v}r_- = \frac{\lambda\bar{v}}{\eta^2} \left( \sqrt{\beta^2 - 2\eta^2\alpha} - \beta \right) . \end{equation*}

Note we are evaluating it at $u-i/2$, so
\begin{align*}
\alpha &= -\frac{1}{2}\left(u^2+\frac{1}{4}\right) \\
\beta &= \left(\lambda-\frac{\rho\eta}{2}\right)-\rho\eta iu .
\end{align*}

Denote $\xi=\eta^2/\lambda\bar{v}$, and differentiate it to get
\begin{equation*} \psi'(u) = \frac{1}{\xi}\left( \frac{\beta \frac{\partial \beta}{\partial u} - \eta^2 \frac{\partial \alpha}{\partial u}}{\sqrt{\beta^2 - 2\eta^2\alpha}} - \frac{\partial \beta}{\partial u} \right) = \frac{1}{\xi}\left( \frac{-i\rho\eta\beta+\eta^2u}{\sqrt{\beta^2 - 2\eta^2\alpha}} + i\rho\eta \right) = -ix \end{equation*}
yielding the following expression 
\begin{equation*} -(\xi x + \rho\eta)^2 (\beta^2 - 2\eta^2\alpha) = \left(i\rho\eta\left(\lambda-\frac{\rho\eta}{2}-\rho\eta iu\right)-\eta^2u\right)^2 = \left(i\rho\eta\left(\lambda-\frac{\rho\eta}{2}\right)-\eta^2(1-\rho^2)u\right)^2 . \end{equation*}

On LHS we expand $\beta^2-2\eta^2\alpha$ to get
\begin{equation*} \left(\lambda-\frac{\rho\eta}{2}\right)^2 + \frac{\eta^2}{4} - 2\rho\eta\left(\lambda-\frac{\rho\eta}{2}\right)iu + \eta^2(1-\rho^2)u^2 . \end{equation*}

Define characteristic constants:
\begin{align*}
A^2 &= \eta^2(1-\rho^2) \\
B &= \rho\eta\left(\lambda-\frac{\rho\eta}{2}\right) \\
C^2 &= \left(\lambda-\frac{\rho\eta}{2}\right)^2 + \frac{\eta^2}{4} \\
D^2 &= \left(\frac{1}{A}\right)^2 + \left(\frac{C}{B}\right)^2 \\
m &= -\frac{\rho\eta}{\xi} \\
a &= \frac{\rho\eta}{\lambda} .
\end{align*}

Plugging back in,
\begin{equation*} -\xi^2(x-m)^2\left( C^2 - 2Biu + A^2u^2 \right) = (iB - A^2u)^2 , \end{equation*}
a quadratic equation in $u$:
\begin{equation*} \left(C^2\xi^2(x-m)^2-B^2\right) - 2Biu\left(A^2+\xi^2(x-m)^2\right) + A^2\left(A^2+\xi^2(x-m)^2\right)u^2 = 0 \end{equation*}
where we define constants involving strike $x$, abbreviated ``\textit{strike constants}'':
\begin{align*}
\Theta(x)^2 &= 1-\xi^2\left(\frac{C}{B}\right)^2(x-m)^2 \\
\Sigma(x)^2 &= 1+\xi^2\left(\frac{1}{A}\right)^2(x-m)^2 .
\end{align*}

Rewriting the quadratic equation,
\begin{align*}
B^2 \Theta^2 &= A^2 \Sigma^2 \left( A^2u^2 - 2Biu + \left(i\frac{B}{A}\right)^2 - \left(i\frac{B}{A}\right)^2 \right) \\
\left(\frac{B}{A}\right)^2 \left(\frac{\Theta}{\Sigma}\right)^2 &= \left(Au-i\frac{B}{A}\right)^2 + \left(\frac{B}{A}\right)^2
\end{align*}
we get
\begin{align*}
\tilde{u} &= \frac{1}{A} \left( i\frac{B}{A} \pm \sqrt{\left(\frac{B}{A}\right)^2 \left(\frac{\Theta}{\Sigma}\right)^2 - \left(\frac{B}{A}\right)^2} \right) \\
\hat{u} &\equiv i\tilde{u} = \frac{B}{A^2} \left( -1 \pm \operatorname{sgn} B \sqrt{1 - \left(\frac{\Theta}{\Sigma}\right)^2} \right)
\end{align*}
where plus sign corresponds to call domain $x>x_0$ and minus sign corresponds to put domain $x<x_0$, matching at strike $x_0=m$ which solves $\Theta(x) = \Sigma(x)$.

Assume negative correlation regime: $\rho<0$ hence $B<0$, typical of equity market. This assumption is not necessary but we can remove the sgn functions for clarity.

Then,
\begin{equation*} \hat{u} = \frac{B}{A^2} \left( -1 \mp \sqrt{1-\left(\frac{\Theta}{\Sigma}\right)^2} \right) = \frac{B}{A^2} \left( - \frac{\xi D (x-m)}{\Sigma} - 1 \right) \in \left( -\frac{B}{A^2}(1-AD), -\frac{B}{A^2}(1+AD) \right) . \end{equation*}

Finally we evaluate $\psi(\tilde{u})$:
\begin{align*}
\psi(\tilde{u}) &= \frac{1}{\xi} \left( \sqrt{\left( \lambda-\frac{\rho\eta}{2}-\rho\eta i\tilde{u} \right)^2 + \eta^2 \left( \tilde{u}^2+\frac{1}{4} \right)} - \left( \lambda-\frac{\rho\eta}{2}-\rho\eta i\tilde{u} \right) \right) \\
&= \frac{1}{\xi} \left( \sqrt{C^2 + \left(\frac{\Theta}{\Sigma}\right)^2} - \lambda \left( 1 - \frac{a}{2} - a \cdot \hat{u} \right) \right) \\
&= \frac{1}{\xi} \left( -\frac{BD}{\Sigma} - \lambda \left( 1 - \frac{a}{2} \right) + \lambda a \hat{u} \right) .
\end{align*}

$\hat{u}$ and $\psi(\tilde{u})$ are combined to yield variance quantity $\omega(x) \equiv \hat{u}(x) \cdot x + \psi(\tilde{u}(x))$:
\begin{equation} \begin{aligned}
\omega(x) &= \hat{u} \cdot x + \psi(\tilde{u}) \\
&= \hat{u} \cdot (x-m) + \hat{u} \cdot m + \frac{1}{\xi} \left( - \frac{BD}{\Sigma} - \lambda \left( 1 - \frac{a}{2} \right) \right) + \hat{u} \cdot \frac{\lambda a}{\xi} \\
&= \frac{B}{A^2} \left( - \frac{\xi D (x-m)}{\Sigma} - 1 \right) (x-m) + \frac{1}{\xi} \left( - \frac{BD}{\Sigma} - \lambda \left( 1 - \frac{a}{2} \right) \right) \\
&= - \frac{\lambda}{\xi} \left( 1 - \frac{a}{2} \right) - \frac{B}{A^2} (x-m) - \frac{1}{\Sigma} \frac{BD}{\xi} \left( 1 + \xi^2\left(\frac{1}{A}\right)^2(x-m)^2 \right) \\
&= - \frac{\lambda}{\xi} \left( 1 - \frac{a}{2} \right) - \frac{B}{A^2} (x-m) - \frac{BD}{\xi} \sqrt{ 1 + \xi^2\left(\frac{1}{A}\right)^2(x-m)^2 }
\end{aligned} \label{eq:hesomega} \end{equation}
which is SVI-like.

To reach full smile $v(x)$, we require $|\bar{\omega}(x)| \equiv \sqrt{\omega(x)^2-x^2/4}$, which we have the following proposition.

\begin{proposition} \label{prop:hesomegabar}
Under Heston model, assuming negative correlation $\rho$, $\bar{\omega}(x)$ which fulfills $|\bar{\omega}(x)| \equiv \sqrt{\omega(x)^2-x^2/4}$ is explicitly given by
\begin{equation} \bar{\omega}(x) = - K \frac{\lambda}{\xi} \left( 1 - \frac{a}{2} \right) - K \frac{B}{A^2} (x-m) - \frac{1}{K} \frac{BD}{\xi} \sqrt{ 1 + \xi^2\left(\frac{1}{A}\right)^2(x-m)^2 } \label{eq:hesomegabar} \end{equation}
where
\begin{equation*} K = \sqrt{1 + \frac{aA^2/4B}{1-a/2}} . \end{equation*}
\end{proposition}

\begin{myproof}
See appendix \ref{apdx:hes}.
\end{myproof}

Now, the smile is given by difference of two SVIs:
\begin{equation*} \omega(x) = - \frac{\lambda}{\xi} \left( 1 - \frac{a}{2} \right) - \frac{B}{A^2} (x-m) - \frac{BD}{\xi} \sqrt{ 1 + \xi^2\left(\frac{1}{A}\right)^2(x-m)^2 } \end{equation*}
and 
\begin{equation*} \bar{\omega}(x) = - K \frac{\lambda}{\xi} \left( 1 - \frac{a}{2} \right) - K \frac{B}{A^2} (x-m) - \frac{1}{K} \frac{BD}{\xi} \sqrt{ 1 + \xi^2\left(\frac{1}{A}\right)^2(x-m)^2 } .\end{equation*}

Therefore, the full Heston implied variance smile is explicitly given by
\begin{equation} \begin{aligned}
v(x) &= 4 \left( (K-1) \frac{\lambda}{\xi} \left( 1 - \frac{a}{2} \right) + (K-1) \frac{B}{A^2} (x-m) - \left(1-\frac{1}{K}\right) \frac{BD}{\xi} \sqrt{ 1 + \xi^2\left(\frac{1}{A}\right)^2(x-m)^2 } \right) \\
&= 4(K-1) \left( \frac{\lambda}{\xi} \left( 1 - \frac{a}{2} \right) + \frac{B}{A^2} (x-m) - \frac{1}{K} \frac{BD}{\xi} \sqrt{ 1 + \xi^2\left(\frac{1}{A}\right)^2(x-m)^2 } \right) .
\end{aligned} \label{eq:hesvar} \end{equation}

This echoes with the proof in \cite{gat-convhes} that \textit{large-time Heston implied volatility smile is exactly SVI}.

\subsubsection{$x_\pm$ under Heston}

Recall that $x_\pm$ solves $\omega(x) = \pm x/2$, or equivalently $\bar{\omega}(x) = 0$. By solving with e.g.\ mathematica
\begin{equation*} - K \frac{\lambda}{\xi} \left( 1 - \frac{a}{2} \right) - K \frac{B}{A^2} (x-m) - \frac{1}{K} \frac{BD}{\xi} \sqrt{ 1 + \xi^2\left(\frac{1}{A}\right)^2(x-m)^2 } = 0 , \end{equation*}
we get
\begin{equation} \begin{aligned}
x_- &= -\frac{\bar{v}}{2} \\
x_+ &= \frac{\bar{v}}{2(1-a)} , 
\end{aligned} \label{eq:hesxpm} \end{equation}
consistent with approximation (\ref{eq:xpmapprox}) which yields $x_\pm \sim \pm \bar{v}/2$. This result is also stated in \cite{gat-convhes}.

\subsubsection{SVI Parametrization}

Variance smile $v(x)$ has 4 underlying degrees of freedom arising from Heston: $\bar{v},\rho,\eta,\lambda$. As we derive $v(x)$ out of a risk-neutral stochastic process, coefficients are interlinked in a way that $v(x)$ is arbitrage-free in strike.

We can directly fit $v(x)$ to market smiles optimizing $\bar{v},\rho,\eta,\lambda$, but $v(x)$ may not be flexible enough. Instead, we relax the degrees of freedom removing dependence between parameters in $v(x)$ to reach a parametrization, known as SVI by Gatheral.

SVI parametrizes implied variance $v(x)$ in a time slice with 5 parameters $\{a,b,\sigma,\rho,m\}$ of form
\begin{equation} v(x) \equiv a + b \left( \rho (x-m) + \sqrt{(x-m)^2 + \sigma^2} \right) \label{eq:svi} \end{equation}
which offers extra flexibility in goodness-of-fit but in exchange for arbitrage in strike.

\subsubsection{Numerical Experiment} \label{sec:hesnum}

Assuming the following typical Heston parameters, we illustrate the convergence of volatility smiles computed via FFT into the large-time limit, given by analytic formula (\ref{eq:hesvar}).
\begin{equation*} \bar{v} = 0.04, \quad \rho = -0.7, \quad \eta = 0.1, \quad \lambda = 1 \end{equation*}

We plot the convergence of the overall smiles $\sigma(x,T)$ and ATM implied volatilities $\sigma_0(T) \equiv \sigma(0,T)$ as time to expiry $T$ grows.\footnote{We fix the time-scaled log-strike domain over $x \in (-1,1)$ and for $T$ large, log-strike domain extends to $k \in (-T,T)$ which lies in the far wings and FFT becomes inaccurate, causing the discontinuities in the overall smile. Also, it can be checked that the convergence error of ATM implied volatility decays roughly as $T^{-1}$, which hints the first-order correction to our zeroth-order smile $v(x)$.}

\begin{table}[H]
\centering
\begin{tabular}{|c|c|}
\hline
Overall Smile & ATM Implied Volatility \\ 
\hline \hline
\includegraphics[width=0.4\linewidth]{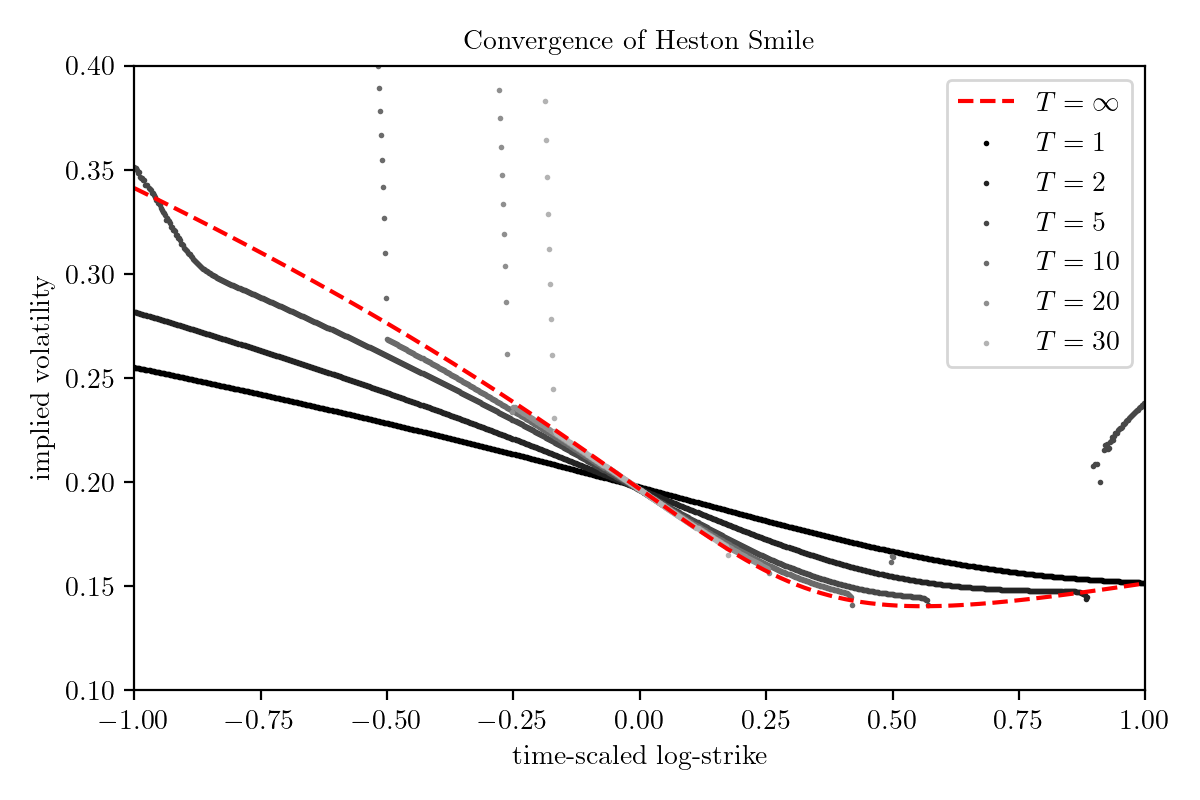} & 
\includegraphics[width=0.4\linewidth]{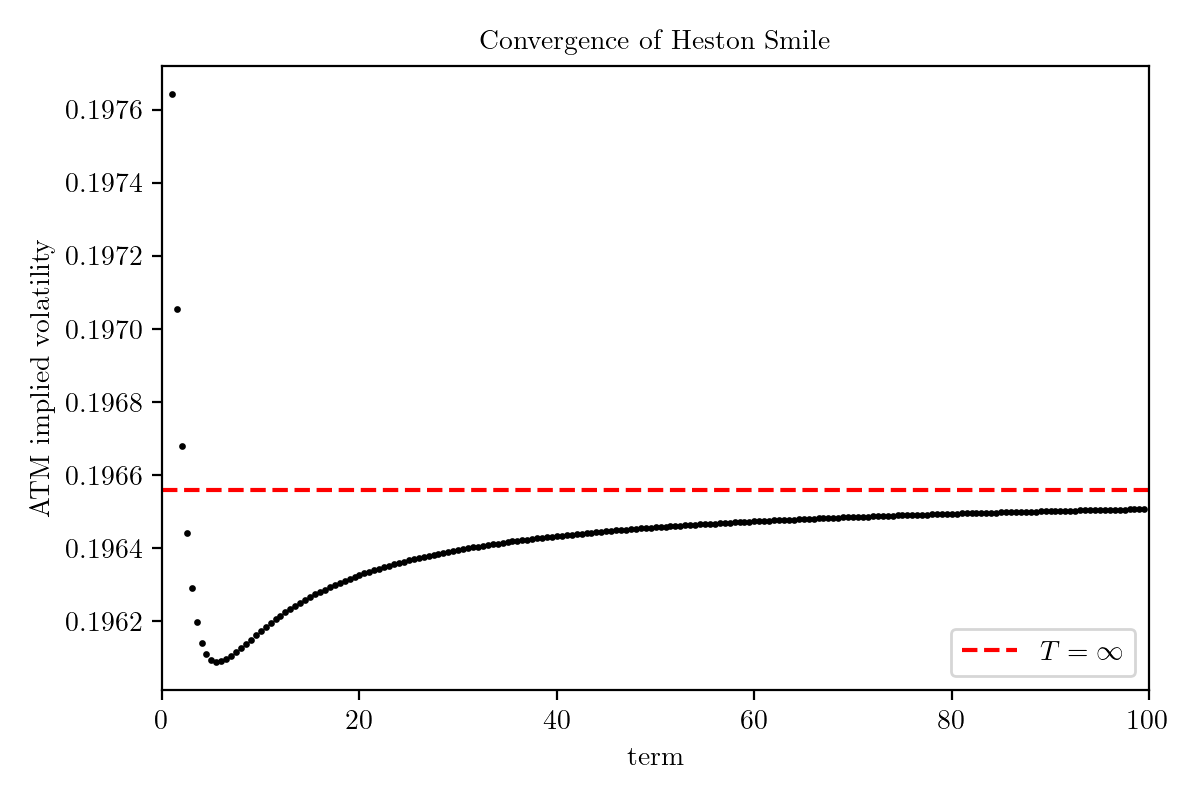} \\ 
\hline
\end{tabular}
\caption{Overall smile and ATM implied volatility under Heston model.}
\end{table}

We also plot variance quantities $\omega(x)$ and $\bar{\omega}(x)$. Notice that (1) $\omega(x)$ tangentially touches $|x/2|$ at points $x_\pm$ and is convex everywhere; (2) $\bar{\omega}(x)$ is roughly $\omega(x)$ shifted down, cutting $x$-axis at $x_\pm$ analytically given by (\ref{eq:hesxpm}): $x_- = -0.0200, x_+ = 0.0187$.

\begin{table}[H]
\centering
\begin{tabular}{|c|c|}
\hline
Zoom in ATM & Zoom out ATM \\ 
\hline \hline
\includegraphics[width=0.4\linewidth]{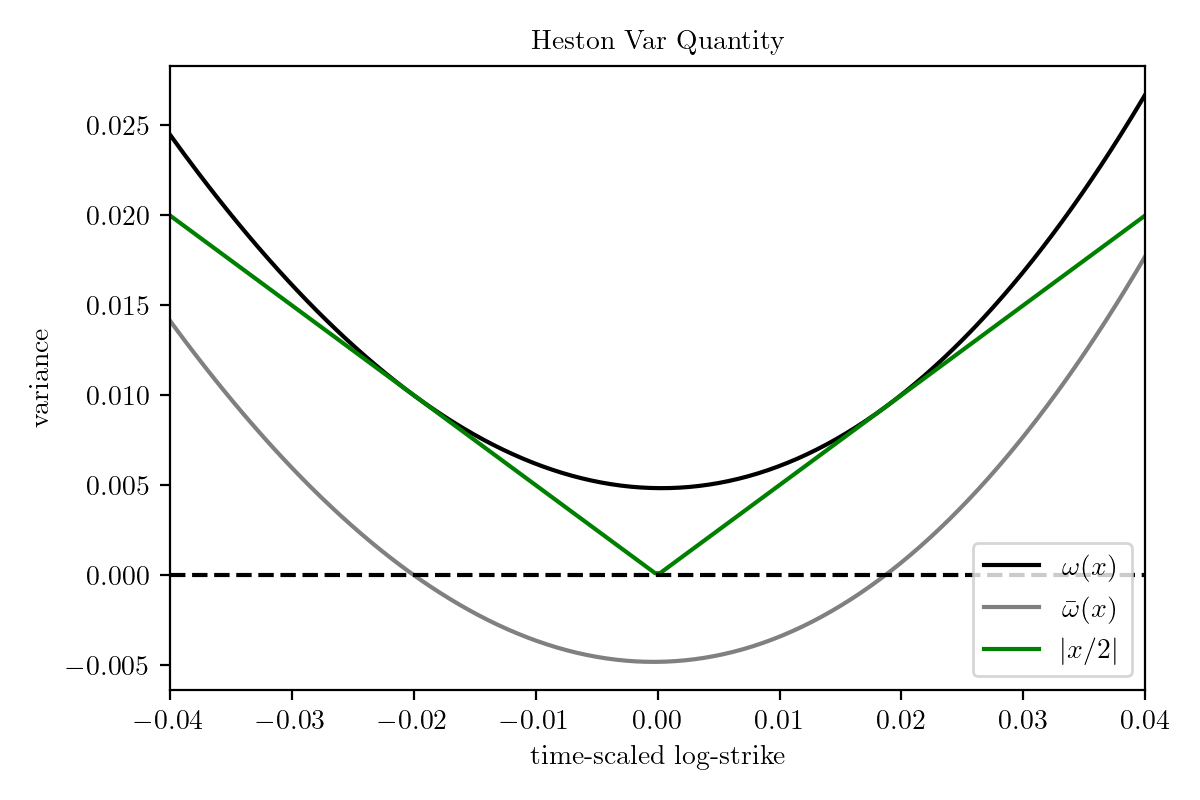} & 
\includegraphics[width=0.4\linewidth]{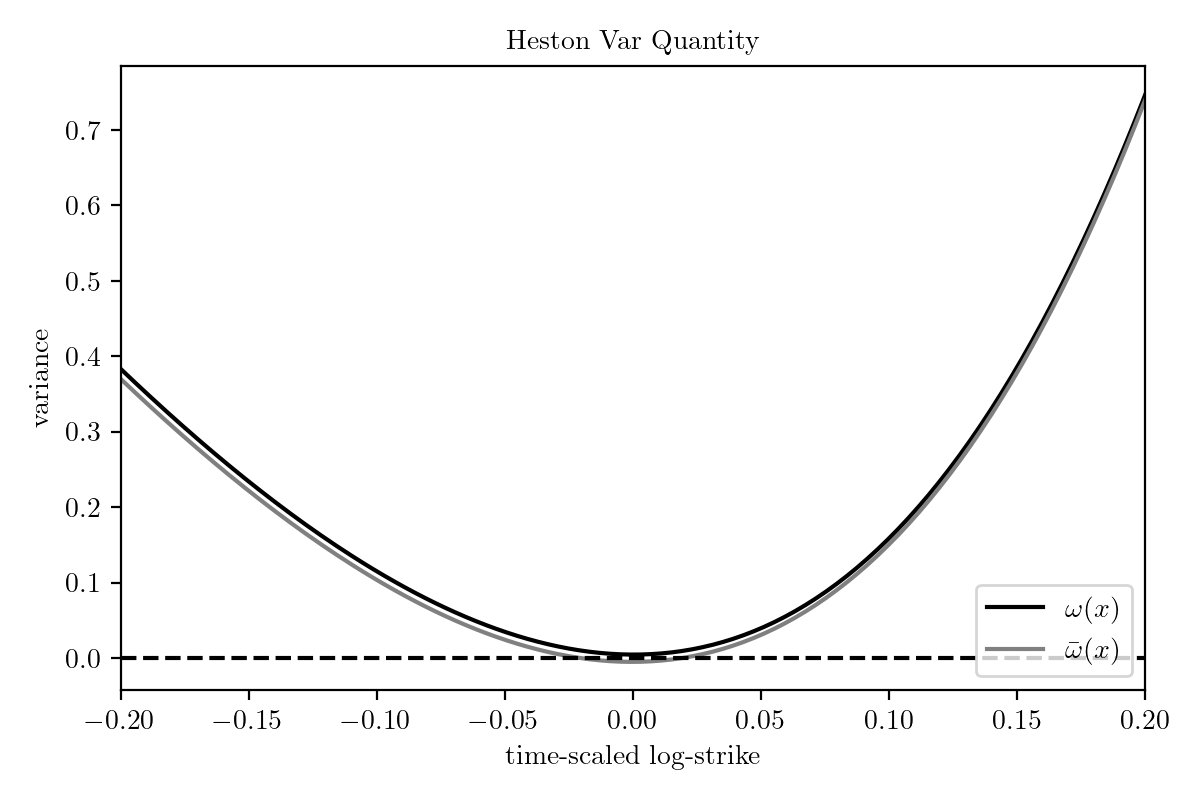} \\ 
\hline
\end{tabular}
\caption{ATM and global behavior of variance quantities under Heston model.}
\end{table}

\subsection{Variance-Gamma Model} \label{sec:vg}

We derive the large-time VG variance smile $v(x)$ as a function of time-scaled log-strike $x$.

VG process for log-spot is a pure-jump process constructed by time-changing a Brownian motion with drift by the gamma time, and is parametrized by drift $\theta$, volatility $\sigma$ and (gamma) time-change variance $\nu$, suitably constraint, with characteristic function
\begin{equation*} \phi_T(u) = \exp \left\{ \frac{T}{\nu} \left[ iu \log \left( 1- \left( \theta + \frac{\sigma^2}{2} \right) \nu \right) - \log \left( 1 - iu \theta\nu + \frac{u^2\sigma^2\nu}{2} \right) \right] \right\} \end{equation*}
following \cite{mad-vg}. Alternatively, one may consider the measure-theoretic ``\textit{CGM}'' formulation, which is a special case discussed in section \ref{sec:redvg}.

As a check, $\phi_T(-i)=1$.

We solve $\psi(u)$ from $\phi_T(u-i/2) \equiv e^{-\psi(u)T}$, which gives
\begin{equation*} \psi(u) = \frac{1}{\nu} \left[ \log \left( 1 - \frac{\theta\nu}{2} - \frac{\sigma^2\nu}{8} - \left( \theta + \frac{\sigma^2}{2} \right) \nu iu + \frac{\sigma^2\nu}{2} u^2 \right) - \left( iu + \frac{1}{2} \right) \log \left( 1- \left( \theta + \frac{\sigma^2}{2} \right) \nu \right) \right] .\end{equation*}

Define characteristic constants:
\begin{align*}
\alpha &= \frac{\sigma^2\nu}{2} \\
\beta &= 1 - \frac{\theta\nu}{2} - \frac{\sigma^2\nu}{8} \\
\xi &= \left( \theta + \frac{\sigma^2}{2} \right) \nu .
\end{align*}

Thus we rewrite
\begin{equation*} \psi(u) = \frac{1}{\nu} \left( \log \left( \beta - \xi iu + \alpha u^2 \right) - \left( iu + \frac{1}{2} \right) \log \left( 1 - \xi \right) \right) . \end{equation*}

Differentiating,
\begin{align*}
&\psi'(u) = \frac{1}{\nu} \left( \frac{- i\xi + 2 \alpha u}{\beta - \xi iu + \alpha u^2} - i \log \left( 1 - \xi \right) \right) = -ix \\
&\frac{\xi + 2 \alpha iu}{\beta - \xi iu + \alpha u^2} = \nu x - \log \left( 1 - \xi \right) .
\end{align*}

Define strike constants:
\begin{equation*} \begin{aligned} 
\Sigma(x) &= \nu x - \log \left( 1 - \xi \right) \\
\frac{1}{\Theta(x)^2} &= \left( \frac{1}{\Sigma(x)} \right)^2 + \left( \frac{\xi}{2\alpha} \right)^2 + \frac{\beta}{\alpha} \\
\frac{1}{\Pi_\pm(x)} &= - \frac{1}{\Sigma(x)} \pm \frac{1}{\Theta(x)}
\end{aligned} \end{equation*}
and solving,
\begin{align*}
\alpha \Sigma (iu)^2 + (\xi \Sigma + 2 \alpha) iu + (\xi - \beta \Sigma) &= 0 \\
\alpha \Sigma \left( (iu)^2 + 2 \left( \frac{1}{\Sigma} + \frac{\xi}{2\alpha} \right) iu + \left( \frac{1}{\Sigma} + \frac{\xi}{2\alpha} \right)^2 - \left( \frac{1}{\Sigma} + \frac{\xi}{2\alpha} \right)^2 \right) &= \beta \Sigma - \xi \\
\left( iu + \frac{1}{\Sigma} + \frac{\xi}{2\alpha} \right)^2 &= \left( \frac{1}{\Sigma} + \frac{\xi}{2\alpha} \right)^2 - \frac{1}{\alpha} \left( \frac{\xi}{\Sigma} - \beta \right)
\end{align*}
thus
\begin{equation*} \begin{aligned}
\hat{u} &\equiv i\tilde{u} = - \left( \frac{1}{\Sigma} + \frac{\xi}{2\alpha} \right) \pm \sqrt{\left( \frac{1}{\Sigma} + \frac{\xi}{2\alpha} \right)^2 - \frac{1}{\alpha} \left( \frac{\xi}{\Sigma} - \beta \right)} \\
&= - \frac{\xi}{2\alpha} - \frac{1}{\Sigma} \pm \sqrt{\left( \frac{1}{\Sigma} \right)^2 + \left( \frac{\xi}{2\alpha} \right)^2 + \frac{\beta}{\alpha}} \\
&= - \frac{\xi}{2\alpha} - \frac{1}{\Sigma} \pm \frac{1}{\Theta} = - \frac{\xi}{2\alpha} + \frac{1}{\Pi_\pm} \\
&\in \left( - \frac{\xi}{2\alpha} - \sqrt{\left( \frac{\xi}{2\alpha} \right)^2 + \frac{\beta}{\alpha}} , - \frac{\xi}{2\alpha} + \sqrt{\left( \frac{\xi}{2\alpha} \right)^2 + \frac{\beta}{\alpha}} \right) . 
\end{aligned} \end{equation*}

Like Heston, plus corresponds to call domain and minus corresponds to put domain, matching at strike $x_0=\log(1-\xi)/\nu$ which solves $\Sigma(x)=0$.

Finally we evaluate
\begin{equation*} \psi(\tilde{u}) = \frac{1}{\nu} \left( \log \left( \beta - \xi \hat{u} - \alpha \hat{u}^2 \right) - \left( \hat{u} + \frac{1}{2} \right) \log \left( 1 - \xi \right) \right) , \end{equation*}
where the argument in log term simplifies as
\begin{align*}
\beta - \xi \hat{u} - \alpha \hat{u}^2 &= \beta + \frac{\xi^2}{4\alpha} - \alpha \left( \hat{u} + \frac{\xi}{2\alpha} \right)^2 \\
&= \beta + \frac{\xi^2}{4\alpha} - \alpha \left( \left( \frac{1}{\Sigma} \right)^2 \mp \frac{2}{\Sigma\Theta} + \left( \frac{1}{\Sigma} \right)^2 + \left( \frac{\xi}{2\alpha} \right)^2 + \frac{\beta}{\alpha} \right) \\
&= \frac{2\alpha}{\Sigma} \left( - \frac{1}{\Sigma} \pm \frac{1}{\Theta} \right) = \frac{2\alpha}{\Sigma\Pi_\pm} .
\end{align*}

$\Pi_\pm(x)$ can be expressed entirely in terms of $\Sigma(x)$:
\begin{equation*} \frac{1}{\Pi_\pm(x)} = \frac{1}{\Sigma(x)} \left( \sqrt{1 + \left( \left( \frac{\xi}{2\alpha} \right)^2 + \frac{\beta}{\alpha} \right) \Sigma(x)^2} - 1 \right) . \end{equation*}

$\hat{u}$ and $\psi(\tilde{u})$ are combined to yield variance quantity $\omega(x) \equiv \hat{u}(x) \cdot x + \psi(\tilde{u}(x))$:
\begin{equation} \begin{aligned}
\omega(x) &= \left( - \frac{\xi}{2\alpha} + \frac{1}{\Pi_\pm(x)} \right) x + \frac{1}{\nu} \left( \log \frac{2\alpha}{\Sigma(x)\Pi_\pm(x)} - \log \left( 1 - \xi \right) \left( \frac{1}{2} - \frac{\xi}{2\alpha} + \frac{1}{\Pi_\pm(x)} \right) \right) \\
&= - x_0 \left( \frac{1}{2} - \frac{\xi}{2\alpha} + \frac{1}{\Pi_\pm(x)} \right) + \left( - \frac{\xi}{2\alpha} + \frac{1}{\Pi_\pm(x)} \right) x + \frac{1}{\nu} \log \frac{2\alpha}{\Sigma(x)\Pi_\pm(x)} \\
&= - \frac{x_0}{2} + \left( \frac{1}{\Pi_\pm(x)} - \frac{\xi}{2\alpha} \right) (x-x_0) + \frac{1}{\nu} \log \frac{2\alpha}{\Sigma(x)\Pi_\pm(x)} \\
&= - \frac{x_0}{2} + \left( \frac{1}{\Sigma(x)} \left( \sqrt{1 + \left( \left( \frac{\xi}{2\alpha} \right)^2 + \frac{\beta}{\alpha} \right) \Sigma(x)^2} - 1 \right) - \frac{\xi}{2\alpha} \right) (x-x_0) + \\
&\quad \frac{1}{\nu} \log \frac{2\alpha}{\Sigma(x)^2} \left( \sqrt{1 + \left( \left( \frac{\xi}{2\alpha} \right)^2 + \frac{\beta}{\alpha} \right) \Sigma(x)^2} - 1 \right) \\
&= - \frac{x_0}{2} - \frac{\xi}{2\alpha} (x-x_0) + \frac{1}{\nu} \left( \sqrt{1 + \nu^2 \left( \left( \frac{\xi}{2\alpha} \right)^2 + \frac{\beta}{\alpha} \right) (x-x_0)^2} - 1 \right) + \\
&\quad \frac{1}{\nu} \log \frac{2\alpha}{\nu^2(x-x_0)^2} \left( \sqrt{1 + \nu^2 \left( \left( \frac{\xi}{2\alpha} \right)^2 + \frac{\beta}{\alpha} \right) (x-x_0)^2} - 1 \right) \\
&\equiv - \frac{x_0}{2} - \frac{\xi}{2\alpha} (x-x_0) + \frac{1}{\nu} \left( \sqrt{1 + \eta^2 (x-x_0)^2} - 1 \right) + \\
&\quad \frac{1}{\nu} \log \frac{2\alpha}{\nu^2(x-x_0)^2} \left( \sqrt{1 + \eta^2 (x-x_0)^2} - 1 \right)
\end{aligned} \label{eq:vgomega} \end{equation}
where we write $\Sigma(x) = \nu x - \log(1-\xi) = \nu (x-x_0)$ and define $\eta^2 = \nu^2 \left( \left( \frac{\xi}{2\alpha} \right)^2 + \frac{\beta}{\alpha} \right)$.

This is SVI corrected by a log term.

To reach full smile $v(x)$, we require $|\bar{\omega}(x)| \equiv \sqrt{\omega(x)^2-x^2/4}$, which we have the following proposition.

\begin{proposition} \label{prop:vgomegabar}
Under variance-gamma model, by matching leading orders in the Taylor-expansions, $\bar{\omega}(x)$ which fulfills $|\bar{\omega}(x)| \equiv \sqrt{\omega(x)^2-x^2/4}$ is approximately given by
\begin{equation} \begin{aligned}
\bar{\omega}(x) &\approx - \frac{x_0}{2K} \left( 1 - \frac{\alpha}{\xi} \right) - \frac{\xi}{2\alpha} K (x-x_0) + \frac{1}{K\nu} \left( \sqrt{1 + \eta^2 (x-x_0)^2} - 1 \right) \\
&\quad + \frac{1}{K\nu} \log \frac{2\alpha}{\nu^2(x-x_0)^2} \left( \sqrt{1 + \eta^2 (x-x_0)^2} - 1 \right)
\end{aligned} \label{eq:vgomegabar} \end{equation}
where
\begin{equation*} K = \frac{1 - \frac{1}{2} \left( 1 - \frac{\alpha}{\xi} \right) \frac{x_0\nu}{\log \frac{\alpha\eta^2}{\nu^2}}}{\sqrt{1 - \frac{x_0\nu}{\log \frac{\alpha\eta^2}{\nu^2}}}} . \end{equation*}
\end{proposition}

\begin{myproof}
See appendix \ref{apdx:vg}.
\end{myproof}

Therefore, the full VG implied variance smile is approximately given by
\begin{equation} \begin{aligned}
v(x) &\approx 4 \left( - \frac{x_0}{2} - \frac{\xi}{2\alpha} (x-x_0) + \frac{1}{\nu} \left( \sqrt{1 + \eta^2 (x-x_0)^2} - 1 \right) + \frac{1}{\nu} \log \frac{2\alpha}{\nu^2(x-x_0)^2} \left( \sqrt{1 + \eta^2 (x-x_0)^2} - 1 \right) \right. \\
&\quad + \frac{x_0}{2K} \left( 1 - \frac{\alpha}{\xi} \right) + \frac{\xi}{2\alpha} K (x-x_0) - \frac{1}{K\nu} \left( \sqrt{1 + \eta^2 (x-x_0)^2} - 1 \right) \\
&\quad \left. - \frac{1}{K\nu} \log \frac{2\alpha}{\nu^2(x-x_0)^2} \left( \sqrt{1 + \eta^2 (x-x_0)^2} - 1 \right) \right) \\
&= 4(K-1) \left( - \frac{x_0}{2K} \left( 1 + \frac{\alpha}{(K-1)\xi} \right) + \frac{\xi}{2\alpha} (x-x_0) + \frac{1}{K\nu} \left( \sqrt{1 + \eta^2 (x-x_0)^2} - 1 \right) \right. \\
&\quad \left. + \frac{1}{K\nu} \log \frac{2\alpha}{\nu^2(x-x_0)^2} \left( \sqrt{1 + \eta^2 (x-x_0)^2} - 1 \right) \right) .
\end{aligned} \label{eq:vgvar} \end{equation}

Note that we matched leading orders in $\bar{\omega}^2$ and $\omega^2-x^2/4$ so the approximation here is only accurate ATM i.e.\ $x \sim x_0$ but starts deviating in wings.

\subsubsection{VG-Inspired Parametrization}

Dependence between parameters in $v(x)$ is relaxed to reach a parametrization for flexibility, which we term variance-gamma-inspired (VGI):
\begin{equation} v(x) = a + b \left( \rho (x-x_0) + \left( \sqrt{1 + \eta^2 (x-x_0)^2} - 1 \right) + \log \frac{\sqrt{1 + \eta^2 (x-x_0)^2} - 1}{(x-x_0)^2} \right) \label{eq:vgi} \end{equation}
with five parameters $a,b,\rho,\eta,x_0$. The $\eta$ inside the log may be further relaxed to an extra parameter.

In the wings, $v(\pm\infty) \sim |x| - \log |x|$ -- sub-linear growth consistent with Lee's bound.

In short, VGI is SVI corrected by a log term. We leave the study of fit quality, parameter stability, variance swap price, arbitrage removal etc.\ under VGI vs.\ SVI as future work. At its core, we should ask: \textit{is the log term actually important from a volatility calibration point of view}?

\subsubsection{Numerical Experiment} \label{sec:vgnum}

Assuming the following typical VG parameters, we illustrate the convergence of volatility smiles computed via FFT into the large-time limit, from solving saddle-point equation (\ref{eq:sdlpt}) plugging in VG's $\omega(x)$. Both plus/minus solutions are plotted causing the seeming explosion ATM.
\begin{equation*} \sigma = 0.12, \quad \theta = -0.14, \quad \nu = 0.17 \end{equation*}

We plot the convergence of the overall smiles $\sigma(x,T)$ and ATM implied volatilities $\sigma_0(T) \equiv \sigma(0,T)$ as time to expiry $T$ grows.

\begin{table}[H]
\centering
\begin{tabular}{|c|c|}
\hline
Overall Smile & ATM Implied Volatility \\ 
\hline \hline
\includegraphics[width=0.4\linewidth]{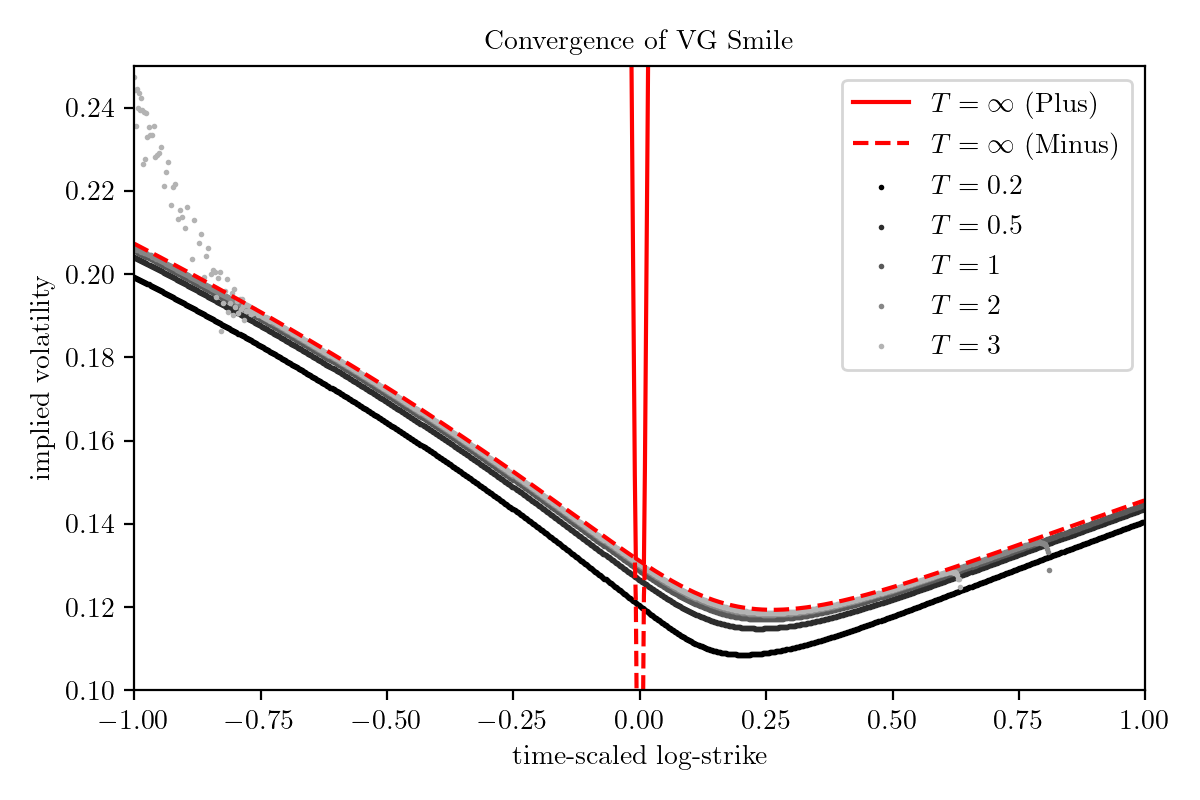} & 
\includegraphics[width=0.4\linewidth]{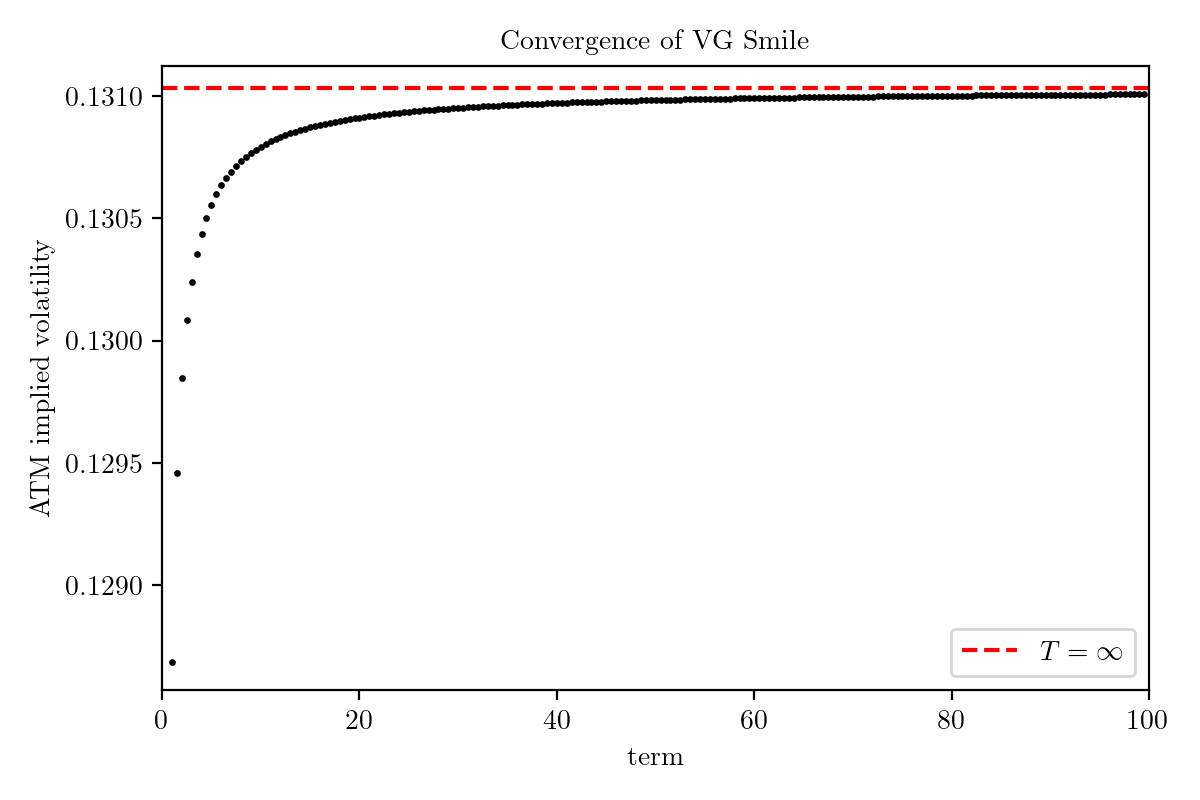} \\ 
\hline
\end{tabular}
\caption{Overall smile and ATM implied volatility under VG model.}
\end{table}

We also plot variance quantities $\omega(x)$ and $\bar{\omega}(x)$ approximated via (\ref{eq:vgomegabar}). Notice that (1) $\omega(x)$ tangentially touches $|x/2|$ at points $x_\pm$ and is convex everywhere; (2) $\bar{\omega}(x)$ is roughly $\omega(x)$ shifted down, cutting $x$-axis at $x_\pm$, which unlike Heston, cannot be analytically solved.

\begin{table}[H]
\centering
\begin{tabular}{|c|c|}
\hline
Zoom in ATM & Zoom out ATM \\ 
\hline \hline
\includegraphics[width=0.4\linewidth]{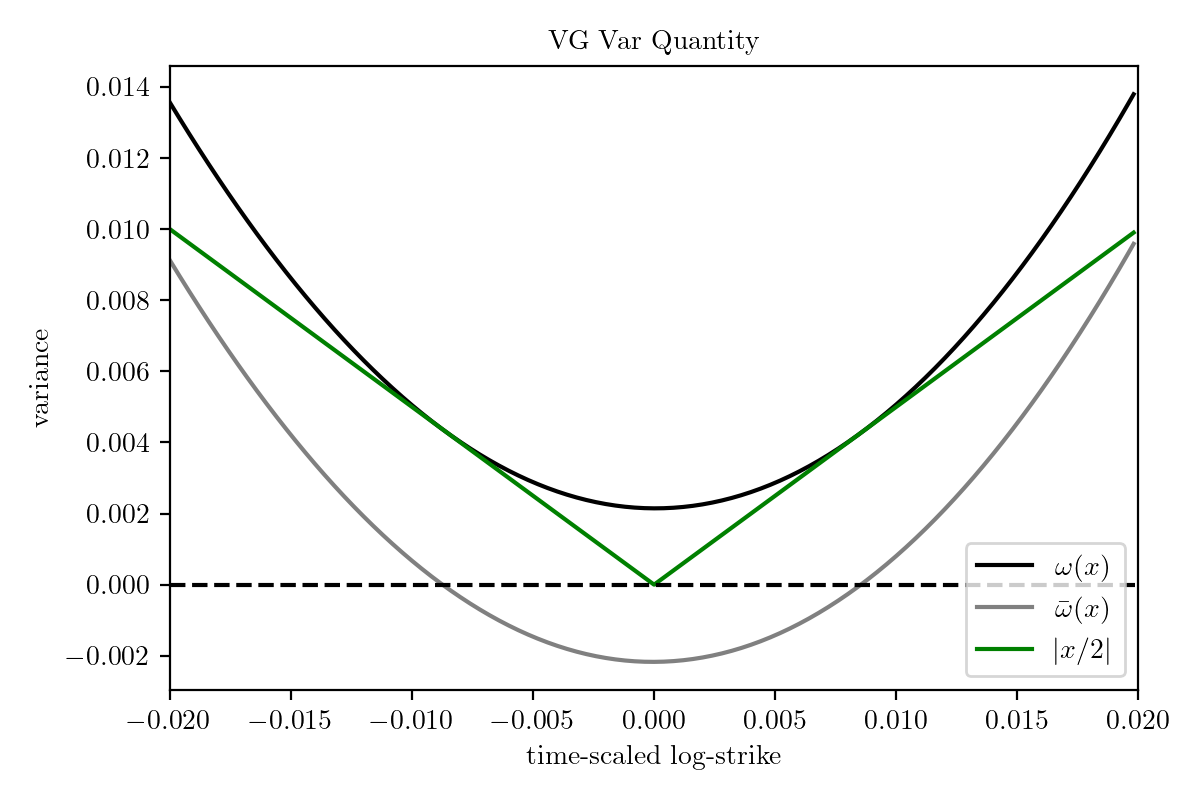} & 
\includegraphics[width=0.4\linewidth]{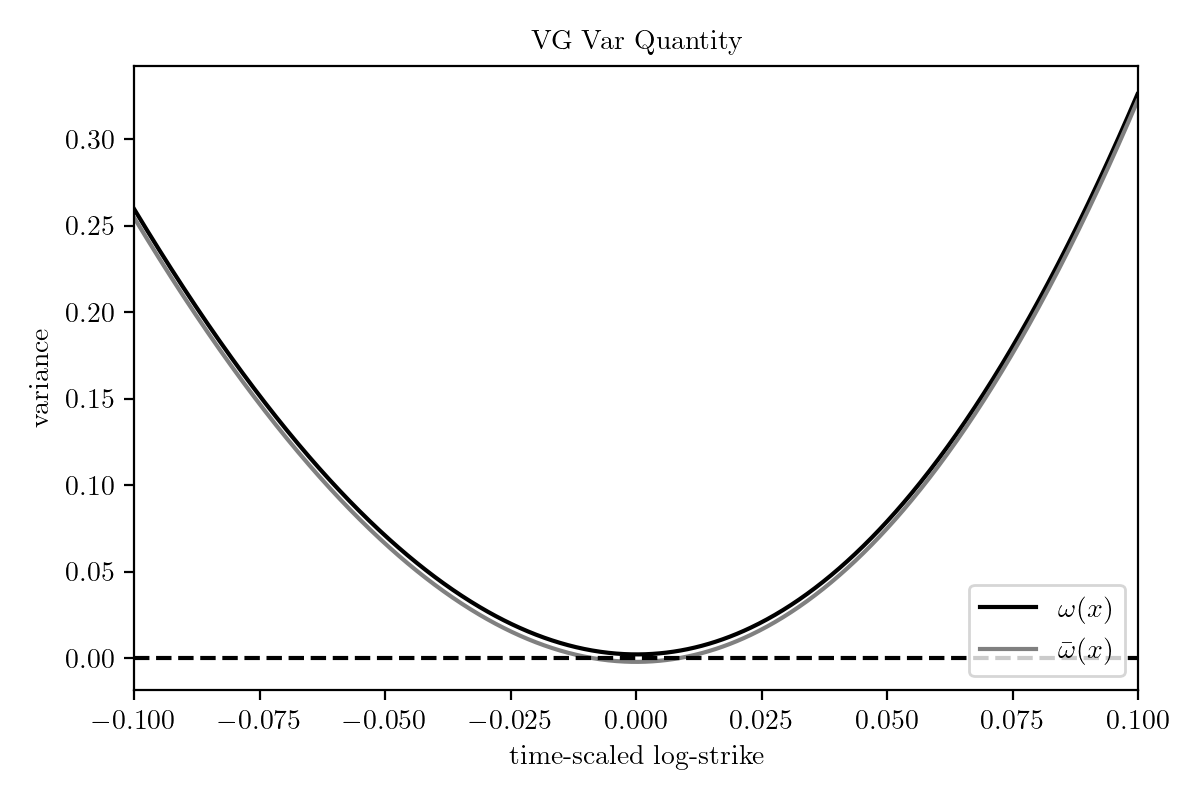} \\ 
\hline
\end{tabular}
\caption{ATM and global behavior of variance quantities under VG model.}
\end{table}

\subsection{Bilateral-Gamma Model} \label{sec:bg}

We derive the large-time BG variance smile $v(x)$ as a function of time-scaled log-strike $x$.

BG process for log-spot is a pure-jump process defined by Lévy measure
\begin{equation*} \mu(x) = \begin{cases} \frac{\alpha_+e^{-\lambda_+x}}{x} & \text{for } x > 0 \\ \frac{\alpha_-e^{-\lambda_-|x|}}{|x|} & \text{for } x < 0 \end{cases} \end{equation*}
with characteristic function, obtained from Lévy-Khintchine formula,
\begin{equation*} \phi_T(u) = \exp \left\{ T \left[ \left( \alpha_+ \log \frac{\lambda_+}{\lambda_+-iu} + \alpha_- \log \frac{\lambda_-}{\lambda_-+iu} \right) - iu \left( \alpha_+ \log \frac{\lambda_+}{\lambda_+-1} + \alpha_- \log \frac{\lambda_-}{\lambda_-+1} \right) \right] \right\} \end{equation*}
following \cite{mad-bg}.

As a sanity check, we see $\phi_T(-i)=1$.

Note that BG covers VG as a special case, by requiring $\alpha_+ = \alpha_-$.

We solve $\psi(u)$ from $\phi_T(u-i/2) \equiv e^{-\psi(u)T}$, which gives
\begin{equation*} \psi(u) = - \alpha_+ \log \frac{\lambda_+}{\lambda_+-1/2-iu} - \alpha_- \log \frac{\lambda_-}{\lambda_-+1/2+iu} + \left( iu + \frac{1}{2} \right) \left( \alpha_+ \log \frac{\lambda_+}{\lambda_+-1} + \alpha_- \log \frac{\lambda_-}{\lambda_-+1} \right) . \end{equation*}

Define characteristic constants:
\begin{align*}
\bar{\lambda}_+ &= \lambda_+ - \frac{1}{2} \\
\bar{\lambda}_- &= \lambda_- + \frac{1}{2} \\
K &= \alpha_+ \log \frac{\lambda_+}{\lambda_+-1} + \alpha_- \log \frac{\lambda_-}{\lambda_-+1} .
\end{align*}

So
\begin{equation*} \psi(u) = - \alpha_+ \log \frac{\lambda_+}{\bar{\lambda}_+-iu} - \alpha_- \log \frac{\lambda_-}{\bar{\lambda}_-+iu} + \left( iu + \frac{1}{2} \right) K . \end{equation*}

Saddle-point $\hat{u}(x) \equiv i\tilde{u}(x)$ fulfills
\begin{equation*} \psi'(\tilde{u}) = - \frac{i\alpha_+}{\bar{\lambda}_+-\hat{u}} + \frac{i\alpha_-}{\bar{\lambda}_-+\hat{u}} + iK = -ix \end{equation*}
leading to a quadratic equation
\begin{align*}
\frac{\alpha_+}{\bar{\lambda}_+-\hat{u}} - \frac{\alpha_-}{\bar{\lambda}_-+\hat{u}} &= K+x \\
\bar{\lambda}_+ \bar{\lambda}_- + (\bar{\lambda}_+ - \bar{\lambda}_-) \hat{u} - \hat{u}^2 &= \frac{\alpha_+\bar{\lambda}_--\alpha_-\bar{\lambda}_++(\alpha_++\alpha_-)\hat{u}}{K+x} \\
\hat{u}^2 + \left( \frac{\alpha_++\alpha_-}{K+x} + \bar{\lambda}_- - \bar{\lambda}_+ \right) \hat{u} &= \bar{\lambda}_+ \bar{\lambda}_- - \frac{\alpha_+\bar{\lambda}_--\alpha_-\bar{\lambda}_+}{K+x} \\
\left( \hat{u} + \frac{1}{2} \left( \frac{\alpha_++\alpha_-}{K+x} + \bar{\lambda}_- - \bar{\lambda}_+ \right) \right)^2 &= \bar{\lambda}_+ \bar{\lambda}_- - \frac{\alpha_+\bar{\lambda}_--\alpha_-\bar{\lambda}_+}{K+x} + \frac{1}{4} \left( \frac{\alpha_++\alpha_-}{K+x} + \bar{\lambda}_- - \bar{\lambda}_+ \right)^2 \\
&= \frac{1}{4} \left( \frac{\alpha_++\alpha_-}{K+x} \right)^2 + \frac{1}{4} (\bar{\lambda}_+ + \bar{\lambda}_-)^2 - \frac{(\alpha_+-\alpha_-)(\bar{\lambda}_++\bar{\lambda}_-)}{2(K+x)} .
\end{align*}

Thus we have
\begin{align*}
\hat{u} &= - \frac{1}{2} \left( \frac{\alpha_++\alpha_-}{K+x} + \bar{\lambda}_- - \bar{\lambda}_+ \right) \pm \frac{1}{2} \sqrt{\left( \frac{\alpha_++\alpha_-}{K+x} \right)^2 - \frac{2(\alpha_+-\alpha_-)(\bar{\lambda}_++\bar{\lambda}_-)}{K+x} + (\bar{\lambda}_+ + \bar{\lambda}_-)^2} \\
&= - \frac{1}{2} \left( \frac{\alpha_++\alpha_-}{K+x} + \bar{\lambda}_- - \bar{\lambda}_+ \right) \pm \frac{1}{2} \sqrt{\frac{4\alpha_+\alpha_-}{(K+x)^2} + \left( \bar{\lambda}_+ + \bar{\lambda}_- - \frac{\alpha_+-\alpha_-}{K+x} \right)^2} \in (-\bar{\lambda}_-,\bar{\lambda}_+)
\end{align*}
where plus/minus correspond to call/put wing, matching at $x_0=-K$. This can be represented using indicators: $(\bbone_{x>-K}-\bbone_{x<-K})$.

Plug back into $\psi$ to get
\begin{align*}
\psi(\tilde{u}) &= - \alpha_+ \log \frac{\lambda_+}{\bar{\lambda}_+-\hat{u}} - \alpha_- \log \frac{\lambda_-}{\bar{\lambda}_-+\hat{u}} + \left( \hat{u} + \frac{1}{2} \right) K \\
&= \left( \frac{K}{2} - \alpha_+ \log \frac{\lambda_+}{\bar{\lambda}_+} - \alpha_- \log \frac{\lambda_-}{\bar{\lambda}_-} \right) + K \hat{u} + \alpha_+ \log \left( 1 - \frac{\hat{u}}{\bar{\lambda}_+} \right) + \alpha_- \log \left( 1 + \frac{\hat{u}}{\bar{\lambda}_-} \right) .
\end{align*}

Combining $\hat{u}$ and $\psi(\tilde{u})$,
\begin{equation} \begin{aligned}
\omega(x) &= \hat{u} x + \psi(\tilde{u}) \\
&= \left( \frac{K}{2} - \alpha_+ \log \frac{\lambda_+}{\bar{\lambda}_+} - \alpha_- \log \frac{\lambda_-}{\bar{\lambda}_-} \right) + (K+x) \hat{u}(x) + \alpha_+ \log \left( 1 - \frac{\hat{u}(x)}{\bar{\lambda}_+} \right) + \alpha_- \log \left( 1 + \frac{\hat{u}(x)}{\bar{\lambda}_-} \right) \\
&= \left( \frac{K}{2} - \alpha_+ \log \frac{\lambda_+}{\bar{\lambda}_+} - \alpha_- \log \frac{\lambda_-}{\bar{\lambda}_-} \right) - \frac{1}{2} \left( \alpha_++\alpha_- + (\bar{\lambda}_- - \bar{\lambda}_+) (K+x) \right) \\
&\quad + \frac{1}{2} \sqrt{4\alpha_+\alpha_- + \left( (\bar{\lambda}_+ + \bar{\lambda}_-) (K+x) - (\alpha_+-\alpha_-) \right)^2} \\
&\quad + \alpha_+ \log \left( 1 - \frac{\hat{u}(x)}{\bar{\lambda}_+} \right) + \alpha_- \log \left( 1 + \frac{\hat{u}(x)}{\bar{\lambda}_-} \right) \\
&= \left( \frac{K}{2} - \alpha_+ \log \frac{\lambda_+}{\bar{\lambda}_+} - \alpha_- \log \frac{\lambda_-}{\bar{\lambda}_-} - \frac{\alpha_++\alpha_-}{2} \right) - \frac{\bar{\lambda}_- - \bar{\lambda}_+}{2} (K+x) \\
&\quad + \sqrt{\alpha_+\alpha_- + \left( \frac{\bar{\lambda}_+ + \bar{\lambda}_-}{2} (K+x) - \frac{\alpha_+-\alpha_-}{2} \right)^2} \\
&\quad + \alpha_+ \log \left( 1 - \frac{\hat{u}(x)}{\bar{\lambda}_+} \right) + \alpha_- \log \left( 1 + \frac{\hat{u}(x)}{\bar{\lambda}_-} \right) 
\end{aligned} \label{eq:bgomega} \end{equation}
which is an SVI corrected by two non-trivial log terms depending on $\hat{u}(x)$, with strictly positive arguments.

\subsubsection{Wing Limit}

Observe that $\hat{u}(x) \in (-\bar{\lambda}_-,\bar{\lambda}_+)$ and $\hat{u}(\pm\infty) = \pm \bar{\lambda}_\pm$. Thus, in call/put-wings, one of the logs in $\omega(x)$ will tend to $\log 0$ -- the same sub-dominant log-correction observed in VG. We now mathematically show this.

We use Taylor-expansion $\hat{u}(x)$ to leading order in $1/(K+x)$, for large $x$ to get
\begin{align*}
\hat{u}(x) &\approx - \frac{\bar{\lambda}_- - \bar{\lambda}_+}{2} - \frac{1}{2} \frac{\alpha_++\alpha_-}{K+x} \pm \frac{\bar{\lambda}_+ + \bar{\lambda}_-}{2} \left( 1 - \frac{\alpha_+-\alpha_-}{\bar{\lambda}_+ + \bar{\lambda}_-} \frac{1}{K+x} \right) \\
&= \begin{cases} +\bar{\lambda}_+ - \frac{\alpha_+}{K+x} & x = +\infty \\ -\bar{\lambda}_- - \frac{\alpha_-}{K+x} & x = -\infty \end{cases} .
\end{align*}

In the wings,
\begin{align*}
\omega(x) &\approx \left( - \frac{\bar{\lambda}_- - \bar{\lambda}_+}{2} \pm - \frac{\bar{\lambda}_+ + \bar{\lambda}_-}{2} \right) (K+x) + \alpha_+ \log \left( \bar{\lambda}_+-\hat{u}(x) \right) + \alpha_- \log \left( \bar{\lambda}_-+\hat{u}(x) \right) \\
&= \begin{cases} \bar{\lambda}_+ (K+x) + \alpha_+ \log \frac{\alpha_+}{K+x} + \alpha_- \log \left( \bar{\lambda}_++\bar{\lambda}_--\frac{\alpha_+}{K+x} \right) & x = +\infty \\ -\bar{\lambda}_- (K+x) + \alpha_+ \log \left( \bar{\lambda}_++\bar{\lambda}_-+\frac{\alpha_-}{K+x} \right) + \alpha_- \log \frac{-\alpha_-}{K+x} & x = -\infty \end{cases} .
\end{align*}

So to leading order,
\begin{equation*} \omega(\pm\infty) \sim \bar{\lambda}_\pm |x| - \alpha_\pm \log |x| \end{equation*}
i.e.\ sub-dominant correction for variance wings is a log -- the same conclusion we reach for VG.

$\bar{\lambda}_\pm$ control wing-skew; $\alpha_\pm$ control log-correction.

\subsubsection{Numerical Experiment} \label{sec:bgnum}

Assuming the following typical BG parameters, we illustrate the convergence of volatility smiles computed via FFT into the large-time limit, from solving saddle-point equation (\ref{eq:sdlpt}) plugging in BG's $\omega(x)$. Both plus/minus solutions are plotted causing the seeming explosion ATM.
\begin{equation*} \alpha_+ = 10, \quad \alpha_- = 0.6, \quad \lambda_+ = 35, \quad \lambda_- = 5 \end{equation*}

We plot the convergence of the overall smiles $\sigma(x,T)$ and ATM implied volatilities $\sigma_0(T) \equiv \sigma(0,T)$ as time to expiry $T$ grows.

\begin{table}[H]
\centering
\begin{tabular}{|c|c|}
\hline
Overall Smile & ATM Implied Volatility \\ 
\hline \hline
\includegraphics[width=0.4\linewidth]{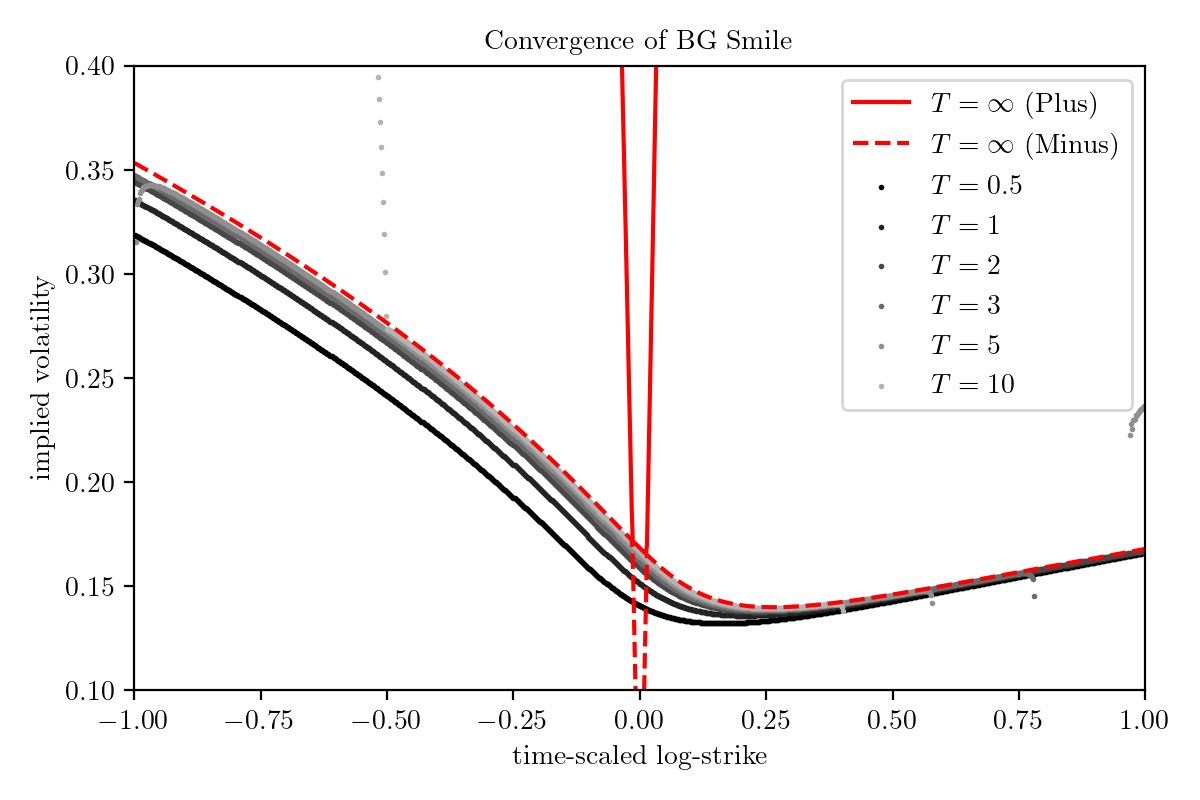} & 
\includegraphics[width=0.4\linewidth]{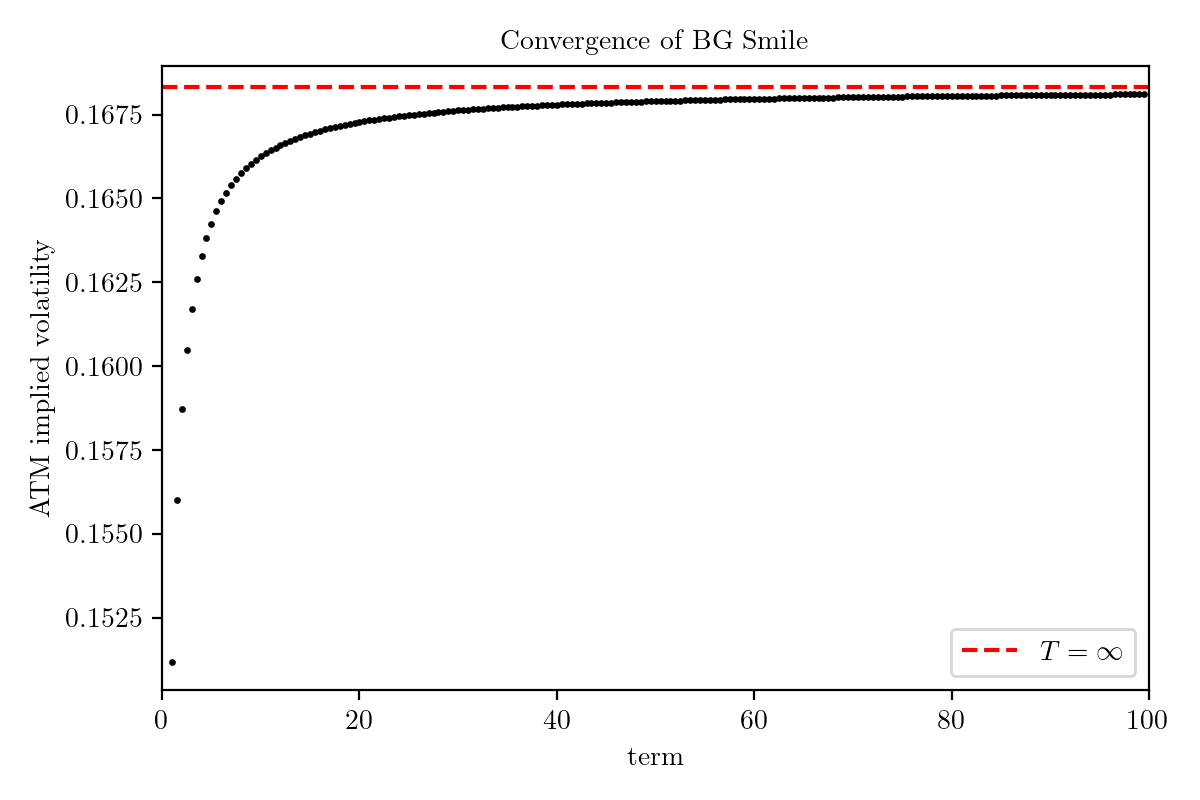} \\ 
\hline
\end{tabular}
\caption{Overall smile and ATM implied volatility under BG model.}
\end{table}

\subsection{CGMY Model} \label{sec:cgmy}

We derive the large-time CGMY variance smile $v(x)$ as a function of time-scaled log-strike $x$.

CGMY process for log-spot is a pure-jump process defined by Lévy measure
\begin{equation*} \mu(x) = \begin{cases} \frac{Ce^{-Mx}}{x^{1+Y}} & \text{for } x > 0 \\ \frac{Ce^{-G|x|}}{|x|^{1+Y}} & \text{for } x < 0 \end{cases} \end{equation*}
with characteristic function, obtained from Lévy-Khintchine formula,
\begin{equation*} \phi_T(u) = \exp \left\{ T C \Gamma(-Y) \left[ (M-iu)^Y - M^Y + (G+iu)^Y - G^Y - iu \left( (M-1)^Y - M^Y + (G+1)^Y - G^Y \right) \right] \right\} \end{equation*}
following \cite{cgmy-assetrtn}.

As a sanity check, we see $\phi_T(-i)=1$.

Note that CGMY covers VG as a special case, by requiring $Y=0$.

Define characteristic constants:
\begin{align*}
\bar{G} &= G + \frac{1}{2} \\
\bar{M} &= M - \frac{1}{2} \\
K &= \frac{1}{Y} \left( (M-1)^Y - M^Y + (G+1)^Y - G^Y \right) .
\end{align*}

We solve $\psi(u)$ from $\phi_T(u-i/2) \equiv e^{-\psi(u)T}$, which gives
\begin{equation*} \psi(u) = - C \Gamma(-Y) \left[ (\bar{M}-iu)^Y - M^Y + (\bar{G}+iu)^Y - G^Y - \left(iu+\frac{1}{2}\right) KY \right] . \end{equation*}

Differentiating $\psi(u)$ to get
\begin{equation*} \psi'(u) = - C \Gamma(-Y) \left( -iY(\bar{M}-iu)^{Y-1} + iY(\bar{G}+iu)^{Y-1} - iKY \right) = -ix \end{equation*}
so $\hat{u} \equiv i\tilde{u}$ satisfies
\begin{equation} \left(\bar{G}+\hat{u}\right)^{Y-1} - \left(\bar{M}-\hat{u}\right)^{Y-1} = K + \frac{x}{CY\Gamma(-Y)} . \label{eq:cgmyeq} \end{equation}

This is a non-linear equation in $\hat{u}$.

\subsubsection{Reduction to Variance-Gamma} \label{sec:redvg}

We first consider special case $Y=0$, which corresponds to VG.

If our derivation is correct, $\hat{u}$ must reduce exactly to 
\begin{equation*} \hat{u} = - \frac{\xi}{2\alpha} - \frac{1}{\Sigma(x)} \pm \sqrt{\left( \frac{1}{\Sigma(x)} \right)^2 + \left( \frac{\xi}{2\alpha} \right)^2 + \frac{\beta}{\alpha}} \end{equation*}
where $\Sigma(x) = \nu x - \log(1-\xi)$, which we now show.

In the limit $Y \approx 0$, $Y\Gamma(-Y) \approx -1$ hence
\begin{align*}
\frac{1}{\bar{G}+\hat{u}} - \frac{1}{\bar{M}-\hat{u}} &= K - \frac{x}{C} \\
\bar{M} - \bar{G} - 2\hat{u} &= \left(K - \frac{x}{C}\right) \left(\bar{G}\bar{M}+(\bar{M}-\bar{G})\hat{u}-\hat{u}^2\right) \\
\left(K - \frac{x}{C}\right) \left(\hat{u}^2 - \left(\bar{M}-\bar{G}+\frac{2}{K - \frac{x}{C}}\right)\hat{u}\right) &= \bar{G}\bar{M} \left(K - \frac{x}{C}\right) - (\bar{M}-\bar{G}) \\
\left(\hat{u}^2 - \left(\frac{\bar{M}-\bar{G}}{2}+\frac{1}{K - \frac{x}{C}}\right)\right)^2 &= \left(\frac{\bar{M}-\bar{G}}{2}+\frac{1}{K - \frac{x}{C}}\right)^2 + \bar{G}\bar{M} - \frac{\bar{M}-\bar{G}}{K - \frac{x}{C}} \\
\hat{u} = -\frac{\bar{G}-\bar{M}}{2}-\frac{1}{\frac{x}{C} - K} &\pm \sqrt{\left(\frac{1}{\frac{x}{C} - K}\right)^2 + \left(\frac{\bar{G}-\bar{M}}{2}\right)^2 + \bar{G}\bar{M}}
\end{align*}
taking identical form as VG, with correspondence
\begin{align*}
\frac{\xi}{2\alpha} &\equiv \frac{\bar{G}-\bar{M}}{2} \\
\frac{\beta}{\alpha} &\equiv \bar{G}\bar{M} \\
\Sigma(x) &\equiv \frac{x}{C}-K
\end{align*}
from which we back out relations between $C$, $G$, $M$ and $\theta$, $\sigma$, $\nu$.

\subsubsection{Approximation of $\hat{u}$}

Back to our non-linear equation for $\hat{u}$. We obtain an approximation based on Taylor-expansion for small $Y>0$.

Taylor-expansion of LHS of equation (\ref{eq:cgmyeq}) gives:
\begin{align*}
&\bar{G}^{Y-1} \left(1+\frac{\hat{u}}{\bar{G}}\right)^{Y-1} - \bar{M}^{Y-1} \left(1-\frac{\hat{u}}{\bar{M}}\right)^{Y-1} \\
=& \; \left(\bar{G}^{Y-1}-\bar{M}^{Y-1}\right) + (Y-1) \left(\bar{G}^{Y-2}+\bar{M}^{Y-2}\right) \hat{u} + \frac{(Y-1)(Y-2)}{2} \left(\bar{G}^{Y-3}-\bar{M}^{Y-3}\right) \hat{u}^2 + O(\hat{u}^3) .
\end{align*}

Constant term cancels roughly with $K \approx \frac{Y\bar{G}^{Y-1}-Y\bar{M}^{Y-1}}{Y} = \bar{G}^{Y-1}-\bar{M}^{Y-1}$. Denote $\xi \equiv \bar{M}/\bar{G}$ so
\begin{align*}
\frac{x}{CY\Gamma(-Y)} &\approx (Y-1) \bar{G}^{Y-1} \hat{u} \left[ (1+\xi^{Y-2}) + \frac{Y-2}{2!} (1-\xi^{Y-3}) \left(\frac{\hat{u}}{\bar{G}}\right) + \frac{(Y-2)(Y-3)}{3!} (1+\xi^{Y-4}) \left(\frac{\hat{u}}{\bar{G}}\right)^2... \right] \\
&= (Y-1) \bar{G}^{Y-1} \hat{u} \left[ 1 + \frac{Y-2}{2!} \left(\frac{\hat{u}}{\bar{G}}\right) + \frac{(Y-2)(Y-3)}{3!} \left(\frac{\hat{u}}{\bar{G}}\right)^2... \right. \\
&\quad \left. + \xi^{Y-2} \left( 1 - \frac{Y-2}{2!} \left(\frac{\hat{u}}{\bar{M}}\right) + \frac{(Y-2)(Y-3)}{3!} \left(\frac{\hat{u}}{\bar{M}}\right)^2... \right) \right] \\
&\approx (Y-1) \bar{G}^{Y-1} \hat{u} \left[ 1 - \left(\frac{\hat{u}}{\bar{G}}\right) + \left(\frac{\hat{u}}{\bar{G}}\right)^2... + \xi^{Y-2} \left( 1 + \left(\frac{\hat{u}}{\bar{M}}\right) + \left(\frac{\hat{u}}{\bar{M}}\right)^2... \right) \right] \\
&= (Y-1) \bar{G}^{Y-1} \hat{u} \left[ \frac{1}{1+\hat{u}/\bar{G}} + \frac{\xi^{Y-2}}{1-\hat{u}/\bar{M}} \right]
\end{align*}
which gives
\begin{equation*} \Sigma(x) \equiv \frac{x}{CY(Y-1)\Gamma(-Y)\bar{G}^{Y-2}} \approx \frac{1}{1/\hat{u}+1/\bar{G}} + \frac{\xi^{Y-2}}{1/\hat{u}-1/\bar{M}} . \end{equation*}

We cast this into a quadratic equation
\begin{equation*} \left(\frac{1}{\hat{u}}\right)^2 + \left(\frac{1}{\bar{G}}-\frac{1}{\bar{M}}-\frac{1+\xi^{Y-2}}{\Sigma(x)}\right)\left(\frac{1}{\hat{u}}\right) \approx \left(\frac{\xi^{Y-2}}{\bar{G}}-\frac{1}{\bar{M}}\right)\frac{1}{\Sigma(x)} + \frac{1}{\bar{G}\bar{M}} \end{equation*}
thus
\begin{equation} \begin{aligned} \frac{1}{\hat{u}(x)} &\approx - \frac{1}{2} \left(\frac{1}{\bar{G}}-\frac{1}{\bar{M}}-\frac{1+\xi^{Y-2}}{\Sigma(x)}\right) + (\bbone_{x \geq 0}-\bbone_{x < 0}) \cdot \\
&\quad \; \sqrt{\left(\frac{\xi^{Y-2}}{\bar{G}}-\frac{1}{\bar{M}}\right)\frac{1}{\Sigma(x)} + \frac{1}{\bar{G}\bar{M}} + \frac{1}{4} \left(\frac{1}{\bar{G}}-\frac{1}{\bar{M}}-\frac{1+\xi^{Y-2}}{\Sigma(x)}\right)^2} . \end{aligned} \label{eq:cgmyuhat} \end{equation}

\subsubsection{Approximation of $\psi(\tilde{u})$}

It remains to evaluate
\begin{equation*} \psi(\tilde{u}) = - C \Gamma(-Y) \left( (\bar{M}-\hat{u})^Y - M^Y + (\bar{G}+\hat{u})^Y - G^Y - \left(\hat{u}+\frac{1}{2}\right) KY \right) \end{equation*}
which we Taylor-expand the bracket:
\begin{align*}
&(\bar{M}-\hat{u})^Y - M^Y + (\bar{G}+\hat{u})^Y - G^Y - \left(\hat{u}+\frac{1}{2}\right) KY \\
=& \; \bar{M}^Y \left(1-\frac{\hat{u}}{\bar{M}}\right)^Y + \bar{G}^Y \left(1+\frac{\hat{u}}{\bar{G}}\right)^Y - \left(M^Y+G^Y+\frac{KY}{2}\right) - KY\hat{u} \\
=& \; \bar{M}^Y \left(1-Y\left(\frac{\hat{u}}{\bar{M}}\right)+\frac{Y(Y-1)}{2}\left(\frac{\hat{u}}{\bar{M}}\right)^2...\right) + \bar{G}^Y \left(1+Y\left(\frac{\hat{u}}{\bar{G}}\right)+\frac{Y(Y-1)}{2}\left(\frac{\hat{u}}{\bar{G}}\right)^2...\right) \\
& \; - \left(M^Y+G^Y+\frac{KY}{2}\right) - KY\hat{u} \\
=& \; \left(\bar{M}^Y+\bar{G}^Y-M^Y-G^Y-\frac{KY}{2}\right) + Y \left(\bar{G}^{Y-1}-\bar{M}^{Y-1}-K\right) \hat{u} + \frac{Y(Y-1)}{2} \left(\bar{M}^{Y-2}+\bar{G}^{Y-2}\right) \hat{u}^2 + O(\hat{u}^3) .
\end{align*}

Note $K \approx \bar{G}^{Y-1}-\bar{M}^{Y-1}$ and consider the series in $\hat{u}$. Factoring quadratic term out to get
\begin{align*}
&\frac{Y(Y-1)}{2} \bar{G}^{Y-2} \hat{u}^2 \left[ (1+\xi^{Y-2}) + \frac{Y-2}{3} (1-\xi^{Y-3}) \left(\frac{\hat{u}}{\bar{G}}\right) + \frac{(Y-2)(Y-3)}{3 \cdot 4} (1+\xi^{Y-4}) \left(\frac{\hat{u}}{\bar{G}}\right)^2... \right] \\
=& \; \frac{Y(Y-1)}{2} \bar{G}^{Y-2} \hat{u}^2 \left\{ 1 + \frac{Y-2}{3} \left(\frac{\hat{u}}{\bar{G}}\right) + \frac{(Y-2)(Y-3)}{3 \cdot 4} \left(\frac{\hat{u}}{\bar{G}}\right)^2... \right. \\
& \; \left. + \xi^{Y-2} \left[ 1 - \frac{Y-2}{3} \left(\frac{\hat{u}}{\bar{M}}\right) + \frac{(Y-2)(Y-3)}{3 \cdot 4} \left(\frac{\hat{u}}{\bar{M}}\right)^2... \right] \right\} .
\end{align*}

Again, assume small $Y>0$, thus we approximate the series term
\begin{align*}
&2 \left\{ \frac{1}{2} - \frac{1}{3} \left(\frac{\hat{u}}{\bar{G}}\right) + \frac{1}{4} \left(\frac{\hat{u}}{\bar{G}}\right)^2... + \xi^{Y-2} \left[ \frac{1}{2} + \frac{1}{3} \left(\frac{\hat{u}}{\bar{M}}\right) + \frac{1}{4} \left(\frac{\hat{u}}{\bar{M}}\right)^2... \right] \right\} \\
=& \; 2 \left\{ \frac{\hat{u}/\bar{G}-\log(1+\hat{u}/\bar{G})}{(\hat{u}/\bar{G})^2} + \xi^{Y-2} \frac{-\hat{u}/\bar{M}-\log(1-\hat{u}/\bar{M})}{(\hat{u}/\bar{M})^2} \right\} \\
=& \; 2 \left(\frac{\bar{G}}{\hat{u}}\right)^2 \left\{ \frac{\hat{u}}{\bar{G}} - \log\left(1+\frac{\hat{u}}{\bar{G}}\right) + \xi^Y \left\{ - \frac{\hat{u}}{\bar{M}} - \log\left(1-\frac{\hat{u}}{\bar{M}}\right) \right) \right\} .
\end{align*}

Plugging back in to the above equation to get
\begin{equation} \begin{aligned} \psi(\tilde{u}) &\approx -C\Gamma(-Y) \left[ Y(Y-1)\bar{G}^Y \left\{ \left( \frac{1}{\bar{G}} - \frac{\xi^Y}{\bar{M}} \right) \hat{u} - \left( \log\left(1+\frac{\hat{u}}{\bar{G}}\right) + \xi^Y \log\left(1-\frac{\hat{u}}{\bar{M}}\right) \right) \right\} \right. \\
&\quad \left. + \left\{ \bar{M}^Y+\bar{G}^Y-M^Y-G^Y-\frac{KY}{2} \right\} \right] . \label{eq:cgmypsi} \end{aligned} \end{equation}

\subsubsection{$\omega(x)$ under CGMY}

We invert $\hat{u}$ in (\ref{eq:cgmyuhat}) to get
\begin{align*}
\hat{u} &\approx \frac{\frac{1}{2} \left(\frac{1}{\bar{G}}-\frac{1}{\bar{M}}-\frac{1+\xi^{Y-2}}{\Sigma(x)}\right) + \operatorname{sgn}{\Sigma(x)} \sqrt{\left(\frac{\xi^{Y-2}}{\bar{G}}-\frac{1}{\bar{M}}\right)\frac{1}{\Sigma(x)} + \frac{1}{\bar{G}\bar{M}} + \frac{1}{4} \left(\frac{1}{\bar{G}}-\frac{1}{\bar{M}}-\frac{1+\xi^{Y-2}}{\Sigma(x)}\right)^2}}{\left(\frac{\xi^{Y-2}}{\bar{G}}-\frac{1}{\bar{M}}\right)\frac{1}{\Sigma(x)} + \frac{1}{\bar{G}\bar{M}}} \\
&= \frac{- \frac{1+\xi^{Y-2}}{2} + \left(\frac{1}{\bar{G}}-\frac{1}{\bar{M}}\right) \frac{\Sigma(x)}{2} + \sqrt{\left(\frac{\xi^{Y-2}}{\bar{G}}-\frac{1}{\bar{M}}\right)\Sigma(x) + \frac{\Sigma(x)^2}{\bar{G}\bar{M}} + \frac{\Sigma(x)^2}{4} \left(\frac{1}{\bar{G}}-\frac{1}{\bar{M}}-\frac{1+\xi^{Y-2}}{\Sigma(x)}\right)^2}}{\frac{\xi^{Y-2}}{\bar{G}}-\frac{1}{\bar{M}} + \frac{\Sigma(x)}{\bar{G}\bar{M}}} \\
&= \frac{- \frac{1+\xi^{Y-2}}{2} + \left(\frac{1}{\bar{G}}-\frac{1}{\bar{M}}\right) \frac{\Sigma(x)}{2} + \sqrt{\frac{1}{4} \left[\left(\frac{1}{\bar{G}}+\frac{1}{\bar{M}}\right)\Sigma(x)-(1-\xi^{Y-2})\right]^2 + \xi^{Y-2}}}{\frac{\xi^{Y-2}}{\bar{G}}-\frac{1}{\bar{M}} + \frac{\Sigma(x)}{\bar{G}\bar{M}}} \in (-\bar{G},\bar{M}) .
\end{align*}

Notice that numerator is an SVI.

Denote
\begin{equation*} \bar{K} = \frac{\bar{M}^Y+\bar{G}^Y-M^Y-G^Y-\frac{KY}{2}}{Y(Y-1)\bar{G}^Y} . \end{equation*}

Now we simplify
\begin{equation} \begin{aligned}
\omega(x) &= \hat{u} \cdot x + \psi(\tilde{u}) \\
&= \hat{u} \cdot CY(Y-1)\Gamma(-Y)\bar{G}^{Y-2} \Sigma(x) + \psi(\tilde{u}) \\
&\approx CY(Y-1)\Gamma(-Y)\bar{G}^Y \left\{ -\bar{K} + \left[ \frac{\Sigma(x)}{\bar{G}^2} - \left(\frac{1}{\bar{G}}-\frac{\xi^Y}{\bar{M}}\right) \right] \hat{u}(x) + \left[ \log\left(1+\frac{\hat{u}(x)}{\bar{G}}\right) + \xi^Y \log\left(1-\frac{\hat{u}(x)}{\bar{M}}\right) \right] \right\} \\
&= CY(Y-1)\Gamma(-Y)\bar{G}^Y \Bigg\{ -\bar{K} + \xi \Bigg[ - \frac{1+\xi^{Y-2}}{2} + \left(\frac{1}{\bar{G}}-\frac{1}{\bar{M}}\right) \frac{\Sigma(x)}{2} + \\
&\quad \left. \left. \sqrt{\frac{1}{4} \left[\left(\frac{1}{\bar{G}}+\frac{1}{\bar{M}}\right)\Sigma(x)-(1-\xi^{Y-2})\right]^2 + \xi^{Y-2}} \right] + \left[ \log\left(1+\frac{\hat{u}(x)}{\bar{G}}\right) + \xi^Y \log\left(1-\frac{\hat{u}(x)}{\bar{M}}\right) \right] \right\}
\end{aligned} \label{eq:cgmyomega} \end{equation}
which like BG, is an SVI corrected by two non-trivial log terms.

\subsubsection{Numerical Experiment}

Assuming the following typical CGMY parameters, we illustrate the convergence of volatility smiles computed via FFT into the large-time limit, from solving saddle-point equation (\ref{eq:sdlpt}) plugging in CGMY's $\omega(x)$. Both plus/minus solutions are plotted causing the seeming explosion ATM.
\begin{equation*} C = 20, \quad G = 80, \quad M = 120, \quad Y = 0.25 \end{equation*}

We plot the convergence of the overall smiles $\sigma(x,T)$ and ATM implied volatilities $\sigma_0(T) \equiv \sigma(0,T)$ as time to expiry $T$ grows. Note that our large-time CGMY smile here is not exact as we have made approximations in $\hat{u}$ and $\psi(\tilde{u})$ -- small $Y>0$.

\begin{table}[H]
\centering
\begin{tabular}{|c|c|}
\hline
Overall Smile & ATM Implied Volatility \\ 
\hline \hline
\includegraphics[width=0.4\linewidth]{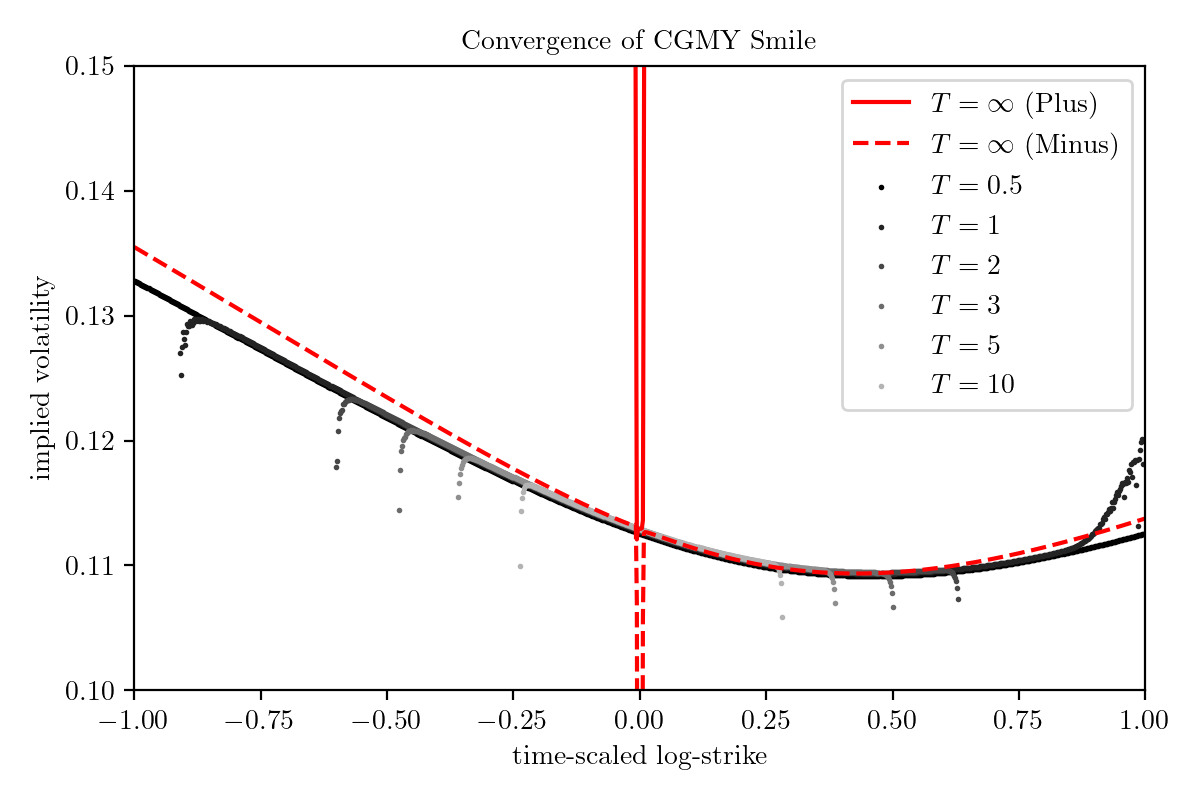} & 
\includegraphics[width=0.4\linewidth]{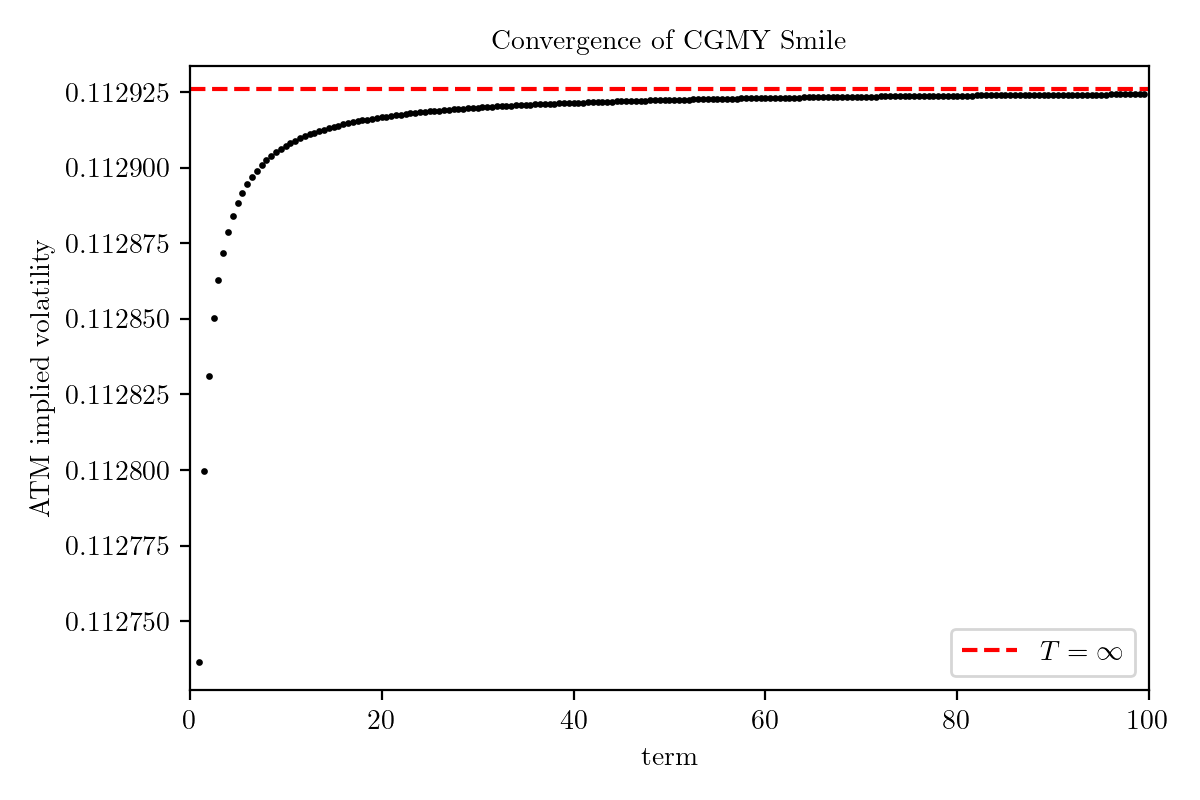} \\ 
\hline
\end{tabular}
\caption{Overall smile and ATM implied volatility under CGMY model.}
\end{table}

\subsection{Merton Jump-Diffusion Model} \label{sec:merton}

We derive the large-time Merton-jump variance smile $v(x)$ as a function of time-scaled log-strike $x$.

Merton-jump model is characteristized by a diffusion process in spot $(S_t)_{t \geq 0}$ accompanied by log-normal jumps $J_t$ modeled by a Poisson process $q_t$ of rate $\lambda$, with Brownian motion $Z_t$, parametrized by diffusion volatility $\sigma$, jump rate $\lambda$, jump-size mean $\alpha$ and jump-size s.d.\ $\delta$, suitably constraint:
\begin{align*}
&dS_t = \sigma S_t dZ_t + (J_t-1) S_t dq_t \\
&\log J_t \sim N(\alpha,\delta^2)
\end{align*}
where $S_t$ is to be appropriately compensated under the risk-neutral measure.

The characteristic function, obtained from Lévy-Khintchine formula, reads
\begin{equation*} \phi_T(u) = \exp\left\{ T \left[ - \frac{\sigma^2}{2}u(u+i) + \lambda \left( e^{iu\alpha-u^2\delta^2/2}-1-iu\left(e^{\alpha+\delta^2/2}-1\right) \right) \right] \right\} \end{equation*}
following \cite{gat-volsurf}.

As a sanity check, we see $\phi_T(-i)=1$.

Define characteristic constants:
\begin{align*}
\theta &= e^{\alpha+\delta^2/2}-1 \\
\bar{\sigma^2} &= \frac{\sigma^2}{\lambda} .
\end{align*}

We solve $\psi(u)$ from $\phi_T(u-i/2) \equiv e^{-\psi(u)T}$, which gives
\begin{equation*} \psi(u) = \frac{\sigma^2}{2} \left( \frac{1}{4}-(iu)^2 \right) - \lambda \left( e^{\alpha(iu+1/2)+\delta^2(iu+1/2)^2/2}-1-\left(iu+\frac{1}{2}\right)\theta \right) . \end{equation*}

Differentiating,
\begin{equation*} \psi'(u) = i \left[ -\sigma^2 \cdot iu - \lambda \left( e^{\alpha(iu+1/2)+\delta^2(iu+1/2)^2/2} \left( \alpha+\delta^2\left(iu+\frac{1}{2}\right) \right) - \theta \right) \right] = -ix \end{equation*}
so $\hat{u} \equiv i\tilde{u}$ satisfies
\begin{equation*} \sigma^2 \cdot \hat{u} + \lambda \left( e^{\alpha(\hat{u}+1/2)+\delta^2(\hat{u}+1/2)^2/2} \left( \alpha+\delta^2\left(\hat{u}+\frac{1}{2}\right) \right) - \theta \right) = x . \end{equation*}

Denote $\bar{u} \equiv \hat{u}+1/2$. Then,
\begin{equation} \bar{\sigma^2} \cdot \bar{u} + e^{\alpha\bar{u}+\delta^2\bar{u}^2/2} \left( \alpha+\delta^2\bar{u} \right) = \frac{x}{\lambda} + \frac{\bar{\sigma^2}}{2} + \theta . \label{eq:mereq} \end{equation}

This is a non-linear equation in $\bar{u}$, which spans the whole real line.

Unlike CGMY's equation which lends itself to Taylor-expansion, approximation here is not trivial -- we leave the study of Merton-jump model as future work. In the numerical experiment section, we show (1) like all previous models, Merton-jump exhibits some convergent large-time smile, and (2) how $\hat{u}$ varies as a function of $x$.

\subsubsection{Reduction to Black-Scholes}

As a sanity check, without jump component, the equation must lead to a $\hat{u}$ consistent with BS (\ref{eq:bsutilde}).

Suppose $\alpha=\delta=0$. Then,
\begin{equation*} \hat{u} = \frac{x}{\sigma^2} \end{equation*}
which agrees with BS.

\subsubsection{Numerical Experiment}

Assuming the following typical Merton-jump parameters, we illustrate the convergence of volatility smiles computed via FFT to illustrate the existence of some large-time smile.
\begin{equation*} \sigma = 0.1, \quad \lambda = 0.1, \quad \alpha = -0.4, \quad \delta = 0.4 \end{equation*}

We plot the convergence of the volatility smiles $\sigma(x,T)$ and variance smiles $v(x,T)$ as time to expiry $T$ grows. Notice that variance appears to tend to linear in the wings, consistent with Lee's bound, but appears to contradict equation (\ref{eq:betapm}) which predicts that wing skews $\beta_\pm \rightarrow 0$ must flatten out as $\hat{u}(\pm\infty) \rightarrow \pm\infty$. This is because $\hat{u}$ grows like the product-log function $W(x) \sim o(\log x)$, so slow that over reasonable observed strike domain $(L_-,L_+)$, $\hat{u}(L_\pm)$ remains bounded, hence non-zero $\beta_\pm$. Readers may contrast this with BS.

\begin{table}[H]
\centering
\begin{tabular}{|c|c|}
\hline
Volatility Smile & Variance Smile \\ 
\hline \hline
\includegraphics[width=0.4\linewidth]{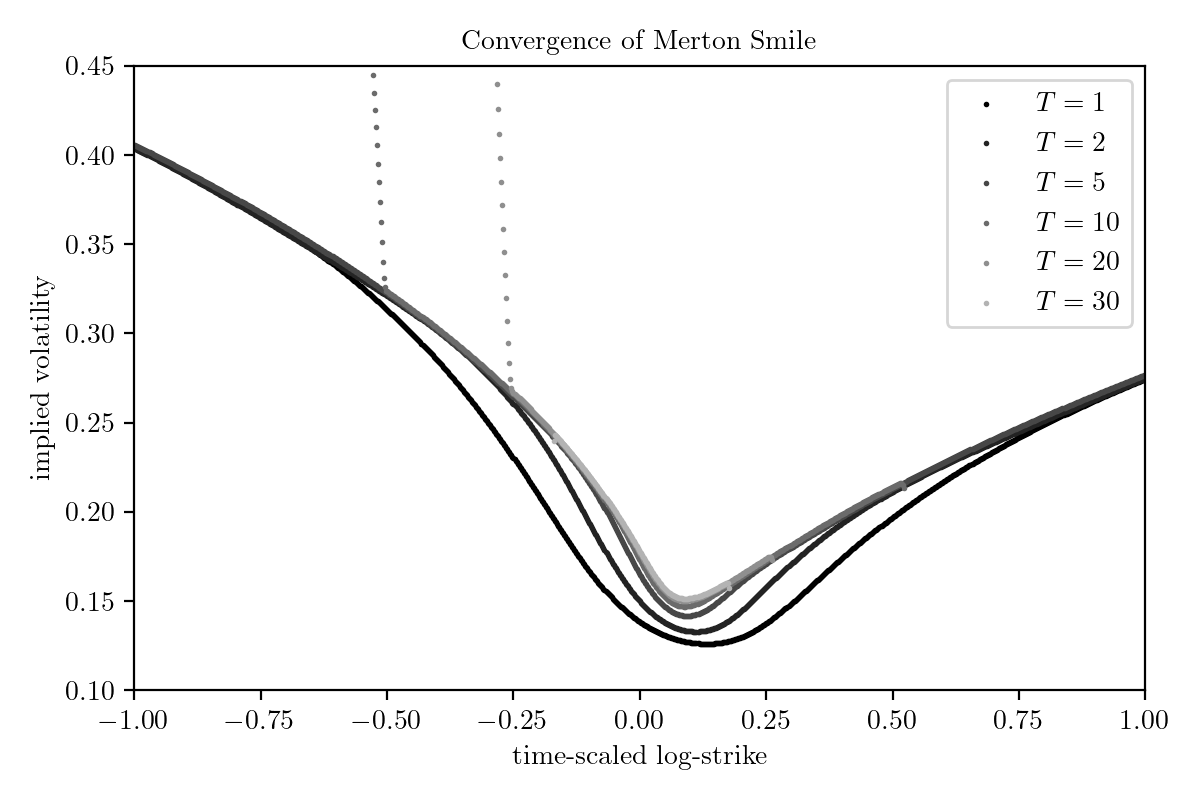} & 
\includegraphics[width=0.4\linewidth]{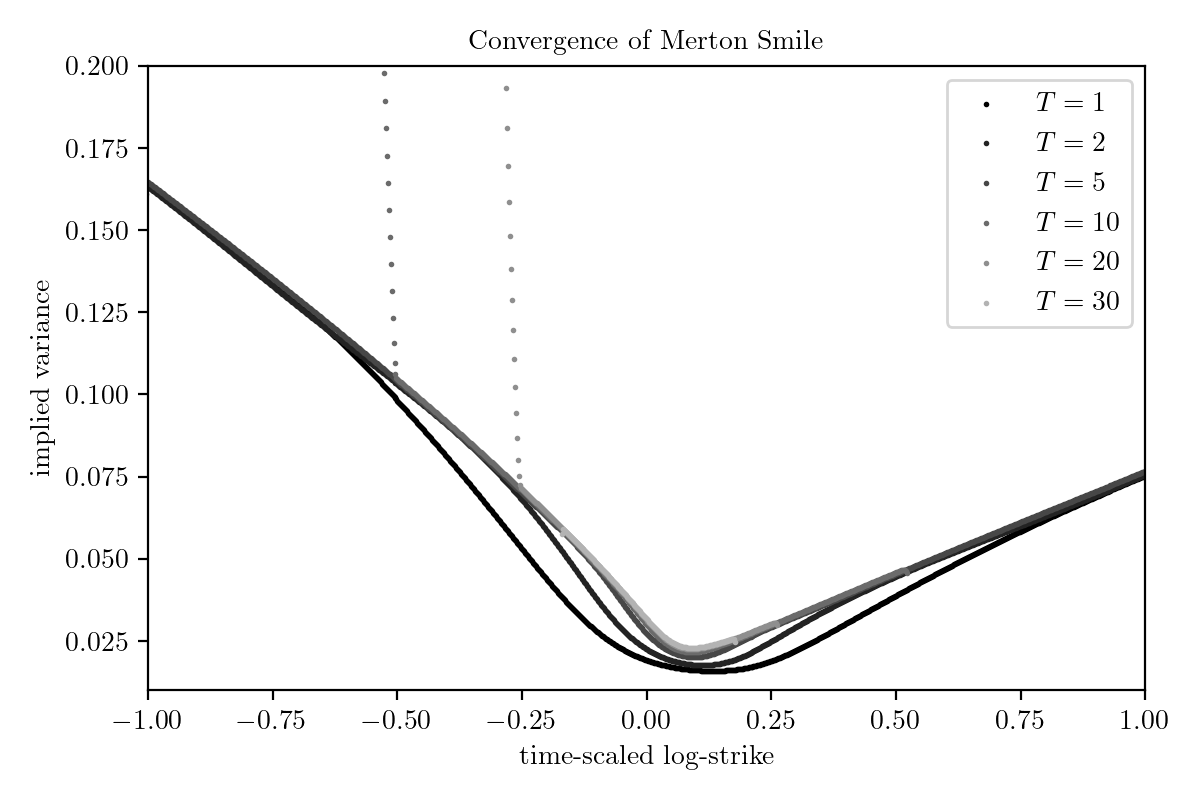} \\ 
\hline
\end{tabular}
\caption{Volatility and variance smiles under Merton-jump model.}
\end{table}

We plot $\hat{u} \equiv \bar{u}-1/2$ as a function of $x$. Locally at $x=0$, assuming small $\alpha$ and $\delta$,
\begin{equation*} \bar{u}(x) \approx \frac{x/\lambda + \bar{\sigma^2}/2 + \theta - \alpha}{\bar{\sigma^2} + \delta^2} \end{equation*}
with
\begin{equation*} \bar{u}(0) \approx \frac{\bar{\sigma^2}/2 + \delta^2/2}{\bar{\sigma^2} + \delta^2} = \frac{1}{2} , \end{equation*}
while for large $x$, $\bar{u}(x)$ solves
\begin{equation*} e^{\alpha\bar{u}(x)+\delta^2\bar{u}(x)^2/2} (\alpha+\delta^2\bar{u}(x)) \approx \frac{x}{\lambda} \Rightarrow \bar{u}(x) \approx \frac{1}{\delta^2} \left( \pm \delta \sqrt{W\left(e^{(\alpha/\delta)^2}\left(\frac{x}{\lambda\delta}\right)^2\right)} - \alpha \right) \end{equation*}
respectively for $x \rightarrow \pm \infty$ and $W$ is the product-log function.

A global approximation for $\bar{u}(x)$ that fulfills these limits is desired.

\begin{table}[H]
\centering
\begin{tabular}{|c|c|}
\hline
Zoom in ATM & Zoom out ATM \\ 
\hline \hline
\includegraphics[width=0.4\linewidth]{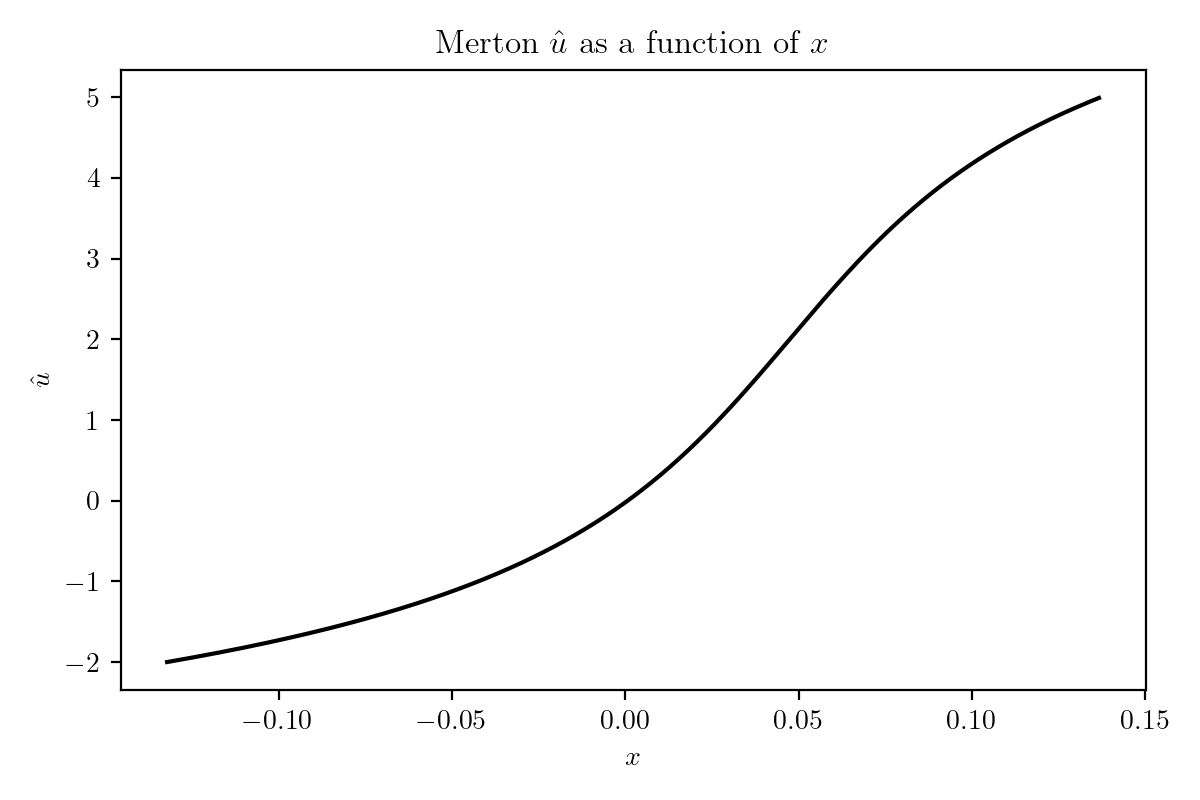} & 
\includegraphics[width=0.4\linewidth]{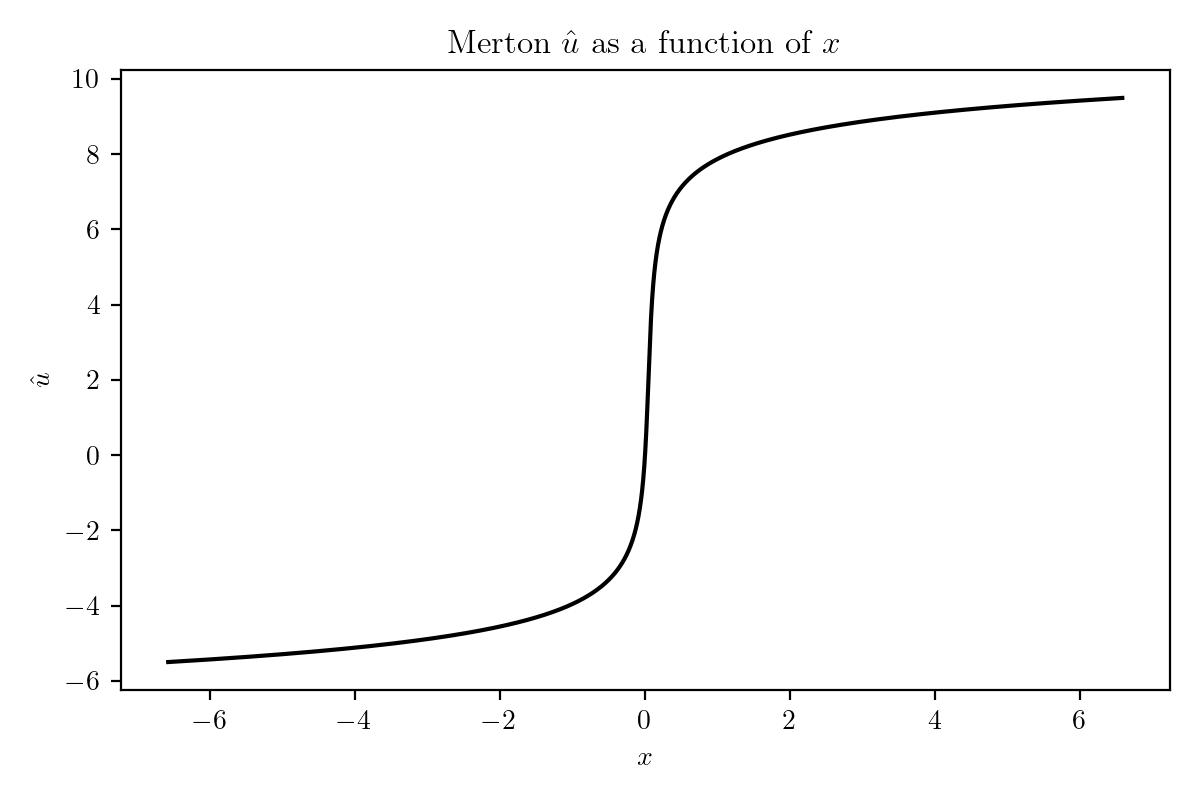} \\ 
\hline
\end{tabular}
\caption{ATM and global behavior of $\bar{u}(x)$ under Merton-jump model.}
\end{table}

\subsection{Model-Inspired Parametrization} \label{sec:modinsp}

Variance smile $v(x)$ is the large-time limit of some specific model. To inspire a parametrization, we relax the large-time condition, obtaining an implicit parametrization for total implied variance $w(k,T)$ at time slice $T<\infty$ in terms of model parameters hidden in (model-specific) $\omega(x)$:
\begin{equation} \frac{w(k,T)}{8} + \frac{k^2}{2w(k,T)} \equiv \omega\left(\frac{k}{T}\right) \cdot T , \label{eq:modinsp} \end{equation}
in which our parametrized $w(k,T)$ has the advantages of being (1) motivated from model (thus expressible entirely in terms of model parameters), (2) arbitrage-free along strike, and (3) fast to compute and calibrate, given closed-form $\omega(x)$. SVI may be derived in this manner plugging in Heston's $\omega(x)$.

What about calendar arbitrage? How does it constrain the term-structures of parameters?

To be free of calendar arbitrage, we require $\partial_T w(k,T) \geq 0$. Consider the wings i.e.\ large $k$ which is where calendar arbitrage typically occurs in practice. Differentiating, we obtain a necessary condition on time-derivative of $\omega(x)$, with implicit time-dependence in the model parameters:
\begin{align*}
\left(\frac{1}{8} - \frac{1}{2}\left(\frac{k}{w}\right)^2\right) \partial_T w = \omega - \frac{k}{T} \partial_T \omega \leq 0 \Rightarrow x \partial_T \log \omega(x) \geq 1
\end{align*}
where the first bracketed expression is always negative due to Lee's bound: $w(k,T) < 2|k|$ for large $k$ \cite{lee-moment}. One may derive specific conditions on model parameters in order to fulfill this calendar inequality -- we leave this for future work. Intuitively, the inequality says, if calendar arbitrage is absent, for each fixed $x$, $x\log\omega(x)$ plotted over time to expiry $T$ grows at a slope not smaller than unity.

\subsection{Summary} \label{sec:summary}

The large-time model-implied variance smile $v(x)$ as a function of time-scaled log-strike $x$ is formulated as the solution to the (quadratic) saddle-point equation (\ref{eq:sdlpt}) with model-specific $\omega(x)$. Here, we summarize the model parameters, characteristic constants and $\omega(x)$ of Heston, VG, BG and CGMY, whose calculations we completed. For Heston and VG, we also present the smile $v(x)$.

\vspace{0.5in}

\begin{itemize}
\item \textbf{Heston model}
\begin{itemize}
\item Model parameters: long-run mean variance $\bar{v}$, mean-reversion rate $\lambda$, volatility of volatility $\eta$ and correlation $\rho$
\item Characteristic constants:
\begin{align*}
A^2 &= \eta^2(1-\rho^2) \\
B &= \rho\eta\left(\lambda-\frac{\rho\eta}{2}\right) \\
C^2 &= \left(\lambda-\frac{\rho\eta}{2}\right)^2 + \frac{\eta^2}{4} \\
D^2 &= \left(\frac{1}{A}\right)^2 + \left(\frac{C}{B}\right)^2 \\
\xi &= \frac{\eta^2}{\lambda\bar{v}} \\
m &= -\frac{\rho\eta}{\xi} \\
a &= \frac{\rho\eta}{\lambda} \\
K &= \sqrt{1 + \frac{aA^2/4B}{1-a/2}}
\end{align*}
\item Variance smile: assuming negative correlation $\rho<0$,
\begin{align*}
\omega(x) &= - \frac{\lambda}{\xi} \left( 1 - \frac{a}{2} \right) - \frac{B}{A^2} (x-m) - \frac{BD}{\xi} \sqrt{ 1 + \xi^2\left(\frac{1}{A}\right)^2(x-m)^2 } \\
v(x) &= 4(K-1) \left( \frac{\lambda}{\xi} \left( 1 - \frac{a}{2} \right) + \frac{B}{A^2} (x-m) - \frac{1}{K} \frac{BD}{\xi} \sqrt{ 1 + \xi^2\left(\frac{1}{A}\right)^2(x-m)^2 } \right)
\end{align*}
\end{itemize}

\item \textbf{Variance-Gamma model}
\begin{itemize}
\item Model parameters: drift (skew) $\theta$, volatility $\sigma$ and variance of gamma time-change (kurtosis) $\nu$
\item Characteristic constants:
\begin{align*}
\alpha &= \frac{\sigma^2\nu}{2} \\
\beta &= 1 - \frac{\theta\nu}{2} - \frac{\sigma^2\nu}{8} \\
\xi &= \left( \theta + \frac{\sigma^2}{2} \right) \nu \\
\eta^2 &= \nu^2 \left( \left( \frac{\xi}{2\alpha} \right)^2 + \frac{\beta}{\alpha} \right) \\
x_0 &= \frac{\log(1-\xi)}{\nu} \\
K &= \frac{1 - \frac{1}{2} \left( 1 - \frac{\alpha}{\xi} \right) \frac{x_0\nu}{\log \frac{\alpha\eta^2}{\nu^2}}}{\sqrt{1 - \frac{x_0\nu}{\log \frac{\alpha\eta^2}{\nu^2}}}}
\end{align*}
\item Variance smile:
\begin{align*}
\omega(x) &= - \frac{x_0}{2} - \frac{\xi}{2\alpha} (x-x_0) + \frac{1}{\nu} \left( \sqrt{1 + \eta^2 (x-x_0)^2} - 1 \right) \\
&\quad + \frac{1}{\nu} \log \frac{2\alpha}{\nu^2(x-x_0)^2} \left( \sqrt{1 + \eta^2 (x-x_0)^2} - 1 \right) \\
v(x) &\approx 4(K-1) \left( - \frac{x_0}{2K} \left( 1 + \frac{\alpha}{(K-1)\xi} \right) + \frac{\xi}{2\alpha} (x-x_0) + \frac{1}{K\nu} \left( \sqrt{1 + \eta^2 (x-x_0)^2} - 1 \right) \right. \\
&\quad \left. + \frac{1}{K\nu} \log \frac{2\alpha}{\nu^2(x-x_0)^2} \left( \sqrt{1 + \eta^2 (x-x_0)^2} - 1 \right) \right)
\end{align*}
\end{itemize}

\item \textbf{Bilateral-Gamma model}
\begin{itemize}
\item Model parameters: $\alpha_+, \alpha_-, \lambda_+, \lambda_-$, which characterize the Lévy measure
\item Characteristic constants:
\begin{align*}
\bar{\lambda}_+ &= \lambda_+ - \frac{1}{2} \\
\bar{\lambda}_- &= \lambda_- + \frac{1}{2} \\
K &= \alpha_+ \log \frac{\lambda_+}{\lambda_+-1} + \alpha_- \log \frac{\lambda_-}{\lambda_-+1}
\end{align*}
\item Variance smile:
\begin{align*}
\hat{u}(x) &= - \frac{1}{2} \left( \frac{\alpha_++\alpha_-}{K+x} + \bar{\lambda}_- - \bar{\lambda}_+ \right) + \frac{1}{2} \frac{\sqrt{4\alpha_+\alpha_- + \left( (\bar{\lambda}_++\bar{\lambda}_-)(K+x) - (\alpha_+-\alpha_-) \right)^2}}{K+x} \\
\omega(x) &= \left( \frac{K}{2} - \alpha_+ \log \frac{\lambda_+}{\bar{\lambda}_+} - \alpha_- \log \frac{\lambda_-}{\bar{\lambda}_-} - \frac{\alpha_++\alpha_-}{2} \right) - \frac{\bar{\lambda}_- - \bar{\lambda}_+}{2} (K+x) \\
&\quad + \sqrt{\alpha_+\alpha_- + \left( \frac{\bar{\lambda}_+ + \bar{\lambda}_-}{2} (K+x) - \frac{\alpha_+-\alpha_-}{2} \right)^2} \\
&\quad + \alpha_+ \log \left( 1 - \frac{\hat{u}(x)}{\bar{\lambda}_+} \right) + \alpha_- \log \left( 1 + \frac{\hat{u}(x)}{\bar{\lambda}_-} \right)
\end{align*}
\end{itemize}

\item \textbf{CGMY model}
\begin{itemize}
\item Model parameters: $C,G,M,Y$, which characterize the Lévy measure
\item Characteristic constants:
\begin{align*}
\bar{G} &= G + \frac{1}{2} \\
\bar{M} &= M - \frac{1}{2} \\
K &= \frac{1}{Y} \left( (M-1)^Y - M^Y + (G+1)^Y - G^Y \right) \\
\bar{K} &= \frac{\bar{M}^Y+\bar{G}^Y-M^Y-G^Y-\frac{KY}{2}}{Y(Y-1)\bar{G}^Y} \\
\xi &= \frac{\bar{M}}{\bar{G}}
\end{align*}
\item Variance smile: assuming small $Y>0$,
\begin{align*}
\Sigma(x) &\equiv \frac{x}{CY(Y-1)\Gamma(-Y)\bar{G}^{Y-2}} \\
\hat{u}(x) &\approx \frac{- \frac{1+\xi^{Y-2}}{2} + \left(\frac{1}{\bar{G}}-\frac{1}{\bar{M}}\right) \frac{\Sigma(x)}{2} + \sqrt{\frac{1}{4} \left[\left(\frac{1}{\bar{G}}+\frac{1}{\bar{M}}\right)\Sigma(x)-(1-\xi^{Y-2})\right]^2 + \xi^{Y-2}}}{\frac{\xi^{Y-2}}{\bar{G}}-\frac{1}{\bar{M}} + \frac{\Sigma(x)}{\bar{G}\bar{M}}} \\
\omega(x) &\approx CY(Y-1)\Gamma(-Y)\bar{G}^Y \Bigg\{ -\bar{K} + \xi \Bigg[ - \frac{1+\xi^{Y-2}}{2} + \left(\frac{1}{\bar{G}}-\frac{1}{\bar{M}}\right) \frac{\Sigma(x)}{2} + \\
&\quad \left. \left. \sqrt{\frac{1}{4} \left[\left(\frac{1}{\bar{G}}+\frac{1}{\bar{M}}\right)\Sigma(x)-(1-\xi^{Y-2})\right]^2 + \xi^{Y-2}} \right] + \left[ \log\left(1+\frac{\hat{u}(x)}{\bar{G}}\right) + \xi^Y \log\left(1-\frac{\hat{u}(x)}{\bar{M}}\right) \right] \right\}
\end{align*}
\end{itemize}
\end{itemize}

\vspace{0.5in}


\section{Calibration Example} \label{sec:calibration}

From equation (\ref{eq:modinsp}), based on a model-specific $\omega(x)$, we inspire a parametrization for total implied variance $w(k,T)$, calibrated to market smiles slice by slice. Here we study in particular a \textbf{bilateral-gamma-inspired (BGI) parametrization}, without relaxation of parameters thus preserving its original form (\ref{eq:bgomega}). Like SVI, BGI variance takes a V-shape, which was adequate for fitting equity index smiles a decade ago, but now not so anymore. Still, we adopt BGI as the case study here, because (1) our BG calculations are exact (unlike CGMY), (2) parameter controls over the variance smile are independent and physically interpretable, which is an improvement over SVI, and (3) calendar arbitrage constraints are easily enforceable, thus one may calibrate a BGI volatility surface arbitrage-free over both strike and time. Similar analysis carries over to other candidate parametrizations.

\subsection{BG-Inspired Parametrization}

BG is parametrized by four parameters: $\alpha_\pm,\lambda_\pm>0$, appearing in the Lévy measure.

Under BG, $\hat{u}(x)$ and $\omega(x)$ take the following forms. Note that $\hat{u}(x)$ is rewritten so that indicators are removed.
\begin{align*}
\hat{u}(x) &= - \frac{1}{2} \left( \frac{\alpha_++\alpha_-}{K+x} + \bar{\lambda}_- - \bar{\lambda}_+ \right) + \frac{1}{2} \frac{\sqrt{4\alpha_+\alpha_- + \left( (\bar{\lambda}_++\bar{\lambda}_-)(K+x) - (\alpha_+-\alpha_-) \right)^2}}{K+x} \\
\omega(x) &= \left( \frac{K}{2} - \alpha_+ \log \frac{\lambda_+}{\bar{\lambda}_+} - \alpha_- \log \frac{\lambda_-}{\bar{\lambda}_-} - \frac{\alpha_++\alpha_-}{2} \right) - \frac{\bar{\lambda}_- - \bar{\lambda}_+}{2} (K+x) \\
&\quad + \sqrt{\alpha_+\alpha_- + \left( \frac{\bar{\lambda}_+ + \bar{\lambda}_-}{2} (K+x) - \frac{\alpha_+-\alpha_-}{2} \right)^2} \\
&\quad + \alpha_+ \log \left( 1 - \frac{\hat{u}(x)}{\bar{\lambda}_+} \right) + \alpha_- \log \left( 1 + \frac{\hat{u}(x)}{\bar{\lambda}_-} \right)
\end{align*}

For time slice $T$ (fixed), total implied variance $w(k,T)$ as a function of log-strike $k \equiv \log(K/F)$ is implicitly \textit{defined} as the solution to the following (quadratic) saddle-point equation:
\begin{equation*} \frac{w(k,T)}{8} + \frac{k^2}{2w(k,T)} \equiv \omega\left(\frac{k}{T}\right) \cdot T . \end{equation*}

We numerically study the impacts of model parameters on variance smiles assuming these typical parameters. Specifically, we bump $\alpha_-$ and $\lambda_-$, which correspond to the put wing, and observe how the smile moves. Likewise, one may consider $\alpha_+$ and $\lambda_+$ for the call wing.
\begin{equation*} \alpha_+ = 10, \quad \alpha_- = 0.6, \quad \lambda_+ = 35, \quad \lambda_- = 5 \end{equation*}

From plots that follow, we observe that
\begin{itemize}
\item $\alpha_-$ controls the overall level of put wing, and larger $\alpha_-$ lifts the put wing up (in a parallel manner) biasing the minimum variance point to the right;
\item $\lambda_-$ controls the skew (slope) of put wing, and smaller $\lambda_-$ skews the put wing up roughly preserving the minimum variance point;
\item $\alpha_\pm,\lambda_\pm$ serve as four \textit{independent} dials that control the level and skew of call/put wings.
\end{itemize}

\begin{table}[H]
\centering
\begin{tabular}{|c|c|}
\hline
Bump $\alpha_-$ & Bump $\lambda_-$ \\ 
\hline \hline
\includegraphics[width=0.4\linewidth]{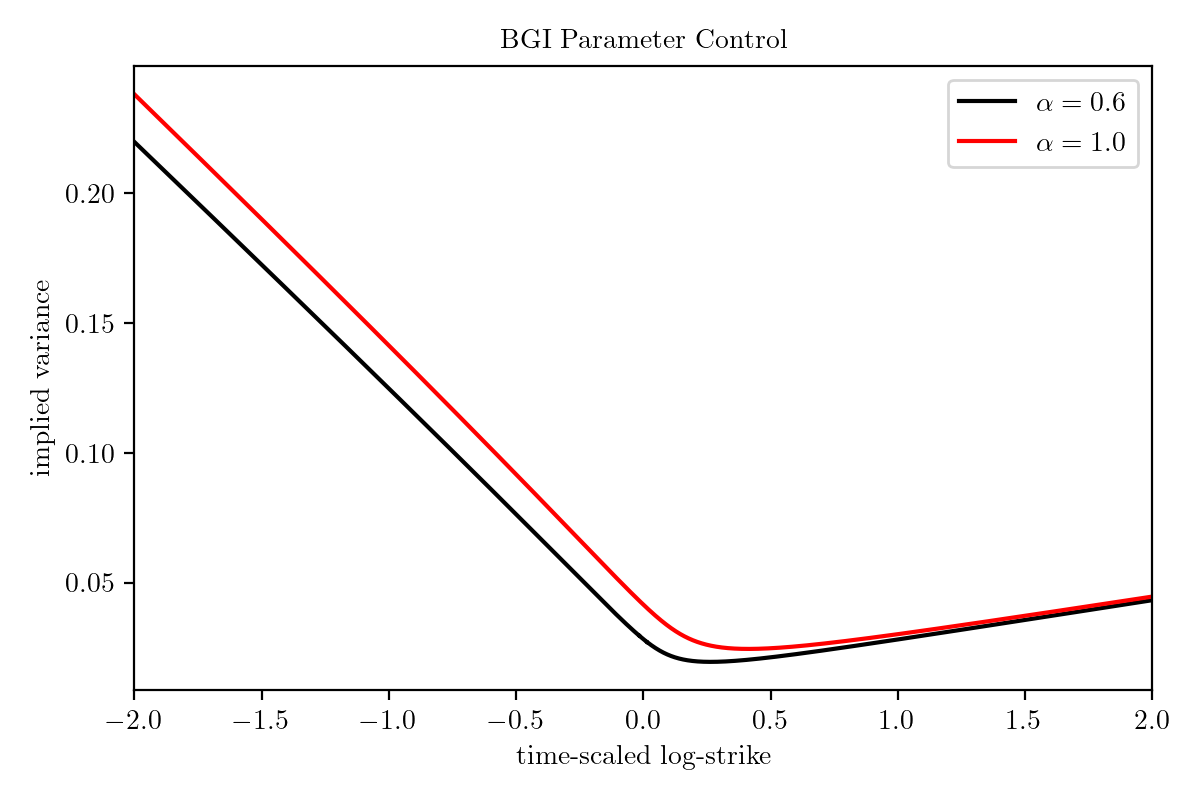} & 
\includegraphics[width=0.4\linewidth]{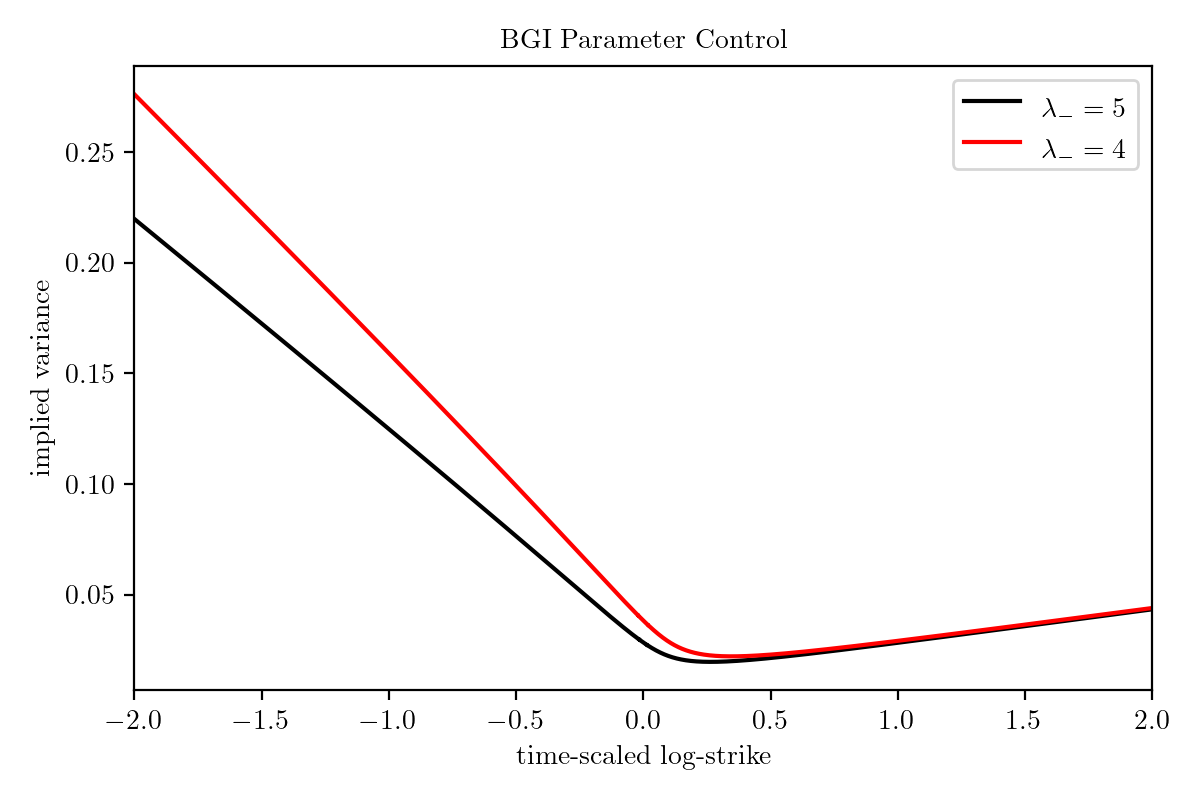} \\ 
\hline
\end{tabular}
\caption{Impacts of put-wing parameters on variance smile.}
\end{table}

Physically, this can be argued as follows. Consider downward jumps: $\alpha_-$ signifies the jump arrival rate and $\lambda_-$ signifies the jump size, roughly. With higher $\alpha_-$, downward jumps arrive more frequently so the effective volatility of log-spot is higher, causing the overall level of put wing to rise. With lower $\lambda_-$, downward jumps tend to be larger, so implied density of log-spot becomes more heavy-tailed on the left, skewing the put wing up.

We give the wing skew (slope) under BGI:
\begin{equation*}
\frac{\partial w}{\partial k}(\pm\infty) = \pm 4 \bar{\lambda}_\pm \left( 1 - \sqrt{1-\frac{1}{4\bar{\lambda}_\pm^2}} \right) ,
\end{equation*}
completely independent of $\alpha_\pm$. This can be derived from equation (\ref{eq:betapm}), or by taking derivatives on the saddle-point equation and sending it to the large-strike limit.

The at-the-money (ATM) skew, however, is not so trivial, as it is a highly non-linear function of model parameters. So a disadvantage is the non-intuitive ATM behavior coupled with the wings -- practitioners generally desire decoupled ATM and wing fits.

We have seen that $\alpha_\pm$ lift the wings while $\lambda_\pm$ skew the wings. Consider calendar arbitrage between two total variance slices $w_1(k,T_1)$ and $w_2(k,T_2)$ under BGI, with $T_1<T_2$. The two slices are free of calendar arbitrage if $w_{1,2}$ do not intersect for any $k$, with $w_2$ lying entirely above $w_1$. That $w_1(k,T_1)$ and $w_2(k,T_1)$ do not intersect is \textit{sufficient} to guarantee that $w_1(k,T_1)$ and $w_2(k,T_2)$ do not intersect, therefore we consider the two slices $v_{1,2}(x)$ in time-scaled log-strike space. Higher $\alpha_\pm$ lift the variance wings up and lower $\lambda_\pm$ skew the variance wings up, avoiding intersection between $v_1$ and $v_2$. We conjecture that term-structure constraints $\partial_TT\alpha_\pm(T) \geq 0$ and $\partial_T\lambda_\pm(T) \leq 0$ are \textit{sufficient} to guarantee the absense of calendar arbitrage. This echoes with \cite{mad-bg} whose result states that these constraints deliver additivity of the log-spot process, thus we still have a consistent (non-stationary) risk-neutral process with an arbitrage-free volatility surface.

\begin{namedthm}[Conjecture]
Under BGI, term-structure constraints $\partial_TT\alpha_\pm(T) \geq 0$ and $\partial_T\lambda_\pm(T) \leq 0$ are sufficient to guarantee the absense of calendar arbitrage.
\end{namedthm}

These constraints are used to restrict the optimization domain of model parameters during calibration.

\subsection{Calibration Routine}

We present a piece of Python pseudo-code that instructs the calibration of BGI parametrization given the volatility smile quoted in the market with bids and asks over discrete strikes $\{K_i\}_{1:n}$, at time slice $T$ with forward price $F$ (presumably implied from the smile). The idea is to optimize model parameters $\alpha_\pm,\lambda_\pm$ which minimize some least-sqaure function that quantifies deviation from bid-ask spreads.

\begin{lstlisting}[language=Python]
import numpy as np
import pandas as pd
from scipy.optimize import fmin, minimize

# BS vega omitting S and T
def vegaBS(k, w):
    # k = log-strike
    # w = total implied variance
    d1 = -k/np.sqrt(w)+np.sqrt(w)/2
    return np.exp(-d1**2/2)

# implied volatility under BGI
def volBGI(x, Ap, Am, Lp, Lm):
    # x = time-scaled log-strike
    # Ap, Am, Lp, Lm = BGI parameters
    Lp0 = Lp-0.5
    Lm0 = Lm+0.5
    K0 = Ap*np.log(Lp/(Lp-1))+Am*np.log(Lm/(Lm+1))
    u = lambda x: -0.5*((Ap+Am)/(K0+x)+Lm0-Lp0)+
        0.5*np.sqrt(4*Ap*Am+((Lp0+Lm0)*(K0+x)-(Ap-Am))**2)/(K0+x)
    w = lambda x: (K0/2-Ap*np.log(Lp/Lp0)-
        Am*np.log(Lm/Lm0)-(Ap+Am)/2)-(Lm0-Lp0)/2*(K0+x)+
        np.sqrt(Ap*Am+((Lp0+Lm0)/2*(K0+x)-(Ap-Am)/2)**2)+
        Ap*np.log(1-u(x)/Lp0)+Am*np.log(1+u(x)/Lm0)
    xp = fmin(lambda x: w(x)-x/2, 0, disp=False)[0] # or fsolve
    xm = fmin(lambda x: w(x)+x/2, 0, disp=False)[0]
    v = 4*(w(x)+(2*(x>xm)*(x<xp)-1)*np.sqrt(w(x)**2-x**2/4))
    return np.sqrt(v)

######################################################################
############ Local Variables & Functions at Time Slice T #############
######################################################################

# dataframe of OTM options at expiry T
df = ...
#################################################
# | Expiry | Forward | Strike | BidVol | AskVol |
# | 0.5000 | 100.20  | 65.00  | 0.1900 | 0.2100 |
# | 0.5000 | 100.20  | 70.00  | 0.1800 | 0.2000 |
# | 0.5000 | 100.20  | 75.00  | 0.1700 | 0.1900 |
# | 0.5000 | 100.20  | 80.00  | 0.1600 | 0.1800 |
# | ........................................... |

# market inputs
T   = df['Expiry']
F   = df['Forward']
K   = df['Strike']
bid = df['BidVol']
ask = df['AskVol']
mid = (bid+ask)/2 # mid-vol
k   = np.log(K/F) # log-strike
x   = k/T         # time-scaled log-strike

# BS vega weights
w0 = vegaBS(k, mid**2*T)

# initial guess & bounds for minimization
guess = (10, 0.6, 35, 5)
bounds = ((0, 1000), (0, 1000), (0, 1000), (0, 1000))

# objective function for minimization
def objective(params):
    sig = volBGI(x, *params)
    return np.sum(w0*((sig-bid)**2+(ask-sig)**2)) # or other variations

opt = minimize(objective, x0=guess, bounds=bounds)
Ap, Am, Lp, Lm = opt.x
\end{lstlisting}

We start from the shortest expiry progressively to the longest. After calibrating time slice $T_i$ with optimal parameters $\alpha_+^i,\alpha_-^i,\lambda_+^i,\lambda_-^i$, the optimization domain of the next slice $T_{i+1}$ is restricted to 
\begin{equation*} \left(\left(\frac{T_i\alpha_+^i}{T_{i+1}},1000\right),\left(\frac{T_i\alpha_-^i}{T_{i+1}},1000\right),(0,\lambda_+^i),(0,\lambda_-^i)\right) , \end{equation*}
following the term-structure conjecture above.

\subsection{SPX Calibration}

We calibrate BGI parametrization to S\&P500 (SPX) vanilla options on September 15, 2005,\footnote{This was the date chosen for the fitting of SVI in \cite{gat-volsurf}.} the day before triple witching, which contains 7 expiries totaling 244 out-of-the-money options data points, omitting the front term. The data are obtained from OptionMetrics.\footnote{For each time slice, from the raw bid and ask price data, first the forward price and discount factor are imputed from put-call parity applied to near-the-money options via a least-square optimization, then the options bid and ask volatilities are implied from Black-Scholes 
(Black-76) formula.} On this date, the market smiles have the iconic V-shape and BGI qualifies as a potential candidate. We examine (1) parameter term-structures and (2) goodness-of-fit.

Calibration yields the following parameters over terms, guaranteeing increasing $T\alpha_\pm(T)$ and decreasing $\lambda_\pm(T)$:
\begin{table}[H]
\centering
\begin{tabular}{|c|c|c|c|c|}
\hline
Term & $\alpha_+$ & $\alpha_-$ & $\lambda_+$ & $\lambda_-$ \\
\hline \hline
0.098563 & 978.563336 & 2.223200 & 410.809992 & 18.941492 \\
0.175337 & 996.689851 & 1.922465 & 409.309786 & 14.505594 \\
0.251996 & 752.980032 & 1.341641 & 331.890287 & 11.736468 \\
0.501141 & 656.500331 & 1.022875 & 331.711796 & 8.636279 \\
0.750171 & 658.434076 & 0.686101 & 326.238741 & 6.687754 \\
1.248574 & 397.891981 & 0.456478 & 259.855153 & 4.989019 \\
1.746749 & 284.781058 & 0.326300 & 215.649502 & 3.998353 \\
\hline
\end{tabular}
\caption{BGI parameter term-structures of SPX calibrated as of September 15, 2005.}
\end{table}

Term-structures for the put-wing parameters are plotted:
\begin{table}[H]
\centering
\begin{tabular}{|c|c|}
\hline
$T\alpha_-(T)$ & $\lambda_-(T)$ \\ 
\hline \hline
\includegraphics[width=0.4\linewidth]{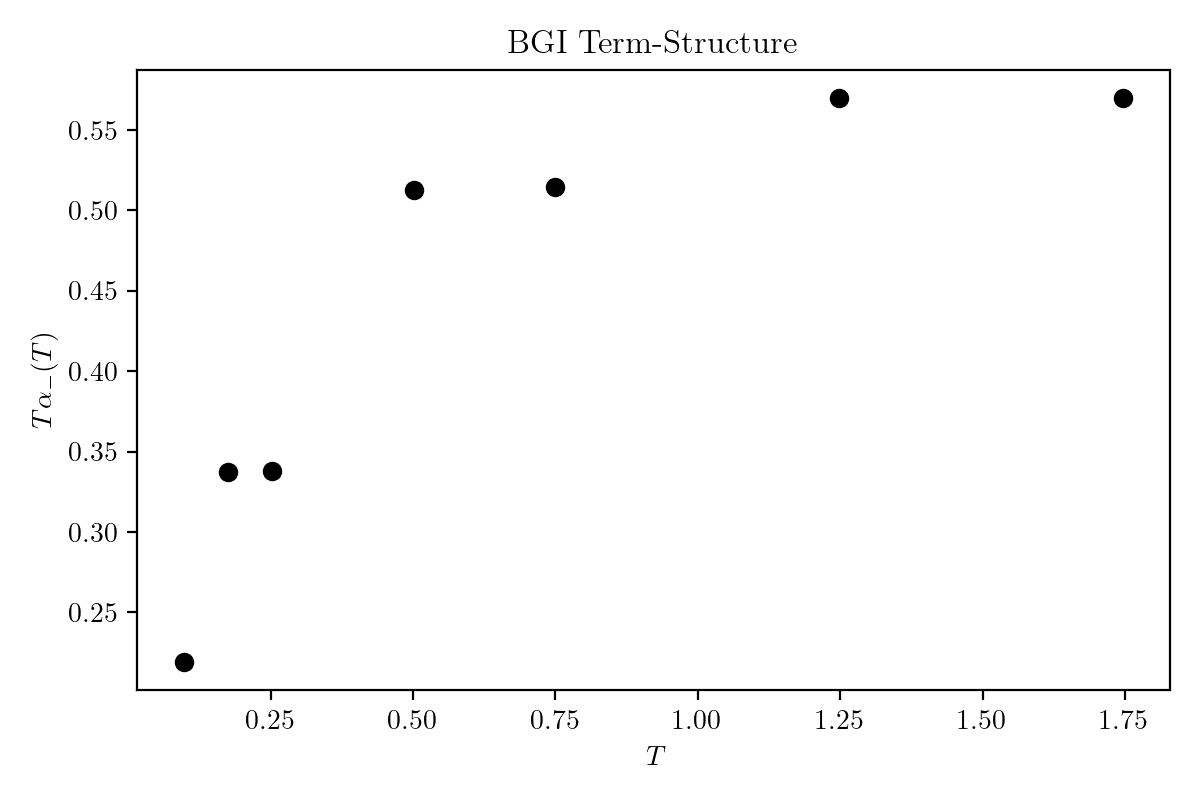} & 
\includegraphics[width=0.4\linewidth]{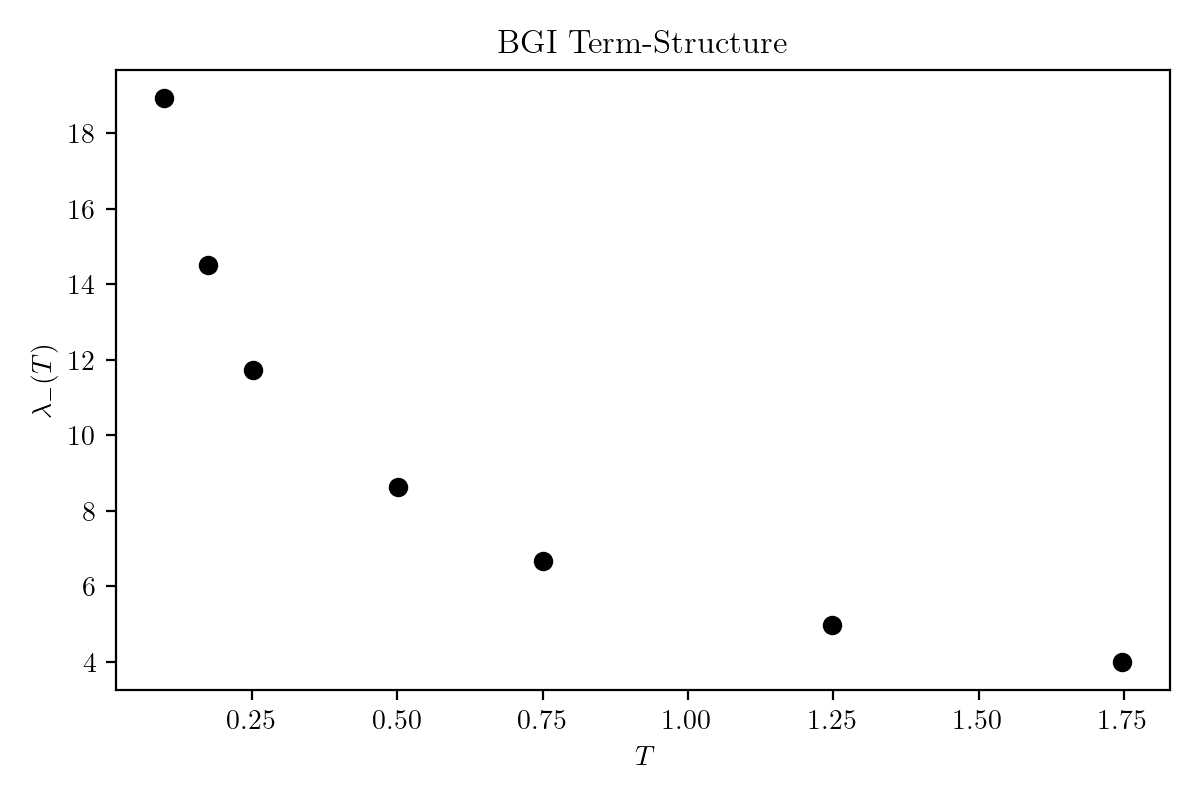} \\ 
\hline
\end{tabular}
\caption{BGI put-wing parameter term-structures.}
\end{table}

The calibrated smiles are plotted in black against the bid and ask vols, with the spreads labeled in green. It is observed that BGI performs well for long expiries but wings flatten out too quickly for short expiries, in particular the call wings.
\begin{figure}[H]
\centering
\includegraphics[width=0.8\linewidth]{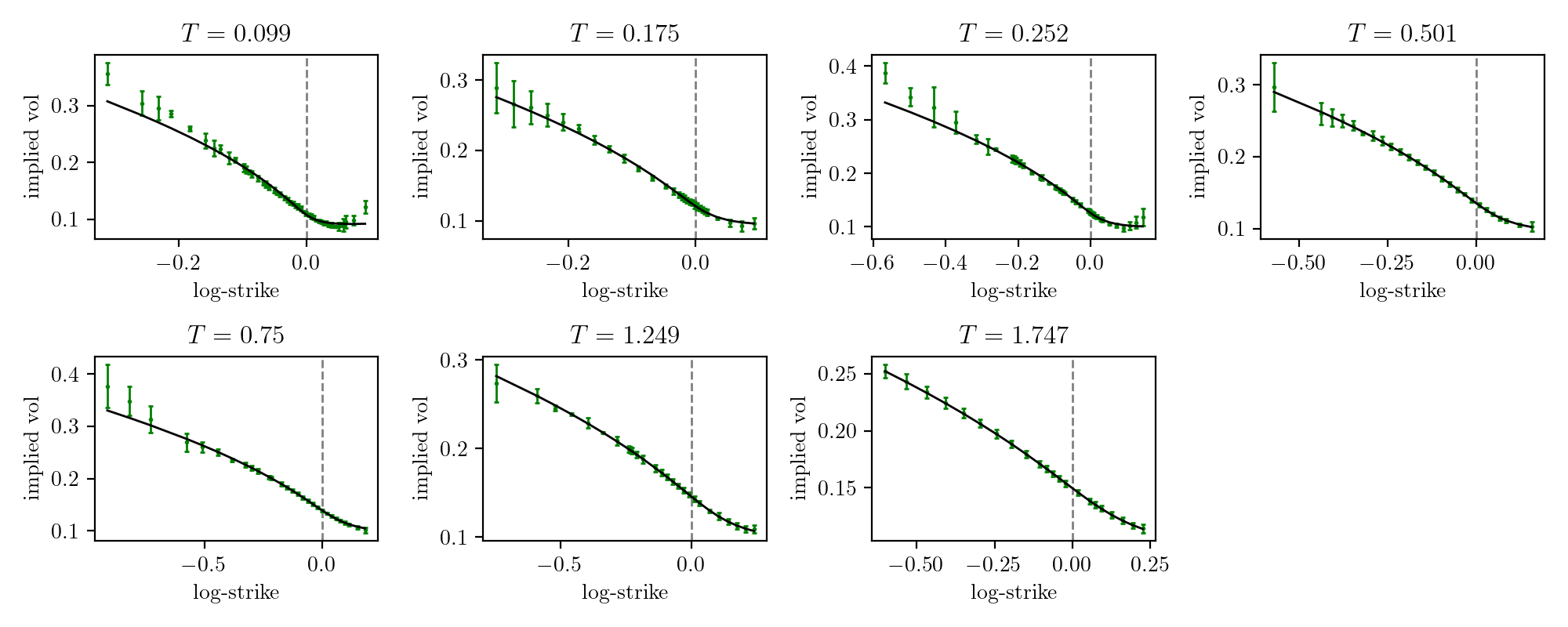}
\caption{BGI smiles calibrated to bid-ask volatilities.}
\end{figure}

\subsection{Arbitrage Check}

We numerically show the absense of butterfly and calendar arbitrage in our BGI calibration. Let $w(k,T)$ be a parametrization for total implied variance e.g.\ BGI in our case. From \cite{gat-svi}, arbitrage constraints read:

\begin{namedthm}[Butterfly Arbitrage]
Parametrization $w(k,T)$ at time slice $T$ is free of butterfly arbitrage if
\begin{equation*} \left( 1 - \frac{kw'(k,T)}{2w(k,T)} \right)^2 - \frac{w'(k,T)^2}{4} \left( \frac{1}{w(k,T)} + \frac{1}{4} \right) + \frac{w''(k,T)}{2} \geq 0 , \end{equation*}
with implied density given by
\begin{equation*} p(X_T=k) = \left[ \left( 1 - \frac{kw'(k,T)}{2w(k,T)} \right)^2 - \frac{w'(k,T)^2}{4} \left( \frac{1}{w(k,T)} + \frac{1}{4} \right) + \frac{w''(k,T)}{2} \right] \frac{e^{-(k-w(k,T)/2)^2/2w(k,T)}}{\sqrt{2\pi w(k,T)}} . \end{equation*}
\end{namedthm}

\begin{namedthm}[Calendar Arbitrage]
Parametrization $w(k,T)$ at log-strike $k$ is free of calendar arbitrage if
\begin{equation*} \partial_T w(k,T) \geq 0 . \end{equation*}
\end{namedthm}

We plot the BGI implied density and total implied variance over all terms, and it is observed that (1) densities are positive everywhere and (2) variance slices do not cross. With appropriate time-interpolation in parameters $T\alpha_\pm$ and $\lambda_\pm$, one may generate an arbitrage-free volatility surface.

\begin{table}[H]
\centering
\begin{tabular}{|c|c|}
\hline
Butterfly Arbitrage & Calendar Arbitrage \\ 
\hline \hline
\includegraphics[width=0.4\linewidth]{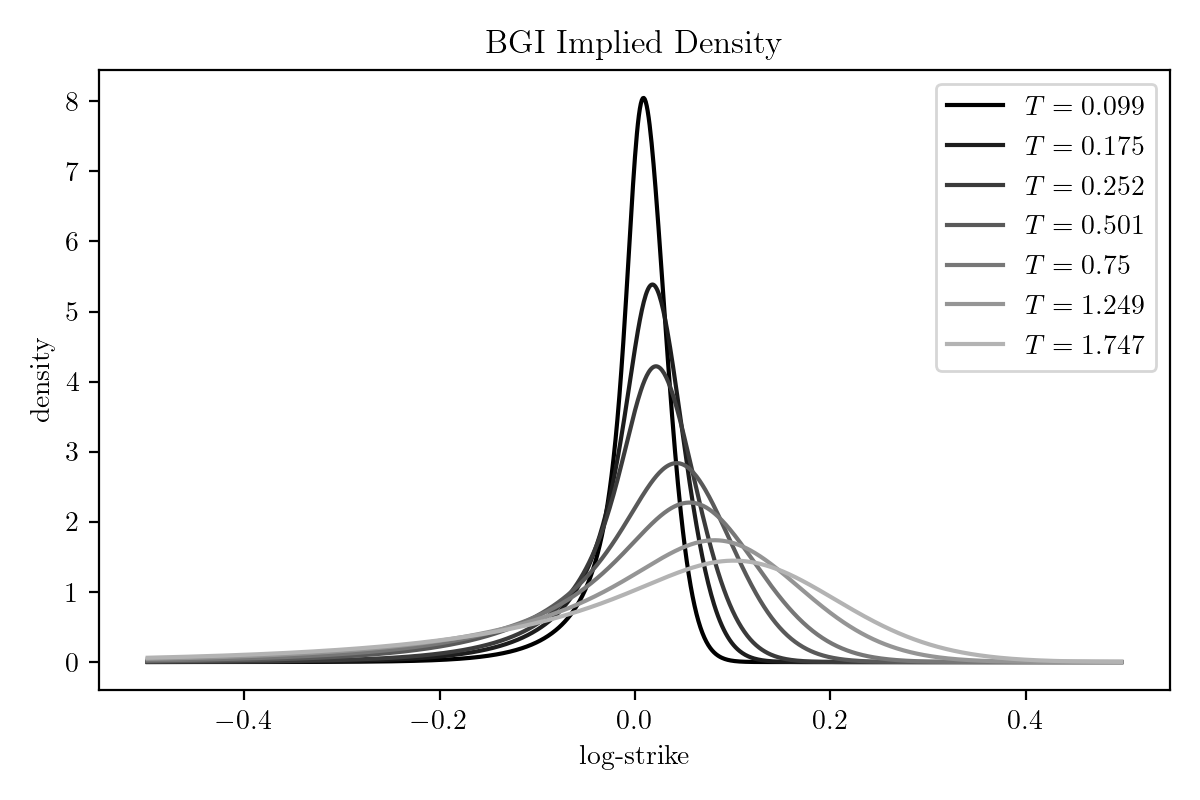} & 
\includegraphics[width=0.4\linewidth]{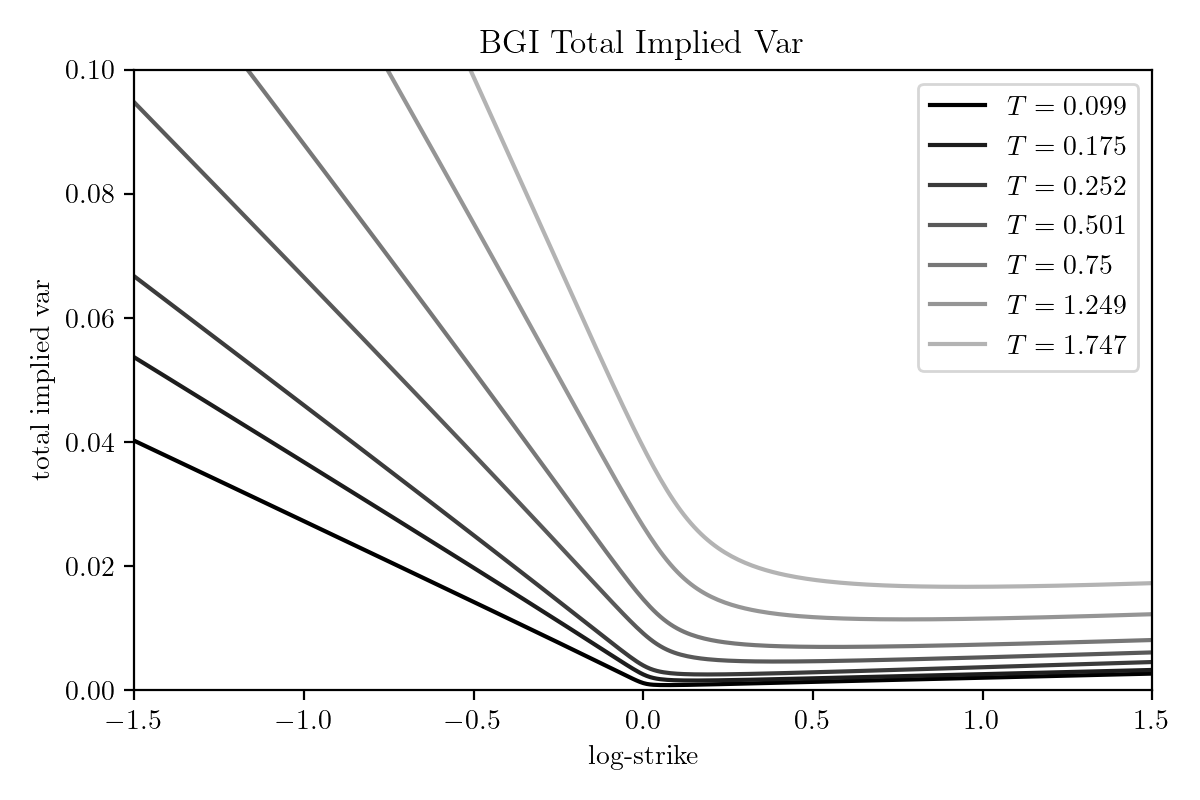} \\ 
\hline
\end{tabular}
\caption{Absence of butterfly and calendar arbitrage under BGI.}
\end{table}

While BGI does not fit well short expiries and call wings, its performance for longer expiries is acceptable. In addition, BGI is arbitrage-free along strike and calendar arbitrage enforcement is straightforward by ensuring appropriate monotonic term-structures in model parameters. We leave the empirical study of other model-inspired candidates as future work.

\begin{remark}
We also calibrate BGI parametrization to SPX vanilla options on a more recent date, November 7, 2022. While we are able to guarantee the full absence of butterfly and calendar arbitrage similarly as above, the fit quality in particular for short expiries is not satisfactory. See appendix \ref{apdx:bgi}, BGI Calibration to SPX 2022, for the relevant discussion.
\end{remark}


\section{Conclusion} \label{sec:concl}

We establish the theoretical foundation of the saddle-point approach, originally pointed out in \cite{gat-convhes} to verify that the large-time Heston variance smile takes the form of SVI. A model-free at-the-money moment expansion of the large-time total variance smile $w(k)$ in log-strike $k$ is obtained, with coefficients in terms of central moments of log-spot under some Esscher measure. The approach is applied to classical models including Heston, variance-gamma, bilateral-gamma, CGMY and Merton-jump, where for the former four, calculations for the model-specific $\omega(x)$ are completed, as summarized above in section \ref{sec:summary}. One may easily extend the approach to other models as long as (1) density of log-spot $X_T$ has support over $\mathbb{R}$, or equivalently terminal spot $S_T$ can end up anywhere in $\mathbb{R}^+$, (2) characteristic function $\phi_T(u)$ has an explicit closed form, and (3) in large time $T \rightarrow \infty$, $\phi_T(u-i/2) \sim e^{-\psi(u)T}$ i.e.\ large-time log-spot evolves like a Lévy process and initial state is damped out. By definition, all Lévy processes (describing the log-spot) are under this umbrella. By choosing a Lévy measure complicated enough, one may be able to inspire an arbitrage-free parametrization for a time slice $T<\infty$, following procedure in section \ref{sec:modinsp}, that fits complex curvature in total implied variance $w(k,T)$, e.g.\ BG entails V-shaped smiles while CGMY entails more or less parabolic smiles. However, $\hat{u}$ under the model may not be easy to solve or approximate, a case we encounter in Merton-jump model. We leave the study of equation (\ref{eq:mereq}) hence its model-implied smile and extensions e.g.\ multiple Merton-jump components as future work. We hope that the saddle-point approach constitutes a systematic way to study the large-time smile behaviors of Lévy-type models and to inspire new arbitrage-free parametrizations or wing extrapolations, motivated from models.

\subsection*{Some Open Questions}

The core assumption behind the saddle-point equation is the Lévy-type scaling in characteristic function in large time, which implicitly demands a Markov log-spot process. In fact, even if one only assumes a Markov process s.t.\ $\phi_T(u-i/2)=e^{-\psi(u,T)}$, one can still obtain the saddle-point equation, except that the saddle-point $\hat{u}$ is now a function of $x$ and $T$. It is easy to see that all results in section \ref{sec:sdlpt} and \ref{sec:varsmile} carry over. Most importantly, $\omega(x,T)$ still tangentially touches $|x/2|$ and is everywhere convex. We start from a Markov risk-neutral process, thus an arbitrage-free model, to $\omega(x,T)$ that fulfills the tangent and convexity property. A natural question would be, by reversing the logic, \textit{does an arbitrary $\omega(x,T)$ that fulfills the tangent and convexity property contain no arbitrage}? If this is the case, one may define a class of convex functions $\omega(x,T)$ that touches $|x/2|$ and calibrate that to the market smiles, by fitting $\omega(x,T)$ to $\omega^*(x,T) \equiv v(x,T)/8+x^2/2v(x,T)$, which potentially constitutes an arbitrage-free parametrization.

Another immediate question would be, \textit{does the market actually exhibit an $\omega^*(x,T)$ that tangentially touches $|x/2|$ and is everywhere convex}? Let us ignore the tangent condition for now. When visualizing $\omega^*(x,T)$, typically we observe for large $T$ strict convexity, or at least a convex curve may be fitted inside the bid-ask spreads. However, for small $T$, while near-the-money $\omega^*(x,T)$ is generally convex, concave wings, for large $|x|$, that gradually tend to linear are observed. See appendix \ref{apdx:bgi} for a plot of $\omega^*(x,T)$ on November 7, 2022, over the 50 expiries quoted. These observations establish that (1) market expects a Markov process in large time, i.e.\ increments in log-spot have little to do with the past and historical impacts have dampened out, and (2) market expects a non-Markov process in small time, i.e.\ increments in log-spot memorize the past and historical impacts persist. Then, \textit{if the log-spot process is non-Markov, what $\omega(x,T)$ will be allowed}? An ad-hoc trick that may fit well is to use a model-inspired $\omega(x,T)$, guaranteed to be touching $|x/2|$ and convex near-the-money, joint to two linear call/put wings at truncation points $x_c,x_p$, appropriately weighted to ensure differentiability. Note that the two linear wings must grow faster than $|x/2|$ by equation (\ref{eq:betapm}), so $|x/2|$ touches $\omega(x,T)$ at only two points $x_\pm$. \textit{Is the variance smile $v(x)$ implicitly defined by $\omega(x,T)$ arbitrage-free}?


\section{Acknowledgement} \label{sec:acknow}

C.Y.\ Yeung is indebted to W.H.\ Levi Poon for assisting with mathematica computations for the moment expansion of smile and some mathematical aspects of the saddle-point equation; K.\ Wang, his supervisor during the internship at Morgan Stanley, for insightful discussions regarding the initial formulation of the saddle-point equation and the implied smile under bilateral-gamma model; W.\ Xu, his team member during the internship, for helpful discussions regarding the calibration aspects of VGI parametrization.

\newpage


\newpage


\appendix

\section{A Fuller Saddle-Point Equation} \label{apdx:sdl}

Here we obtain a fuller version of the saddle-point equation which includes terms of order $T^{-1}$ and higher, still with $T$ large. We consider the integral in Lewis equation, where $x \in \mathbb{R}$
\begin{equation*} \int_{-\infty}^\infty \frac{du}{u^2+\frac{1}{4}} e^{-(iux+\psi(u))T} . \end{equation*} 

We now Taylor-expand both the exponent and factor $1/(u^2+1/4)$ around the saddle-point $\tilde{u}$ that zeros out $u$-derivative of exponent $\psi'(\tilde{u})=-ix$, which gives
\begin{align*}
&\int_{-\infty}^{\infty} du \sum_{n=0}^\infty \frac{(u-\tilde{u})^n}{n!} \left(\frac{d}{du}\right)^n \frac{1}{u^2+\frac{1}{4}} \biggr\rvert_{\tilde{u}} \cdot e^{-(i\tilde{u}x+\psi(\tilde{u}))T - \frac{\psi''(\tilde{u})T}{2}(u-\tilde{u})^2 - T \cdot O(u-\tilde{u})^3} \\
=& \; e^{-(i\tilde{u}x+\psi(\tilde{u}))T} \sum_{n=0}^\infty \frac{1}{n!} \left(\frac{d}{du}\right)^n \frac{1}{u^2+\frac{1}{4}} \biggr\rvert_{\tilde{u}} \int_{-\infty}^{\infty} du (u-\tilde{u})^n e^{- \frac{\psi''(\tilde{u})T}{2}(u-\tilde{u})^2 - T \cdot O(u-\tilde{u})^3} .
\end{align*}

Following similar argument in section \ref{sec:sdlpt}, noting that $\tilde{u}$ is purely imaginary and $\psi''(\tilde{u})>0$, we take only the quadratic term in exponent at large $T$, and approximate the integral as follows:
\begin{align*}
&\int_{-\infty}^{\infty} du (u-\tilde{u})^n e^{- \frac{\psi''(\tilde{u})T}{2}(u-\tilde{u})^2 - T \cdot O(u-\tilde{u})^3} \\
\approx& \; \int_{-\infty}^{\infty} du (u-\tilde{u})^n e^{- \frac{\psi''(\tilde{u})T}{2}(u-\tilde{u})^2} \\
=& \; \int_{-\infty-\tilde{u}}^{\infty-\tilde{u}} du \cdot u^n e^{- \frac{\psi''(\tilde{u})T}{2}u^2} \\
=& \; \int_{-\infty}^{\infty} du \cdot u^n e^{- \frac{\psi''(\tilde{u})T}{2}u^2} \quad \text{(boundary terms are zero)}
\end{align*}
which vanishes when $n$ is odd, and when $n=2m$ is even, the integral simplifies to
\begin{align*}
\frac{\sqrt{2\pi}(2m)!}{2^m m!} \left(\psi''(\tilde{u})T\right)^{-(2m+1)/2} .
\end{align*}

Going back to the original sum, it can be written as
\begin{align*}
&\sqrt{\frac{2\pi}{\psi''(\tilde{u})T}} e^{-(i\tilde{u}x+\psi(\tilde{u}))T} \sum_{m=0}^\infty \frac{1}{(2m)!} \left(\frac{d}{du}\right)^{2m} \frac{1}{u^2+\frac{1}{4}} \biggr\rvert_{\tilde{u}} \cdot \frac{(2m)!}{2^m m!} \left(\psi''(\tilde{u})T\right)^{-m} \\
=& \; \sqrt{\frac{2\pi}{\psi''(\tilde{u})T}} e^{-(i\tilde{u}x+\psi(\tilde{u}))T} \sum_{m=0}^\infty \frac{1}{m!} \left(\frac{1}{2\psi''(\tilde{u})T} \frac{d^2}{du^2}\right)^m \frac{1}{u^2+\frac{1}{4}} \biggr\rvert_{\tilde{u}} .
\end{align*}

Define operator $\partial_{\tilde{u}}^k$ s.t.\
\begin{equation*}
\partial_{\tilde{u}}^k \frac{1}{\tilde{u}^2+\frac{1}{4}} \equiv \left(\frac{d}{du}\right)^k \frac{1}{u^2+\frac{1}{4}} \biggr\rvert_{\tilde{u}}
\end{equation*}
thus the sum can be compactly written as an exponential
\begin{align*}
&\sqrt{\frac{2\pi}{\psi''(\tilde{u})T}} e^{-(i\tilde{u}x+\psi(\tilde{u}))T} \sum_{m=0}^\infty \frac{1}{m!} \left(\frac{\partial_{\tilde{u}}^2}{2\psi''(\tilde{u})T}\right)^m \frac{1}{\tilde{u}^2+\frac{1}{4}} \\
=& \; \sqrt{\frac{2\pi}{\psi''(\tilde{u})T}} \exp \left[ -(i\tilde{u}x+\psi(\tilde{u}))T + \frac{\partial_{\tilde{u}}^2}{2\psi''(\tilde{u})T} \right] \frac{1}{\tilde{u}^2+\frac{1}{4}} .
\end{align*}

Recall our variance quantity $\omega(x) \equiv i\tilde{u}x+\psi(\tilde{u})$, and for Black-Scholes, $\omega_{BS}(x) \equiv v_T(x)/8+x^2/2v_T(x)$. The expression above applies to generic models, and by matching with Black-Scholes, we obtain the equation for implied variance $v(x)$
\begin{equation}
\begin{aligned}
&-\omega(x)T - \frac{1}{2} \log \psi''(\tilde{u}) + \log \left[ \exp \left(\frac{\partial_{\tilde{u}}^2}{2\psi''(\tilde{u})T}\right) \frac{1}{\tilde{u}^2+\frac{1}{4}} \right] \\
=& \; -\omega_{BS}(x)T - \frac{1}{2} \log \psi_{BS}''(\tilde{u}_{BS}) + \log \left[ \exp \left(\frac{\partial_{\tilde{u}_{BS}}^2}{2\psi_{BS}''(\tilde{u}_{BS})T}\right) \frac{1}{\tilde{u}_{BS}^2+\frac{1}{4}} \right] .
\end{aligned} \label{eq:sdlptfull}
\end{equation}

To the first order in $T^{-1}$, we have
\begin{equation}
\begin{aligned}
&-\omega(x)T - \frac{1}{2} \log \psi''(\tilde{u}) + \log \left[\frac{1+O(T^{-1})}{\frac{1}{4}-\hat{u}(x)^2}\right] \\
=& \; -\left( \frac{v_T(x)}{8} + \frac{x^2}{2v_T(x)} \right) T - \frac{1}{2} \log v_T(x) + \log \left[\frac{1+O(T^{-1})}{\frac{1}{4}-\left(\frac{x}{v_T(x)}\right)^2}\right] ,
\end{aligned} \label{eq:sdlptord1}
\end{equation}
whose solution is non-trivial. Denote the time-infinity stationary variance smile $v(x) \equiv v_\infty(x)$, with explicit form given by equation (\ref{eq:vx0}). One may obtain an expansion of $v_T(x)$ as a series of $T^{-1}$, where to the first order, we have the \textit{volatility surface}
\begin{equation}
v_T(x) = v(x) + T^{-1} \frac{8v(x)^2}{4x^2-v(x)^2} \log \frac{\frac{1}{4}-\left(\frac{x}{v(x)}\right)^2}{\frac{1}{4}-\hat{u}(x)^2} \sqrt{\frac{v(x)}{\psi''(\tilde{u})}} + O(T^{-2}) , \label{eq:sdlptexp}
\end{equation}
agreeing with the \textit{form} of equation (1) in \cite{jac-asyhes}, which specializes to the case of Heston model. Note that in general, $\psi''(\tilde{u})$ is a very dirty quantity to compute, but it is in closed form, if the model admits closed-form $\tilde{u}$ and $\psi(u)$. The $T^{-1}$-correction term echoes with the numerical observation that convergence error of at-the-money volatility into the large-time stationary value decays roughly as $T^{-1}$ -- see sections \ref{sec:hesnum} and \ref{sec:vgnum}. We end this section with a remark in \cite{jac-asyhes}: ``\textit{the maturity-dependent strike formulation ... reveals that the implied volatility smile does not flatten but rather spreads out in a very specific way as the maturity increases}''. The ``\textit{specific way}'', as we understand, is precisely through scaling the smile via variable $x=k/T$. This scaling by $T$ may connect to the linear Lévy-type scaling in characteristic exponent, at large time.

\begin{remark}
To inspect the sufficiency of the first order term in approximating the smiles at a non-large $T$, consider the bilateral-gamma model as an example, which we have
\begin{align*}
\psi(\tilde{u}) &= \left( \frac{K}{2} - \alpha_+ \log \frac{\lambda_+}{\bar{\lambda}_+} - \alpha_- \log \frac{\lambda_-}{\bar{\lambda}_-} \right) + K \hat{u}(x) + \alpha_+ \log \left( 1 - \frac{\hat{u}(x)}{\bar{\lambda}_+} \right) + \alpha_- \log \left( 1 + \frac{\hat{u}(x)}{\bar{\lambda}_-} \right) \\
\psi''(\tilde{u}) &= \frac{\alpha_+/\bar{\lambda}_+^2}{(1-\hat{u}(x)/\bar{\lambda}_+)^2} + \frac{\alpha_-/\bar{\lambda}_-^2}{(1+\hat{u}(x)/\bar{\lambda}_-)^2} .
\end{align*}

We repeat similar procedure in the numerical experiment, section \ref{sec:bgnum}, considering the overall smiles and the convergence of at-the-money volatilties, taking into account the first order correction in variance $v_T(x)$, from equation (\ref{eq:sdlptexp}). Assume the same typical parameters. At-the-money, it predicts a decay of the form
\begin{equation*}
v(0)-v_T(0) \sim T^{-1} \cdot 8 \log \frac{1}{1-4\hat{u}(0)^2} \sqrt{\frac{8\psi(\tilde{u}(0))}{\psi''(\tilde{u}(0))}} .
\end{equation*}

The approximation yields the smiles in very high accuracy for $T \gtrsim 1$.
\begin{table}[H]
\centering
\begin{tabular}{|c|c|}
\hline
Overall Smile & ATM Implied Volatility \\ 
\hline \hline
\includegraphics[width=0.4\linewidth]{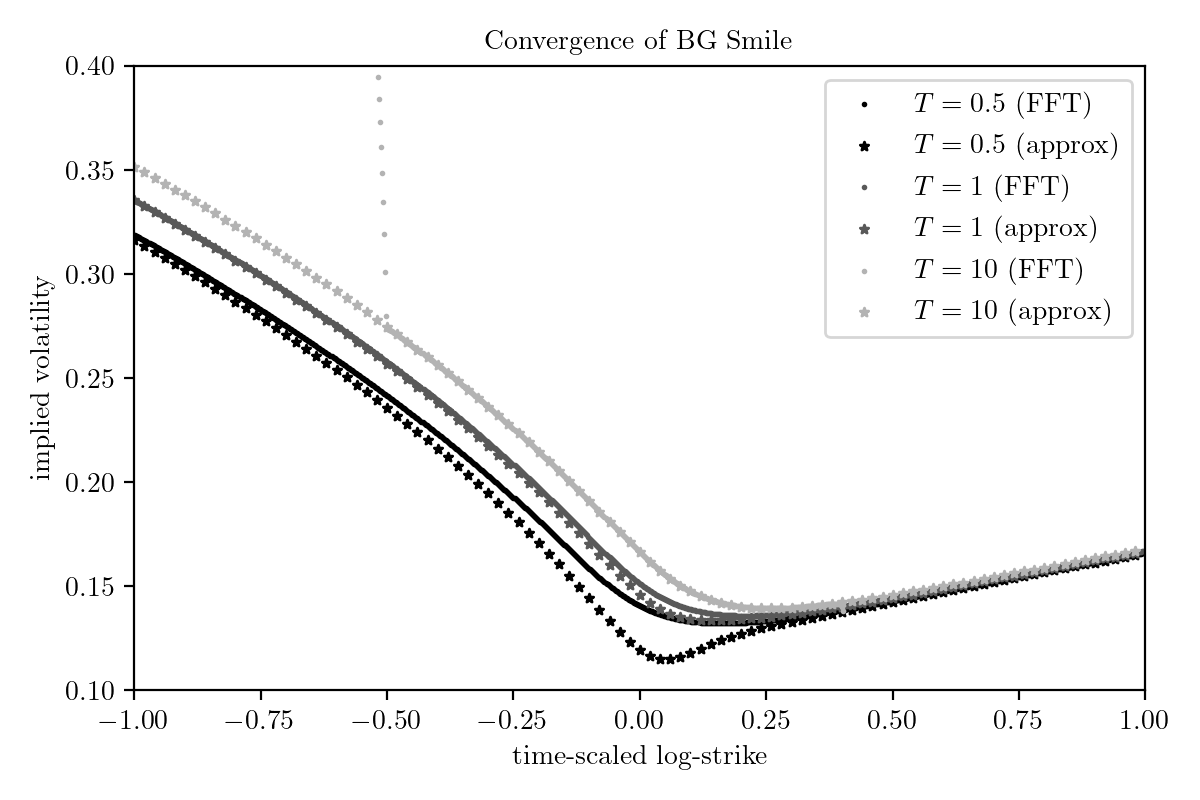} &
\includegraphics[width=0.4\linewidth]{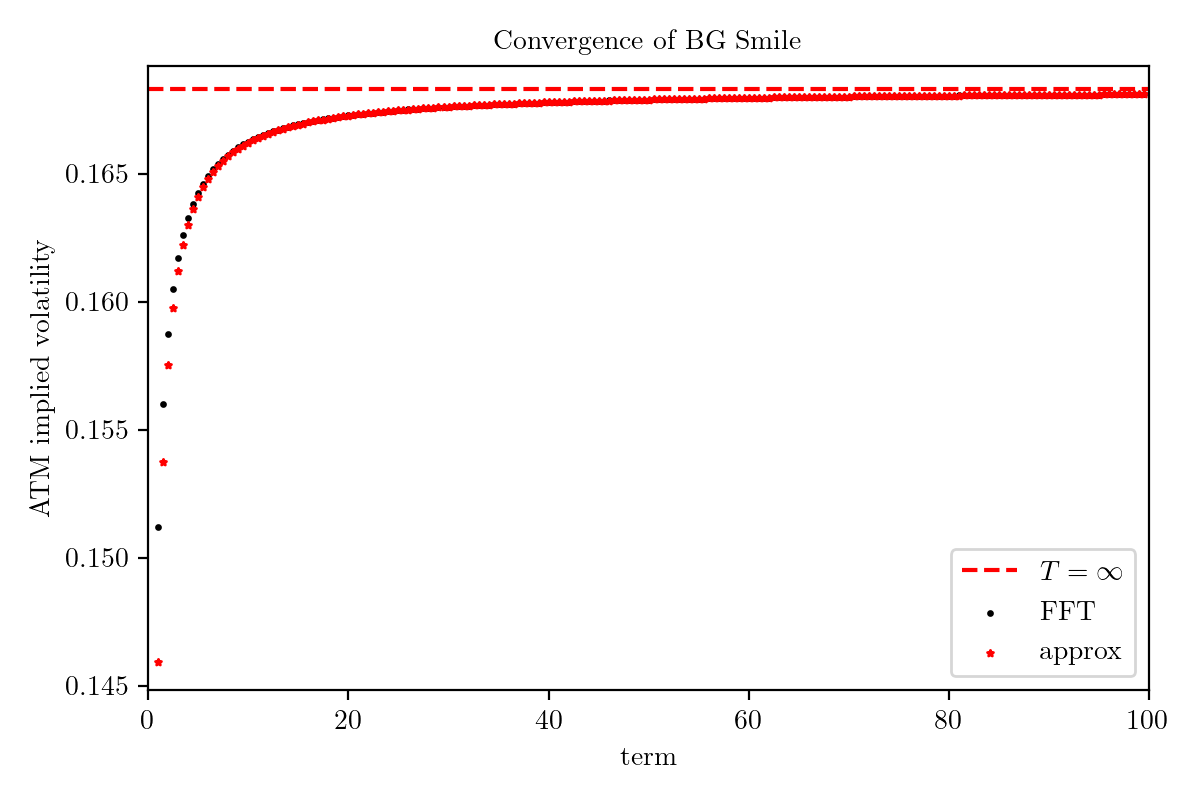} \\
\hline
\end{tabular}
\caption{Overall smile and ATM implied volatility under BG model using the improved saddle-point approximation.}
\end{table}
\end{remark}


\section{Saddle-Point Equation for Small Time} \label{apdx:sdlsmallT}

Here we obtain the analog saddle-point equation for small time $T$, where characteristic function obeys the scaling $\phi_T(u-i/2) \sim e^{-\psi(u)T}$. Consider again the integral in Lewis equation. By a change of variable $u'=uT$ for some fixed small $T$ and Taylor-expanding around an arbitrary complex point $\tilde{u}'$,
\begin{align*} 
&\int_{-\infty}^\infty \frac{du}{u^2+\frac{1}{4}} e^{-(iux+\psi(u))T} \\
=& \; \int_{-\infty}^\infty \frac{du'/T}{\left(\frac{u'}{T}\right)^2+\frac{1}{4}} e^{-\left[\frac{iu'k}{T}+\psi\left(\frac{u'}{T}\right)T\right]} \\
=& \; T \int_{-\infty}^\infty \frac{du'}{(u')^2} \left[1 - \frac{1}{4} \left(\frac{T}{u'}\right)^2...\right] e^{-\left[\frac{i\tilde{u}'k}{T}+\frac{i(u'-\tilde{u}')k}{T}+\psi\left(\frac{\tilde{u}'}{T}\right)T+\psi'\left(\frac{\tilde{u}'}{T}\right)(u'-\tilde{u}')+\frac{\psi''(\tilde{u}'/T)T}{2}(\frac{u'-\tilde{u}'}{T})^2...\right]} . \end{align*} 

Choose $\tilde{u}'$ to be the saddle-point s.t.\ $u'$-derivative of exponent vanishes
\begin{equation*}
\psi'\left(\frac{\tilde{u}'}{T}\right) = -\frac{ik}{T} .
\end{equation*}

Then the integral simplifies to
\begin{align*}
&T e^{-\left[\frac{i\tilde{u}'k}{T}+\psi\left(\frac{\tilde{u}'}{T}\right)T\right]} \int_{-\infty}^\infty \frac{du'}{(u')^2} \left[1 + O(T^2)\right] e^{-\frac{\psi''(\tilde{u}'/T)}{2T}\left(u'-\tilde{u}'\right)^2+\frac{1}{T^2} \cdot O(u'-\tilde{u}')^3} \\
\approx& \; T e^{-\left[\frac{i\tilde{u}'k}{T}+\psi\left(\frac{\tilde{u}'}{T}\right)T\right]} \frac{1}{(\tilde{u}')^2} \sqrt{\frac{2\pi}{\psi''(\tilde{u}'/T)/T}} \\
=& \; \frac{T^{3/2}}{(\tilde{u}')^2} \sqrt{\frac{2\pi}{\psi''(\tilde{u}'/T)}} e^{-\left[\frac{i\tilde{u}'k}{T}+\psi\left(\frac{\tilde{u}'}{T}\right)T\right]} .
\end{align*}

Assume that the exponent contains a term of order $T^{-1}$ so the exponential dominates. Define the time-scaled saddle-point $\tilde{u}=\tilde{u}'/T$ and the exponent reads
\begin{equation*}
\frac{i\tilde{u}'k}{T}+\psi\left(\frac{\tilde{u}'}{T}\right)T = \left(i\tilde{u}\frac{k}{T}+\psi(\tilde{u})\right)T \equiv \omega\left(\frac{k}{T}\right)T ,
\end{equation*}
thus we have the saddle-point equation for small time by matching with Black-Scholes
\begin{equation}
\lim_{T \to 0} \omega\left(\frac{k}{T}\right)T = \lim_{T \to 0} \omega_{BS}\left(\frac{k}{T}\right)T ,
\end{equation}
which is exactly our large-time saddle-point equation should the limit signs disappear.

For Black-Scholes, recall $\omega_{BS}(x)=v(x)/8+x^2/2v(x)$. Consider the smile in the space of total implied variance $w_T(k)=v(k/T)T$ versus log-strike $k$. RHS then tends to
\begin{equation*}
\lim_{T \to 0} \omega_{BS}\left(\frac{k}{T}\right)T = \lim_{T \to 0} \frac{vT}{8}+\frac{k^2}{2vT} = \frac{k^2}{2w_0(k)} .
\end{equation*}

Thus we have the small-time saddle-point equation for total implied variance:
\begin{equation}
w_0(k) = \frac{k^2}{\lim_{T \to 0} 2\omega(k/T)T} .
\end{equation}

As an illustration, consider the variance-gamma model. Taking the small-time limit, we have $\omega(k/T)T \rightarrow \left(-\xi/2\alpha + \operatorname{sgn}k \cdot \eta/\nu\right) k$, thus total implied variance
\begin{equation*} w_0(k) = \begin{cases} \frac{k}{2\left(\frac{\eta}{\nu}-\frac{\xi}{2\alpha}\right)} & k > 0 \\ -\frac{k}{2\left(\frac{\eta}{\nu}+\frac{\xi}{2\alpha}\right)} & k < 0 \end{cases} , \end{equation*}
precisely a tilted V-shape. With the typical parameters as in section \ref{sec:vgnum},
\begin{equation*} \sigma = 0.12, \quad \theta = -0.14, \quad \nu = 0.17 \end{equation*}
we show the convergence of smile $w(k,T)$ as $T$ shrinks. But a caveat is, at short time, out-of-the-money (OTM) options are cheap and their prices are insensitive to magnitude of implied volatilities, so the concept of implied volatility is less meaningful (we actually need time for the volatile stochasticity to act in). For $T$ that is too small, numerical errors, large relative to the OTM prices, dominate, giving insensible smiles, and $T=0.05$ appears to be the furthest we can probe.

\begin{table}[H]
\centering
\begin{tabular}{|c|}
\hline
VG Overall Smile at Small Time \\ 
\hline \hline
\includegraphics[width=0.4\linewidth]{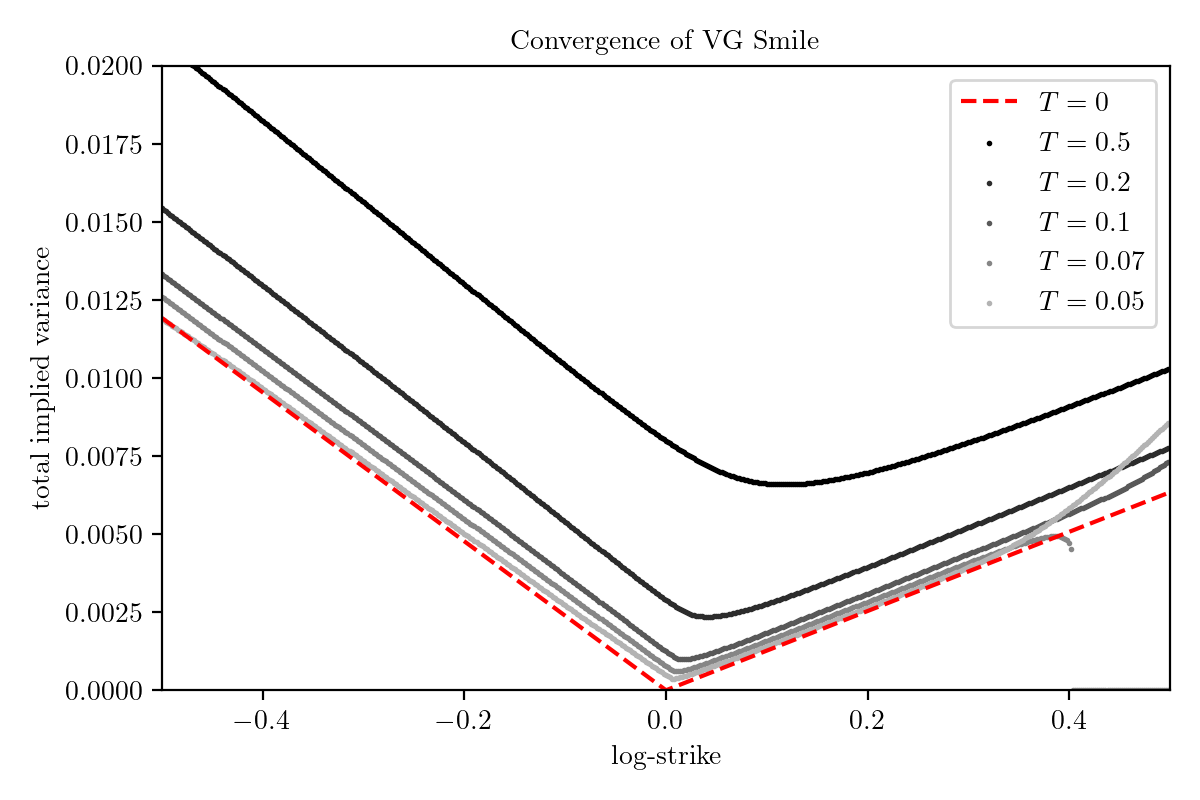} \\
\hline
\end{tabular}
\caption{Overall smile under VG model at small time.}
\end{table}

\begin{remark}
While we do not have the capacity to fully work out the mathematics, we think one can follow the expansion procedure laid out in appendix \ref{apdx:sdl} to derive the higher-order terms in $T$ for the small-time saddle-point equation. An excellent reference is \cite{jac-hessmallT}, which considers the small-time asymptotics of Heston model.
\end{remark}


\section{$\bar{\omega}(x)$ under Heston Stochastic Volatility} \label{apdx:hes}

Here we prove proposition \ref{prop:hesomegabar}. We start with the ansatz that \textit{$\bar{\omega}(x)$ is also SVI-like}, of form
\begin{equation*} \bar{\omega}(x) = - K_0 - K_1 (x-m) - K_2 \sqrt{1 + \xi^2\left(\frac{1}{A}\right)^2(x-m)^2} . \end{equation*}

Expand $\omega(x)^2-x^2/4$ using equation (\ref{eq:hesomega}):
\begin{align*}
&\left( \frac{\lambda}{\xi} \right)^2 \left( 1 - \frac{a}{2} \right)^2 + \left( \frac{B}{A^2} \right)^2 (x-m)^2 + \left( \frac{BD}{\xi} \right)^2 \left( 1 + \xi^2\left(\frac{1}{A}\right)^2(x-m)^2 \right) + \\
&2 \frac{\lambda}{\xi} \frac{B}{A^2} \left( 1 - \frac{a}{2} \right) (x-m) + 2 \frac{\lambda}{\xi} \frac{BD}{\xi} \left( 1 - \frac{a}{2} \right) \sqrt{ 1 + \xi^2\left(\frac{1}{A}\right)^2(x-m)^2 } + \\
&2 \frac{B}{A^2} \frac{BD}{\xi} (x-m) \sqrt{ 1 + \xi^2\left(\frac{1}{A}\right)^2(x-m)^2 } \\
&-\frac{1}{4} \left( (x-m)^2 + 2 m (x-m) + m^2 \right) \\
=& \; \left( \left( \frac{\lambda}{\xi} \right)^2 \left( 1 - \frac{a}{2} \right)^2 + \left( \frac{BD}{\xi} \right)^2 - \frac{m^2}{4} \right) + \left( 2 \frac{\lambda}{\xi} \frac{B}{A^2} \left( 1 - \frac{a}{2} \right) - \frac{m}{2} \right) (x-m) + \\
&\; \left( \left( \frac{B}{A^2} \right)^2 + \xi^2 \left( \frac{BD}{\xi} \right)^2 \left(\frac{1}{A}\right)^2 - \frac{1}{4} \right) (x-m)^2 + \\
&\; 2 \frac{\lambda}{\xi} \frac{BD}{\xi} \left( 1 - \frac{a}{2} \right) \sqrt{...} + 2 \frac{B}{A^2} \frac{BD}{\xi} (x-m) \sqrt{...} \\
\equiv& \; \left( - K_0 - K_1 (x-m) - K_2 \sqrt{1 + \xi^2\left(\frac{1}{A}\right)^2(x-m)^2} \right)^2
\end{align*}
thus the system of equations
\begin{align*}
K_0^2 + K_2^2 &= \left( \frac{\lambda}{\xi} \right)^2 \left( 1 - \frac{a}{2} \right)^2 + \left( \frac{BD}{\xi} \right)^2 - \frac{m^2}{4} \\
K_1^2 + K_2^2 \xi^2 \left(\frac{1}{A}\right)^2 &= \left( \frac{B}{A^2} \right)^2 + \xi^2 \left( \frac{BD}{\xi} \right)^2 \left(\frac{1}{A}\right)^2 - \frac{1}{4} \\
K_0 K_1 &= \frac{\lambda}{\xi} \frac{B}{A^2} \left( 1 - \frac{a}{2} \right) - \frac{m}{4} = \frac{\lambda}{\xi} \frac{B}{A^2} \left( 1 - \frac{a}{2} + \frac{aA^2}{4B} \right) = \frac{\lambda}{\xi} \frac{B}{A^2} \left( 1 - \frac{a}{2} \right) \left( 1 + \frac{aA^2/4B}{1-a/2} \right) \\
K_1 K_2 &= \frac{B}{A^2} \frac{BD}{\xi} \\
K_0 K_2 &= \frac{\lambda}{\xi} \frac{BD}{\xi} \left( 1 - \frac{a}{2} \right) .
\end{align*}

Five equations for three unknowns -- overconstraint! Solving from last three:
\begin{align*}
K_0 &= \frac{\lambda}{\xi} \left( 1 - \frac{a}{2} \right) K \\
K_1 &= \frac{B}{A^2} K \\
K_2 &= \frac{BD}{\xi} \frac{1}{K}
\end{align*}
with
\begin{equation*} K = \sqrt{1 + \frac{aA^2/4B}{1-a/2}} . \end{equation*}

$K_{0,1,2}$ have to automatically satisfy first two equations, if our guess $\omega^2-x^2/4=(...)^2$ is correct. This can be algebraically verified with e.g.\ mathematica. We provide a numerical check here, given our typical Heston parameters in the numerical experiment section \ref{sec:hesnum}:

\begin{verbatim}
Eq. (1): LHS = 50.76705882352938 RHS = 50.76705882352938
Eq. (2): LHS = 613.8985005767010 RHS = 613.8985005767013
\end{verbatim}

Therefore, we have
\begin{equation*} \bar{\omega}(x) = - K \frac{\lambda}{\xi} \left( 1 - \frac{a}{2} \right) - K \frac{B}{A^2} (x-m) - \frac{1}{K} \frac{BD}{\xi} \sqrt{ 1 + \xi^2\left(\frac{1}{A}\right)^2(x-m)^2 } . \end{equation*}


\section{$\bar{\omega}(x)$ under Variance-Gamma} \label{apdx:vg}

Here we prove proposition \ref{prop:vgomegabar}. We start with the ansatz that \textit{$\bar{\omega}(x)$ takes the same form as $\omega(x)$ to leading orders}:
\begin{equation*} \bar{\omega}(x) = - K_0 - K_1 (x-x_0) + K_2 \left( \sqrt{1 + \eta^2 (x-x_0)^2} - 1 \right) + K_3 \log \frac{2\alpha}{\nu^2(x-x_0)^2} \left( \sqrt{1 + \eta^2 (x-x_0)^2} - 1 \right) + ... \end{equation*}
with some higher-order corrections.

Expand $\omega^2-x^2/4$ using equation (\ref{eq:vgomega}):
\begin{align*}
\omega(x)^2 - \frac{x^2}{4} &= \left( - \frac{x_0}{2} - \frac{\xi}{2\alpha} (x-x_0) + \frac{1}{\nu} \left( \sqrt{1 + \eta^2 (x-x_0)^2} - 1 \right) \right. \\
&\quad \left. + \frac{1}{\nu} \log \frac{2\alpha}{\nu^2(x-x_0)^2} \left( \sqrt{1 + \eta^2 (x-x_0)^2} - 1 \right) \right)^2 \\
&\quad - \frac{1}{4} \left( (x-x_0)^2 + 2x_0(x-x_0) + x_0^2 \right) .
\end{align*}

We Taylor-expand $\bar{\omega}^2$ and $\omega^2-x^2/4$. First, consider $\bar{\omega}$ and note
\begin{align*}
\sqrt{1+x} &\approx 1 + \frac{x}{2} - \frac{x^2}{8} + \frac{x^3}{16} \\
\log \frac{\sqrt{1+x}-1}{x} &\approx \log \frac{1}{2} \left( 1 - \frac{x}{4} + \frac{x^2}{8} \right) \\
&\approx \log \frac{1}{2} - \left( \frac{x}{4} - \frac{x^2}{8} \right) - \frac{1}{2} \left( \frac{x}{4} - \frac{x^2}{8} \right)^2 \\
&\approx \log \frac{1}{2} - \frac{x}{4} + \frac{3x^2}{32}
\end{align*}
so
\begin{align*}
\bar{\omega}(x) &\approx - K_0 - K_1 (x-x_0) + K_2 \left( \frac{\eta^2(x-x_0)^2}{2} - \frac{\eta^4(x-x_0)^4}{8} \right) \\
&\quad + K_3 \log \frac{\alpha\eta^2}{\nu^2} + K_3 \left( - \frac{\eta^2(x-x_0)^2}{4} + \frac{3\eta^4(x-x_0)^4}{32} \right) \\
&\approx \left( - K_0 + K_3 \log \frac{\alpha\eta^2}{\nu^2} \right) - K_1 (x-x_0) + \left( K_2 - \frac{K_3}{2} \right) \frac{\eta^2}{2} (x-x_0)^2 + \left( - K_2 + \frac{3}{4} K_3 \right) \frac{\eta^4}{8} (x-x_0)^4
\end{align*}
squared to give
\begin{align*}
\bar{\omega}(x)^2 &\approx \left( - K_0 + K_3 \log \frac{\alpha\eta^2}{\nu^2} \right)^2 - 2 K_1 \left( - K_0 + K_3 \log \frac{\alpha\eta^2}{\nu^2} \right) (x-x_0) + \\
&\quad \left( 2 \left( K_2 - \frac{K_3}{2} \right) \frac{\eta^2}{2} \left( - K_0 + K_3 \log \frac{\alpha\eta^2}{\nu^2} \right) + K_1^2 \right) (x-x_0)^2 + \\
&\quad \left( -2 K_1 \left( K_2 - \frac{K_3}{2} \right) \frac{\eta^2}{2} \right) (x-x_0)^3
\end{align*}
which establishes the system of equations
\begin{align*}
\left( - K_0 + K_3 \log \frac{\alpha\eta^2}{\nu^2} \right)^2 &= \left( - \frac{x_0}{2} + \frac{1}{\nu} \log \frac{\alpha\eta^2}{\nu^2} \right)^2 - \frac{x_0^2}{4} \\
K_1 \left( - K_0 + K_3 \log \frac{\alpha\eta^2}{\nu^2} \right) &= \frac{\xi}{2\alpha} \left( - \frac{x_0}{2} + \frac{1}{\nu} \log \frac{\alpha\eta^2}{\nu^2} \right) + \frac{x_0}{4} \\
\left( K_2 - \frac{K_3}{2} \right) \frac{\eta^2}{2} \left( - K_0 + K_3 \log \frac{\alpha\eta^2}{\nu^2} \right) + \frac{K_1^2}{2} &= \frac{1}{2\nu} \frac{\eta^2}{2} \left( - \frac{x_0}{2} + \frac{1}{\nu} \log \frac{\alpha\eta^2}{\nu^2} \right) + \frac{1}{2} \left( \frac{\xi}{2\alpha} \right)^2 - \frac{1}{8} \\
K_1 \left( K_2 - \frac{K_3}{2} \right) \frac{\eta^2}{2} &= \frac{\xi}{2\alpha} \frac{1}{2\nu} \frac{\eta^2}{2}
\end{align*}
thus
\begin{equation*} - K_0 + K_3 \log \frac{\alpha\eta^2}{\nu^2} = \frac{1}{\nu} \log \frac{\alpha\eta^2}{\nu^2} \sqrt{1 - \frac{x_0\nu}{\log \frac{\alpha\eta^2}{\nu^2}}} \end{equation*}
and
\begin{equation*} K_1 = \frac{\frac{\xi}{2\alpha} \left( - \frac{x_0}{2} + \frac{1}{\nu} \log \frac{\alpha\eta^2}{\nu^2} \right) + \frac{x_0}{4}}{\frac{1}{\nu} \log \frac{\alpha\eta^2}{\nu^2} \sqrt{1 - \frac{x_0\nu}{\log \frac{\alpha\eta^2}{\nu^2}}}} = \frac{\xi}{2\alpha} \frac{1 - \frac{1}{2} \left( 1 - \frac{\alpha}{\xi} \right) \frac{x_0\nu}{\log \frac{\alpha\eta^2}{\nu^2}}}{\sqrt{1 - \frac{x_0\nu}{\log \frac{\alpha\eta^2}{\nu^2}}}} . \end{equation*}

We define
\begin{equation*} K = \frac{1 - \frac{1}{2} \left( 1 - \frac{\alpha}{\xi} \right) \frac{x_0\nu}{\log \frac{\alpha\eta^2}{\nu^2}}}{\sqrt{1 - \frac{x_0\nu}{\log \frac{\alpha\eta^2}{\nu^2}}}} \end{equation*}
so
\begin{equation*} K_1 = \frac{\xi}{2\alpha} K . \end{equation*}

Assume $K_2 = K_3$, motivated by $K_2 = K_3 = 1/\nu$ in $\omega$, giving
\begin{equation*} K_2 = K_3 = \frac{1}{\nu} \frac{1}{K} . \end{equation*}

Finally,
\begin{align*}
K_0 &= K_3 \log \frac{\alpha\eta^2}{\nu^2} - \frac{1}{\nu} \log \frac{\alpha\eta^2}{\nu^2} \sqrt{1 - \frac{x_0\nu}{\log \frac{\alpha\eta^2}{\nu^2}}} \\
&= \frac{1}{\nu} \log \frac{\alpha\eta^2}{\nu^2} \frac{1}{K} - \frac{1}{\nu} \log \frac{\alpha\eta^2}{\nu^2} \frac{1}{K} \left( 1 - \frac{1}{2} \left( 1 - \frac{\alpha}{\xi} \right) \frac{x_0\nu}{\log \frac{\alpha\eta^2}{\nu^2}} \right) \\
&= \frac{1}{\nu} \log \frac{\alpha\eta^2}{\nu^2} \frac{1}{K} \frac{1}{2} \left( 1 - \frac{\alpha}{\xi} \right) \frac{x_0\nu}{\log \frac{\alpha\eta^2}{\nu^2}} \\
&= \frac{x_0}{2} \left( 1 - \frac{\alpha}{\xi} \right) \frac{1}{K} .
\end{align*}

Therefore, we have
\begin{equation*} \begin{aligned}
\bar{\omega}(x) &\approx - \frac{x_0}{2K} \left( 1 - \frac{\alpha}{\xi} \right) - \frac{\xi}{2\alpha} K (x-x_0) + \frac{1}{K\nu} \left( \sqrt{1 + \eta^2 (x-x_0)^2} - 1 \right) \\
&\quad + \frac{1}{K\nu} \log \frac{2\alpha}{\nu^2(x-x_0)^2} \left( \sqrt{1 + \eta^2 (x-x_0)^2} - 1 \right) . 
\end{aligned} \end{equation*}


\section{BGI Calibration to SPX 2022} \label{apdx:bgi}

Here we calibrate BGI parametrization to SPX vanilla options on November 7, 2022, which contains 50 expiries totaling 9578 out-of-the-money options data points, obtained from CBOE delayed quotes as of market close on that day. Smiles of medium to long expiries $T>0.2$ are considered. By ensuring appropriate monotonic term-structures in $T\alpha_\pm(T)$ and $\lambda_\pm(T)$, we guarantee the absence of calendar arbitrage. As BGI is derived out of a risk-neutral stochastic process, arbitrage along strike, namely butterfly arbitrage, is also absent. The parameters for the put wing have the following monotonic trends over terms.
\begin{table}[H]
\centering
\begin{tabular}{|c|c|}
\hline
$T\alpha_-(T)$ & $\lambda_-(T)$ \\ 
\hline \hline
\includegraphics[width=0.4\linewidth]{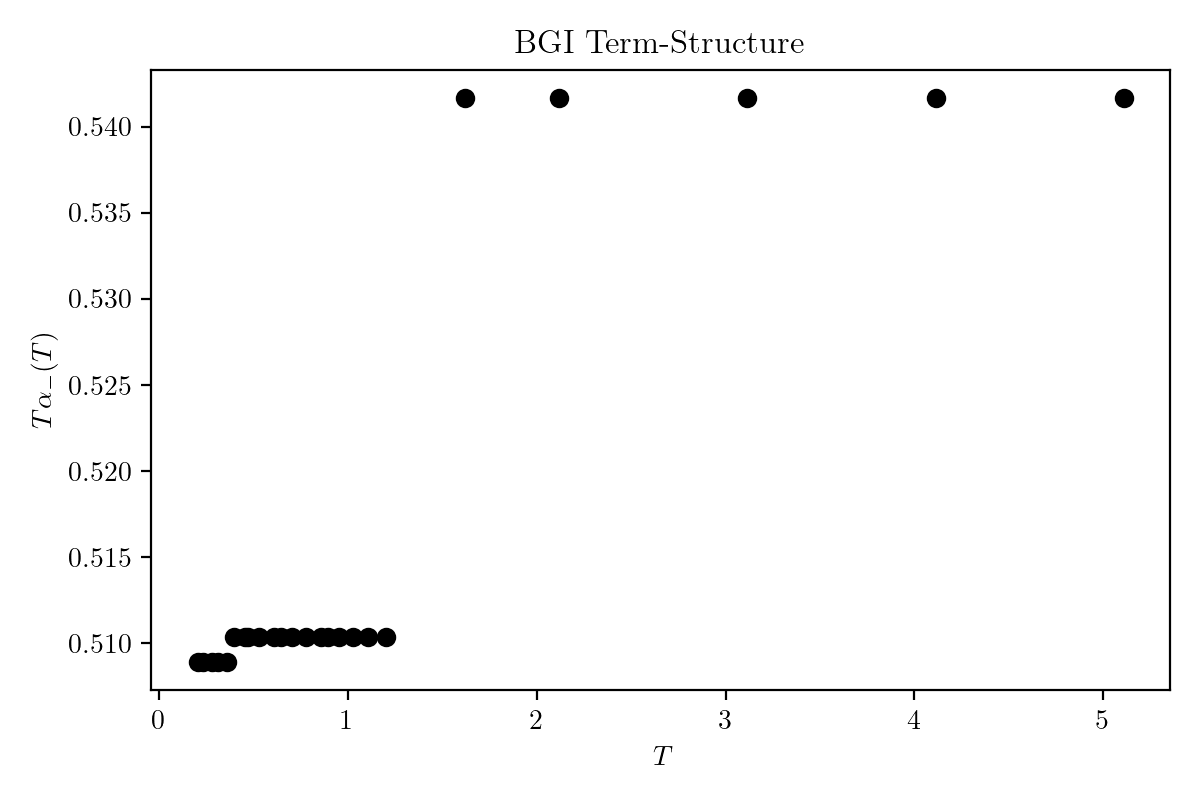} & 
\includegraphics[width=0.4\linewidth]{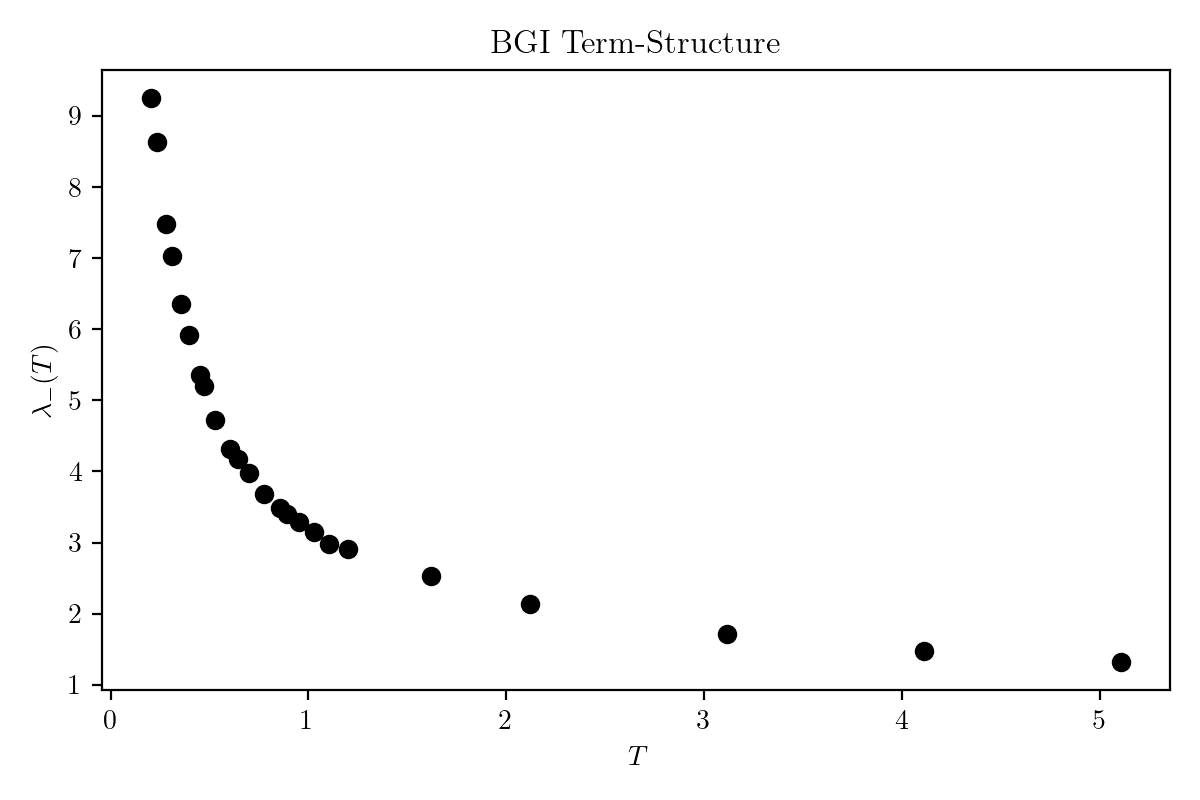} \\ 
\hline
\end{tabular}
\caption{BGI put-wing parameter term-structures.}
\end{table}

As the BGI implied densities are positive everywhere, and the BGI total implied variance slices have no crossing everywhere, butterfly and calendar arbitrage are absent.
\begin{table}[H]
\centering
\begin{tabular}{|c|c|}
\hline
Butterfly Arbitrage & Calendar Arbitrage \\ 
\hline \hline
\includegraphics[width=0.4\linewidth]{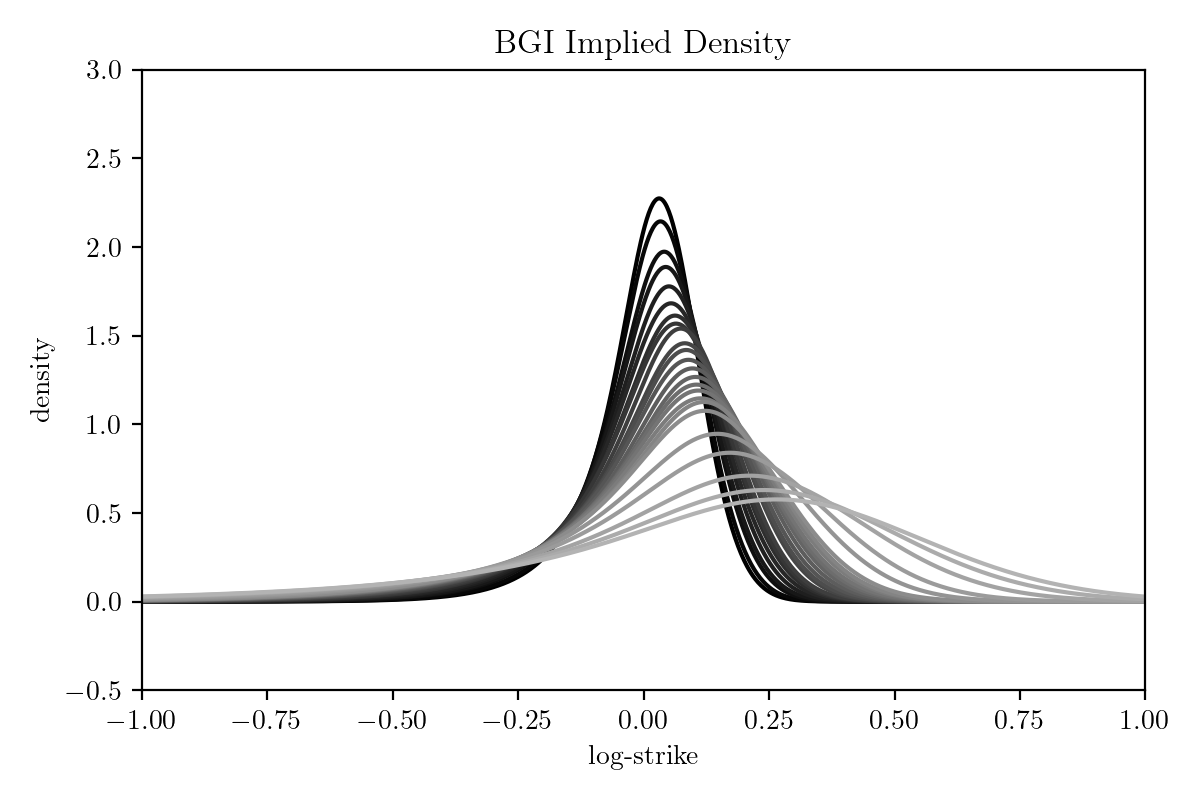} & 
\includegraphics[width=0.4\linewidth]{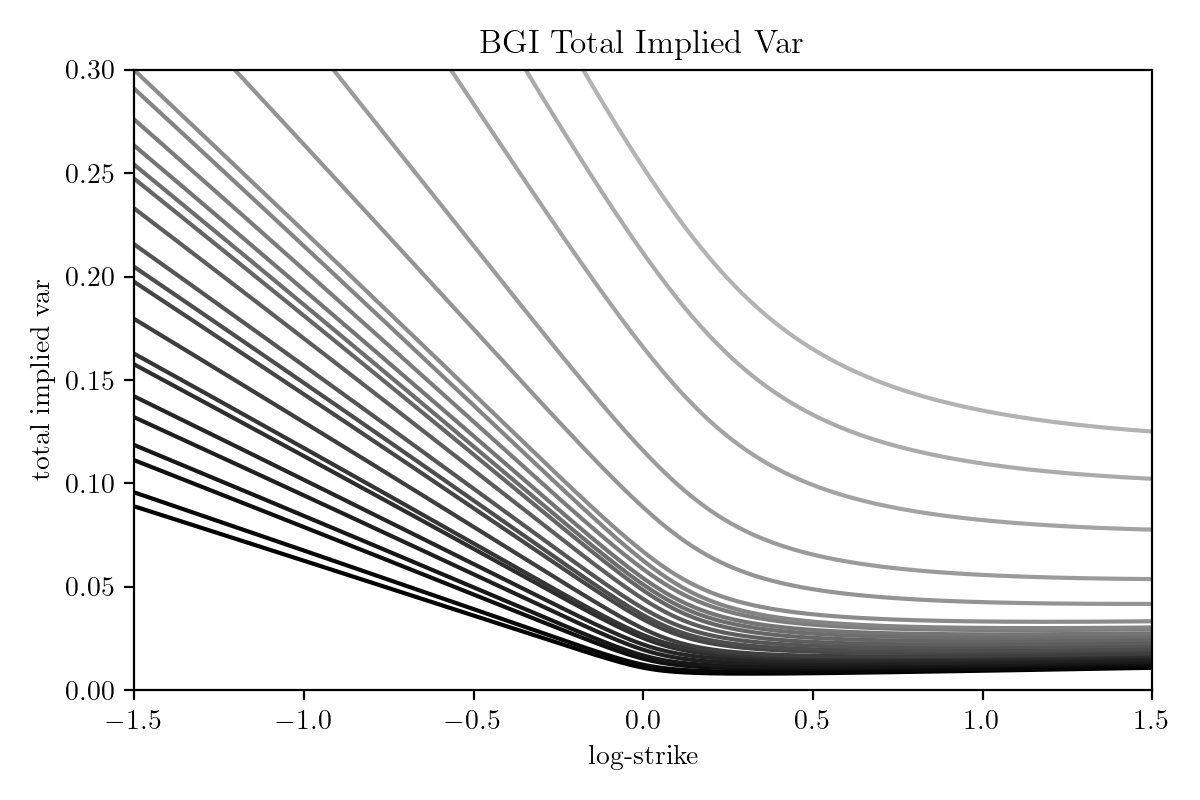} \\ 
\hline
\end{tabular}
\caption{Absence of butterfly and calendar arbitrage under BGI.}
\end{table}

The BGI fits over terms are as follows. It is observed that compared to the market, (1) both the call and put wings flatten out too quickly and the check shape offered by BGI is inadequate for fitting the market smiles, and (2) BGI best describes long expiries but falls apart for short expiries. As pointed out in subsection Some Open Questions in the conclusion, this could be evidence that market expects a Markov evolution of log-spot at large time but strong path-dependence at small time, so the BG smile, derived out of a Lévy-type scaling in characteristic exponent, is no longer adequate.
\begin{figure}[H]
\centering
\includegraphics[width=0.8\linewidth]{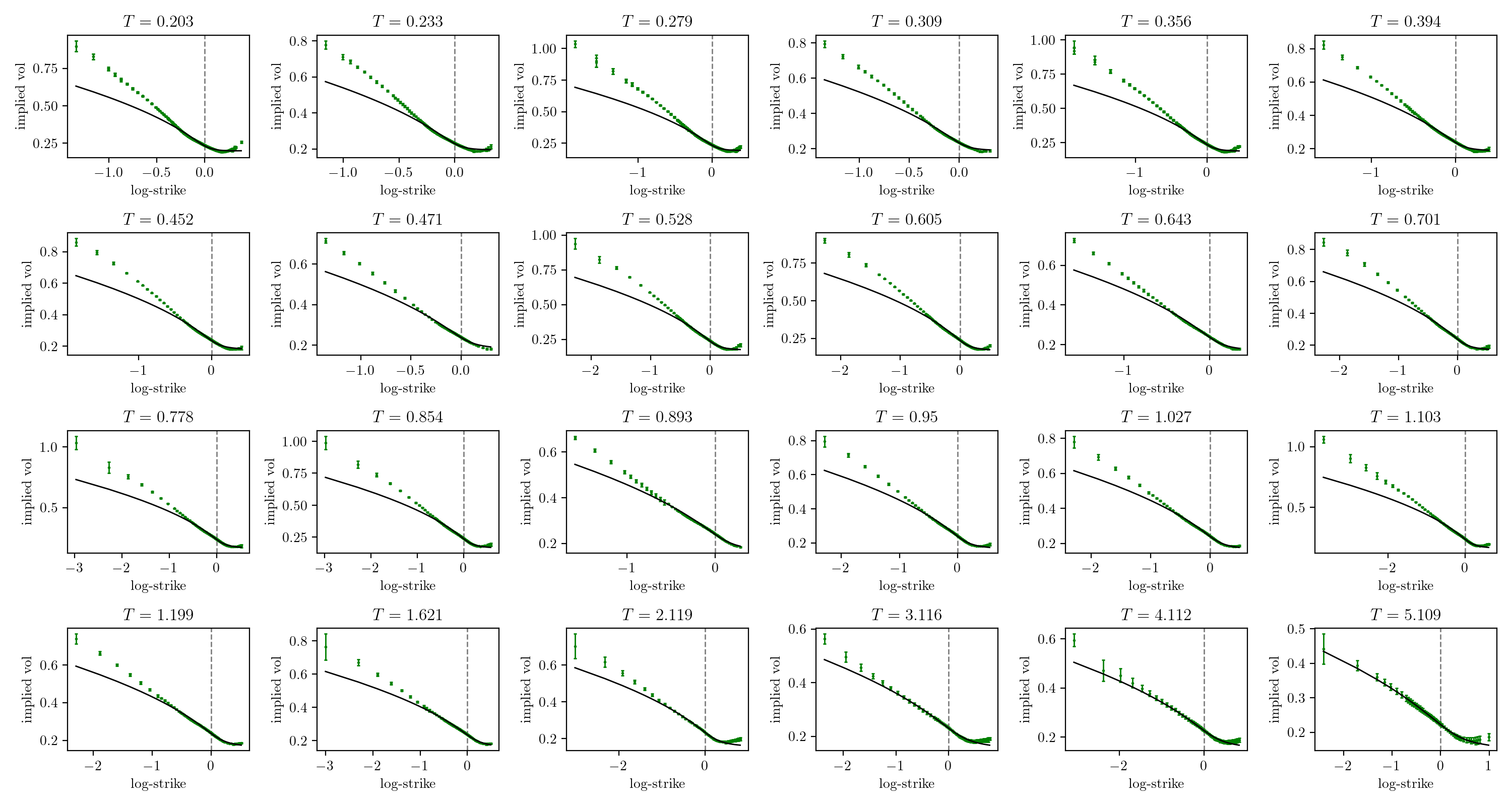}
\caption{BGI smiles calibrated to bid-ask volatilities.}
\end{figure}

Defining $\omega^*(x,T) \equiv v(x,T)/8+x^2/2v(x,T)$, we also consider the smiles both in $\sigma$-$k$ and $\omega^*$-$x$ space. Here, all 50 expiries are plotted, and we observe that (1) for $T$ large, a strictly convex curve that touches $|x/2|$ may be fitted inside the bid-ask spreads, (2) for $T$ small, a curve that is convex near-the-money but has concave wings tending to linear is needed, and (3) for $T$ small, the volatility smiles show some flattening or W-shaped behavior near-the-money but transformation to $\omega^*$ removes that. Is $\omega^*$ a more natural quantity to describe a volatility smile?

\begin{figure}[H]
\centering
\includegraphics[width=\linewidth]{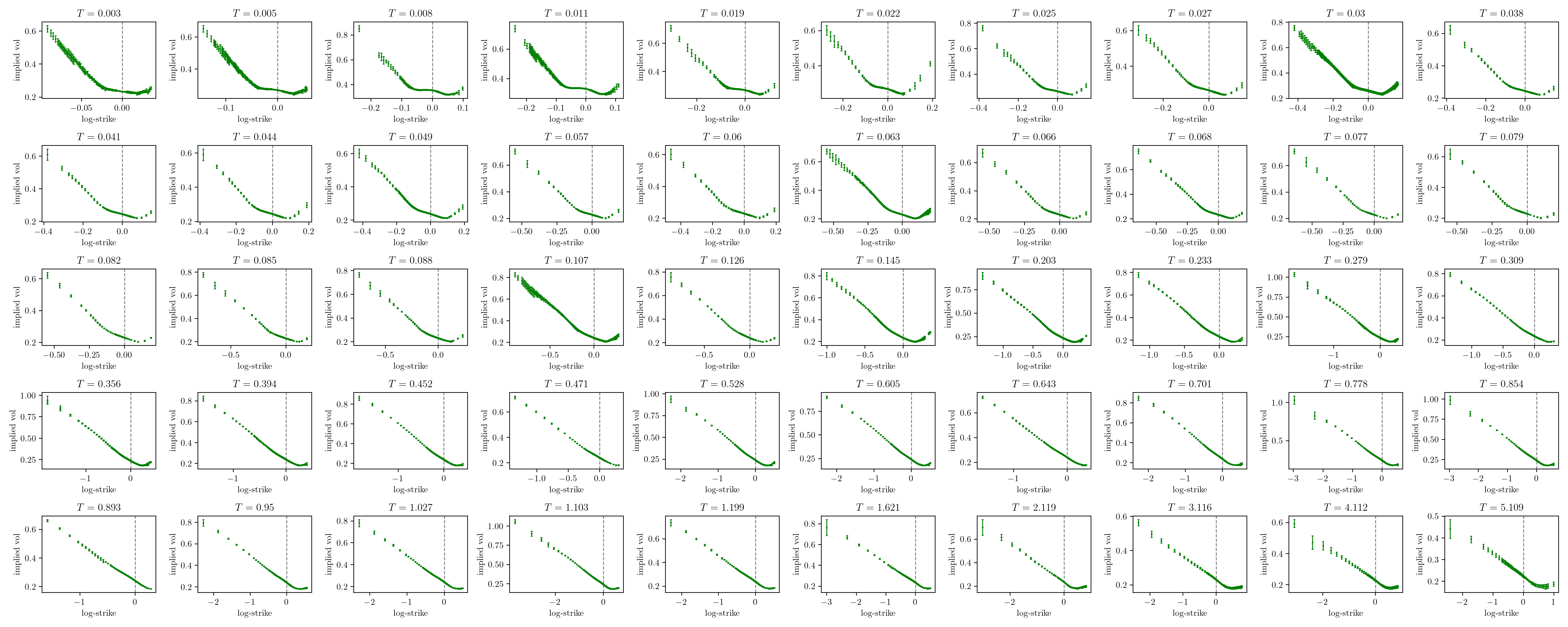}
\end{figure}

\begin{figure}[H]
\centering
\includegraphics[width=\linewidth]{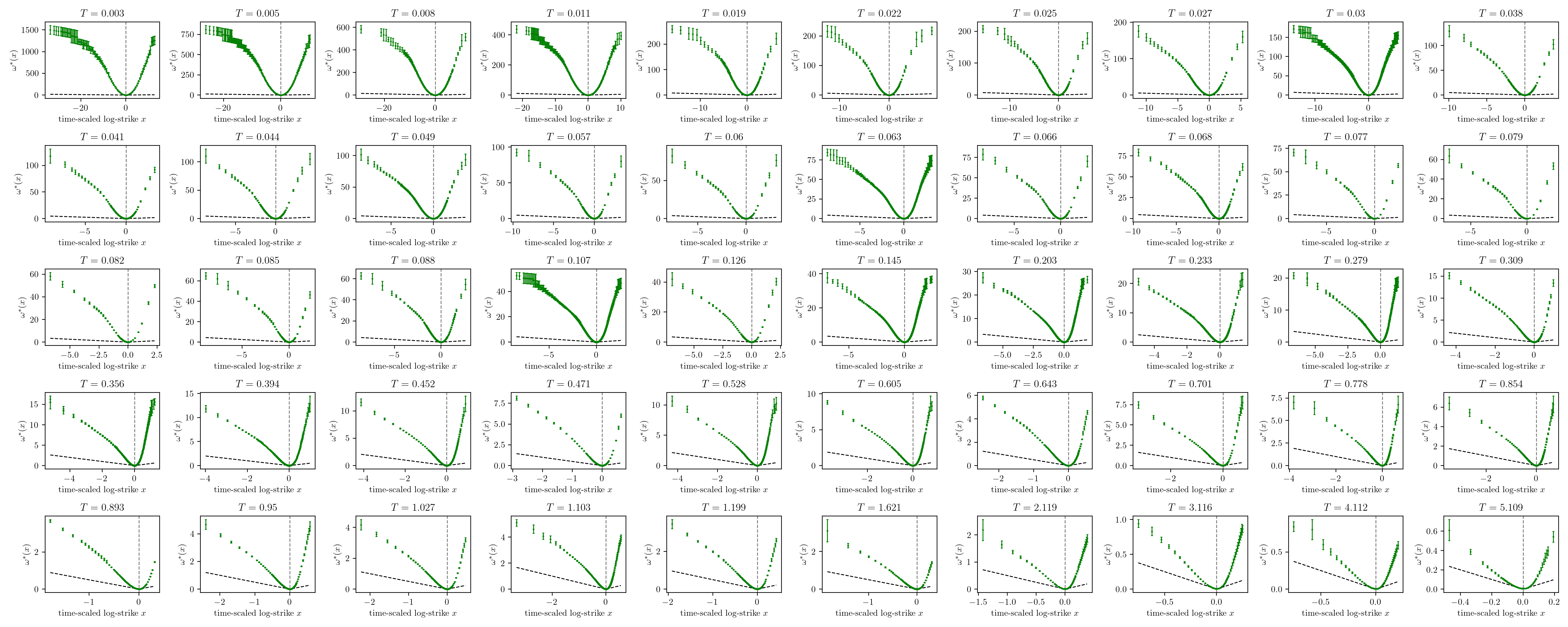}
\end{figure}

\begin{figure}[H]
\centering
\includegraphics[width=\linewidth]{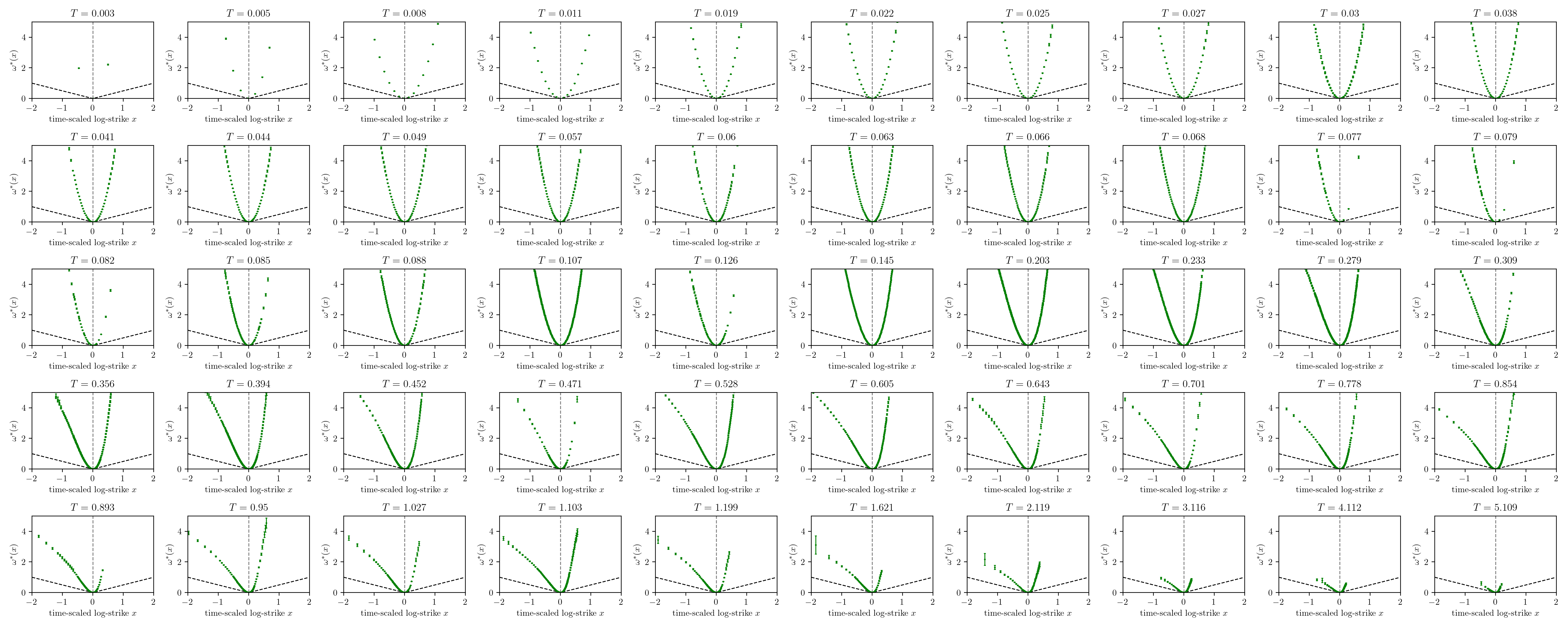}
\end{figure}

\end{document}